\pdfoutput=1
\documentclass[acmsmall,screen,nonacm]{pcqec-arxiv}
\newif\ifappx
\appxtrue

\AtBeginDocument{%
  }

\setcopyright{none}
\setcitestyle{citesep={,}}
\makeatletter
\patchcmd{\ccsdesc@parse}{; }{, }{}{\ClassError{acmart}{CCS punctuation patch failed}{}}
\patchcmd{\ccsdesc@parse}{; }{, }{}{\ClassError{acmart}{CCS punctuation patch failed}{}}
\makeatother

\usepackage{booktabs}
\usepackage{subcaption}
\usepackage{algorithm}
\usepackage[noend]{algpseudocode}
\usepackage{amsmath}
\usepackage{stmaryrd}
\usepackage{xurl}
\usepackage{graphicx}

\newsavebox{\pccdisplaybox}
\newcommand{\pccfitmath}[2][\dimexpr\linewidth-5em\relax]{%
  \begingroup
  \sbox{\pccdisplaybox}{$\displaystyle #2$}%
  \ifdim\wd\pccdisplaybox>#1\relax
    \resizebox{#1}{!}{\usebox{\pccdisplaybox}}%
  \else
    \usebox{\pccdisplaybox}%
  \fi
  \endgroup}

\usepackage{flafter}
\usepackage{placeins}
\usepackage{wrapfig}
\newcommand{\pcctight}{\looseness=-1\relax}

\newcommand{\pccwrapdisplay}[5]{%
 \begin{wrapfloat}{figure}[#1]{r}{#2}%
  \noindent\hbox to\linewidth{\hfil
   \pccfitmath[\dimexpr\linewidth-2.4em\relax]{#5}\hfil
   \refstepcounter{equation}\label{#3}%
   \smash{\raisebox{-#4}{\textup{(\theequation)}}}}%
 \end{wrapfloat}}
\usepackage{fancyvrb}
\usepackage{paralist}
\usepackage{caption}
\usepackage{comment}
\usepackage{xspace}
\usepackage{multirow}
\usepackage{longtable}
\usepackage{booktabs}
\usepackage{enumerate}
\usepackage{enumitem}
\usepackage{caption}
\usepackage{graphicx}
\usepackage{tikz}
\usetikzlibrary{arrows.meta}
\usepackage{pgfplots}
\pgfplotsset{compat=1.18}
\usepgfplotslibrary{groupplots}
\usepackage{ifsym}
\usepackage{circledsteps}
\usepackage{mdframed}
\usepackage{mathpartir}
\usepackage{extarrows}
\usepackage{stmaryrd}
\usepackage{bbding}

\tikzset{
  >=Latex,
  node distance=2.2cm,
  every node/.style={font=\small},
  dot/.style={circle,fill,inner sep=1pt}
}
\usepackage[most]{tcolorbox}

\newenvironment{figureframe}{%
 \begin{tcolorbox}[enhanced,sharp corners,
  colback=white,colframe=black!40,boxrule=.25pt,boxsep=0pt,
  left=0pt,right=0pt,top=.5pt,bottom=.5pt,
  before skip=0pt,after skip=0pt,width=\linewidth]
}{\end{tcolorbox}}

\usetikzlibrary{positioning, calc}
\usepackage{xcolor}

\usepackage{booktabs, array, pifont}

\newcommand{\cmark}{\ding{51}} 
\newcommand{\xmark}{\ding{55}} 
\newcolumntype{C}{>{\centering\arraybackslash}m{2.1em}} 

\usetikzlibrary{arrows}

\usepackage[
    type={CC},
    modifier={by},
    version={4.0},
]{doclicense}

\usepackage{tikz}
\usetikzlibrary{quantikz2}

\definecolor{pcqedCodeKeyword}{RGB}{28,82,148}
\definecolor{pcqedFaultX}{RGB}{35,100,185}
\definecolor{pcqedFaultY}{RGB}{184,108,0}
\definecolor{pcqedFaultZ}{RGB}{191,35,45}
\newcommand{\faultX}{\textcolor{pcqedFaultX}{X}}
\newcommand{\faultY}{\textcolor{pcqedFaultY}{Y}}
\newcommand{\faultZ}{\textcolor{pcqedFaultZ}{Z}}
\newcommand{\faultchoices}{\faultX\,\faultY\,\faultZ}
\newcommand{\faultpanel}[3]{%
 \begin{tikzpicture}[x=1pt,y=1pt,baseline={\dimexpr#2/2-.5ex\relax}]
  \pgfmathsetlengthmacro{\faultthird}{#1/3}
  \fill[pcqedFaultX] (0,0) rectangle (\faultthird,#2);
  \fill[pcqedFaultY] (\faultthird,0) rectangle (2*\faultthird,#2);
  \fill[pcqedFaultZ] (2*\faultthird,0) rectangle (#1,#2);
  \foreach \faultletter/\faultcolumn in {X/0,Y/1,Z/2}
   \node[draw=none,fill=none,text=white,inner sep=0pt,outer sep=0pt,
     minimum width=0pt,minimum height=0pt,
     font=\sffamily\bfseries\fontsize{#3}{#3}\selectfont]
    at ({(\faultcolumn+.5)*\faultthird},.5*#2) {\faultletter};
 \end{tikzpicture}}

\tikzset{fault quantikz/.style={dashed,path picture={
 \fill[pcqedFaultX] (path picture bounding box.south west)
  rectangle ($(path picture bounding box.north west)!1/3!(path picture bounding box.north east)$);
 \fill[pcqedFaultY] ($(path picture bounding box.south west)!1/3!(path picture bounding box.south east)$)
  rectangle ($(path picture bounding box.north west)!2/3!(path picture bounding box.north east)$);
 \fill[pcqedFaultZ] ($(path picture bounding box.south west)!2/3!(path picture bounding box.south east)$)
  rectangle (path picture bounding box.north east);
 \foreach \faultletter/\faultposition in {X/0.166667,Y/0.5,Z/0.833333}
  \node[draw=none,fill=none,text=white,inner sep=0pt,outer sep=0pt,
   minimum width=0pt,minimum height=0pt,
   font=\sffamily\bfseries\fontsize{6.5}{6.5}\selectfont]
   at ($(path picture bounding box.west)!\faultposition!(path picture bounding box.east)$) {\faultletter};
}}}
\tikzset{fault width/.initial=12pt,fault height/.initial=12pt,
 fault font/.initial=5,
 fault box/.style={minimum width=\pgfkeysvalueof{/tikz/fault width},
                  minimum height=\pgfkeysvalueof{/tikz/fault height}}}
\newcommand{\faultmark}[2]{\node[draw=black!65,dashed,fill=white,fault box,
  inner sep=0pt,outer sep=0pt] at (#1,#2)
 {\faultpanel{\pgfkeysvalueof{/tikz/fault width}}
             {\pgfkeysvalueof{/tikz/fault height}}
             {\pgfkeysvalueof{/tikz/fault font}}};}

\newcommand{\SecLink}[1]{\hyperref[#1]{\S\ref*{#1}}}
\newcommand{\ThmLink}[1]{\hyperref[#1]{Thm.~\ref*{#1}}}
\newcommand{\DefLink}[1]{\hyperref[#1]{Def.~\ref*{#1}}}
\newcommand{\FigLink}[1]{\hyperref[#1]{Fig.~\ref*{#1}}}
\newcommand{\JdgLink}[1]{\hyperref[#1]{Judg.~\ref*{#1}}}
\newcommand{\Nref}[2]{\hyperref[#1]{\textcolor{black}{#2}}}

\makeatletter
\algdef{SE}[FOR]{ForEach}{EndForEach}[1]{\algorithmicfor\ \textbf{each} #1 \algorithmicdo}{\algorithmicend\ \algorithmicfor}%
\makeatother

\newcommand{\prog}[1]{\mathtt{#1}}
\newcommand{\kw}[1]{\textcolor{pcqedCodeKeyword}{\mathtt{#1}}}
\newcommand{\assertion}[1]{\mathsf{#1}}
\SetSymbolFont{stmry}{bold}{U}{stmry}{m}{n}
\newcommand{\srcstate}{{\boldsymbol{\sigma}}}
\newcommand{\tgtstate}{{\boldsymbol{\Sigma}}}
\newcommand{\statecfg}{{\boldsymbol{\gamma}}}

\newcommand{\pseq}{\mathbin{;}}

\newcommand{\scheme}{\xi}
\newcommand{\cleanstate}{\tgtstate_{\mathrm{clean}}}
\newcommand{\wt}[2][]{\operatorname{wt}_{#1}(#2)}
\newcommand{\denote}[1]{\llbracket #1\rrbracket}

\newcommand{\circuitcompile}[2]{\mathsf{Compile}_{#1}\llbracket #2\rrbracket}
\newcommand{\nquot}[2]{\lfloor #1/#2\rfloor}
\newcommand{\nrem}[2]{(#1\bmod #2)}
\definecolor{pcqedSigX}{RGB}{63,91,176}
\definecolor{pcqedSigZ}{RGB}{191,35,45}
\makeatletter
\newcommand{\pcc@signatureglyph}[2]{%
  \begingroup
  \ifnum#2=0\relax\def\pccsigcolor{pcqedSigX}%
    \else\def\pccsigcolor{pcqedSigZ}\fi
  \setbox\z@=\hbox{$\m@th#1X$}%
  \dimen@=1.60\ht\z@
  \vcenter{\hbox{\resizebox{!}{\dimen@}{%
    \tikz[x=10pt,y=10pt,line width=.45pt]{%
      \filldraw[fill=pcqedSigZ!18,draw=pcqedSigZ!65]
        (0,0) rectangle (1,-1) (1,-1) rectangle (2,-2);
      \filldraw[fill=pcqedSigX!18,draw=pcqedSigX!65]
        (1,0) rectangle (2,-1) (0,-1) rectangle (1,-2);
      \filldraw[fill=pcqedSigX!18,draw=pcqedSigX!65]
        (0,0) arc[start angle=180,end angle=0,radius=.5] -- cycle;
      \filldraw[fill=pcqedSigX!18,draw=pcqedSigX!65]
        (1,-2) arc[start angle=180,end angle=360,radius=.5] -- cycle;
      \filldraw[fill=pcqedSigZ!18,draw=pcqedSigZ!65]
        (2,0) arc[start angle=90,end angle=-90,radius=.5] -- cycle;
      \filldraw[fill=pcqedSigZ!18,draw=pcqedSigZ!65]
        (0,-1) arc[start angle=90,end angle=270,radius=.5] -- cycle;
      \foreach \pcccut in {0,1,2} {%
        \ifnum#2=0\relax
          \draw[\pccsigcolor,line width=1.1pt] (\pcccut,.12)--(\pcccut,-2.12);
        \else
          \draw[\pccsigcolor,line width=1.1pt] (-.12,-\pcccut)--(2.12,-\pcccut);
        \fi}
      \foreach \pccx in {0,1,2}
        \foreach \pccy in {0,-1,-2}
          \fill[black] (\pccx,\pccy) circle[radius=.085];
    }}}}%
  \endgroup}
\DeclareRobustCommand{\colsig}{\Nref{eq:surface-main-cuts}{\chi_x}}
\DeclareRobustCommand{\rowsig}{\Nref{eq:surface-main-cuts}{\chi_z}}
\makeatother
\definecolor{pcqedParityOne}{RGB}{191,35,45}
\DeclareRobustCommand{\sigstrip}[1]{\mathord{\Nref{eq:surface-cut-values}{%
  \tikz[baseline=-.45ex,x=.19cm,y=.19cm]{%
    \edef\pccstripbits{#1}%
    \foreach \pccbit [count=\pccslot from 0] in \pccstripbits {%
      \ifnum\pccbit=1\relax\def\pccfill{pcqedParityOne}\else\def\pccfill{white}\fi
      \filldraw[fill=\pccfill,draw=black!65,line width=.35pt]
        (\pccslot,0) rectangle +(1,.9);}}}}}
\makeatletter
\let\RuleLabel\label
\newcommand{\RuleDef}[1]{%
  \begingroup
  \ifmeasuring@\else
    \phantomsection\def\@currentlabel{\textsc{#1}}\RuleLabel{rule:#1}%
  \fi
  \textsc{#1}\endgroup}
\newcommand{\Rule}[1]{\ref{rule:#1}}
\makeatother
\allowdisplaybreaks[1]
\newcommand{\PCQEC}{\textsf{PC-QEC}\xspace}
\ifdefined\PCQECSupplement
\usepackage{xr-hyper}
\makeatletter

\renewcommand{\Nref}[2]{%
  \ifcsname r@#1\endcsname\hyperref[#1]{\textcolor{black}{#2}}%
  \else\hyperref[main-#1]{\textcolor{black}{#2}}\fi}
\renewcommand{\Rule}[1]{%
  \ifcsname r@rule:#1\endcsname\ref{rule:#1}%
  \else\ref{main-rule:#1}\fi}

\@for\pccindent:=-1,0,1,2,3\do{\expandafter\let\csname r@tocindent\pccindent\endcsname\relax}
\let\r@TotPages\relax
\makeatother
\fi

\begin{document}
\ifdefined\PCQECSupplement
\title{Supplement: Parametric Proof-Carrying Compilation for Quantum Error-Correction}
\maketitle
\raggedbottom
\else
\title{Parametric Proof-Carrying Compilation for Quantum Error-Correction}
\author{John Zhuoyang Ye}
\affiliation{%
  \institution{University of California Los Angeles}
  \city{Los Angeles}
  \country{USA}}
\email{yezhuoyang@cs.ucla.edu}

\author{Jens Palsberg}
\affiliation{%
  \institution{University of California Los Angeles}
  \city{Los Angeles}
  \country{USA}}
\email{palsberg@cs.ucla.edu}
\renewcommand{\shortauthors}{John Zhuoyang Ye, Jens Palsberg}

\begin{abstract}
We present \PCQEC, a proof-carrying compilation framework for quantum error
correction parameterized by both code distance and syndrome-extraction
scheme. It covers all four schemes---Standard, Flag, Shor and Knill---with
circuit-level distance guarantees for every legal code distance in each
proved family under the declared error and readout contracts. The circuit-level
model permits single-wire Pauli errors at every operation's qubit operands, capturing
data errors, readout errors and ancilla errors that propagate into
multi-qubit hook errors. For each family and scheme, supplied code-algebra
and invariant proofs combine with proved scheme-specific propagation lemmas
into a finite symbolic certificate whose
size is independent of code distance. We demonstrate family-wide guarantees for five code families
across all four schemes. For rotated Surface code, one Standard Scheme
certificate proves a distance lower bound $d$ for every odd $d\ge3$ and checks
in 214\,s on average. Lean 4 mechanizes the metatheory and exact circuit-level
distances for canonical rotated Surface, unrotated Surface and Toric code families.
\end{abstract}


\maketitle
\begingroup
\let\pccMainNormalSize\normalsize
\renewcommand{\normalsize}{\pccMainNormalSize
\setlength{\abovedisplayskip}{3.5pt plus 1pt minus .5pt}
\setlength{\belowdisplayskip}{3.5pt plus 1pt minus .5pt}
\setlength{\abovedisplayshortskip}{0pt plus 1pt}
\setlength{\belowdisplayshortskip}{2pt plus 1pt minus .5pt}}
\normalsize
\section{Introduction}
\label{sec:intro}

A fault-tolerant quantum computer could factor large integers and simulate
quantum systems~\cite{shor1997polynomial,lloyd1996universal}.
However, current quantum computers are all noisy, and long quantum computations, such as Shor's factoring algorithm, must remain
reliable despite noisy operations~\cite{shor1997polynomial,gidney2021factor}.
Quantum error correction (QEC) codes were invented to protect logical qubits by redundantly encoding them across
physical qubits so that errors can in principle be detected and corrected. As a computation
grows longer, it requires a lower logical failure probability under the chosen
noise model~\cite{gottesman2009QECIntro}.

Despite a lot of good QEC code candidates for large-scale fault-tolerant quantum computers, it is still an open problem, in general, how to optimally compile a quantum circuit that implements a quantum error correction code designed in code theory in a fault-tolerant way~\cite{rodatz2025construction,liu2026alphasyndrome,prophunt}. A syndrome extraction circuit is designed to detect errors without revealing the encoded information. In a syndrome extraction circuit, stabilizers are measured by extra ancilla qubits, also called helper qubits.
The resulting bits after measuring these helper qubits form syndrome bits, which a classical decoder uses to choose a
correction~\cite{gottesman2009QECIntro,fowler2012surface}. An extraction scheme specifies the ancilla qubits, gates and
measurements used to measure each stabilizer. A detector tests a parity of
measurement outcomes that is fixed in an error-free run and triggers when this
parity changes~\cite[\S~5.6]{gidney2021stim}.

A big challenge in quantum error correction compilation is that quantum errors can happen anywhere in the circuit, not just on data qubits that are designed to protect quantum information. The syndrome extraction circuit can itself introduce extra errors that a QEC code designer might not expect. As an example, a measurement flip can
cancel evidence of a data error, and a single error on an ancilla qubit can
propagate to several data qubits, producing a hook error~\cite{fowler2012surface,tomita2014low,kishony2026offthehook}. Circuit-level code
distance captures the resulting vulnerability: it is the minimum number of
injected Pauli errors at the circuit's injection sites that can produce a logical
data error without triggering a detector~\cite{rodatz2025construction,webster2026distancefinding}. It is a well-known fact that although quantum error correction codes, such as the famous Surface code, have code distance $d$ by design, a badly compiled circuit with hook errors can reduce the circuit-level code distance by half in the worst case~\cite{kishony2026offthehook}.

We are interested in studying a formal proof of circuit-level code distance, because evidence shows that provable circuit-level distance supports more scalable testing of a quantum error correction circuit.
ScaLER~\cite[\S\S~3, 6.1]{ye2026scalable} takes circuit-level distance $d$
as input and uses it to set a zero-failure cutoff at
$\lfloor(d-1)/2\rfloor$ errors in its extrapolation model.
Justifying this cutoff requires a distance guarantee for the implemented
circuit together with a suitable decoder-correctness argument.
For matching error and readout contracts, a proof of
$d^{\mathrm{circ}}\ge d$ supplies the distance premise. \emph{But how can we verify this property in the QEC compilation stack, given that quantum errors can be injected anywhere?}\pcctight

Computing circuit-level code distance for a given compiled circuit is NP-hard in general. A circuit with
one Pauli injection site per data qubit followed by noiseless extraction recovers
the static code-distance
problem~\cite{kapshikar2023hardness,rodatz2025construction}. Thus, there is a serious scalability barrier to formally verifying code distance at the scale we want to achieve. Code families
introduce a further difficulty. Increasing $d$ changes the
qubits, measurements and emitted gates. A proof for $d=3$, for example,
establishes only that instance, while repeating the argument gate by gate
produces a proof that grows with $d$. To break this scalability barrier, we therefore seek one
finite proof for the whole family that remains valid through compilation,
including the error effects introduced by the extraction circuit. A starting point is to try to take advantage of the knowledge about a code family with a clear geometric structure at the code level, such as the code distance for the Surface code. And we ask the following key research question:

\emph{How can we turn a code family's geometric structure into a finite proof
that its distance lower bound survives syndrome-extraction compilation?} We propose
\PCQEC (parametric proof-carrying quantum error correction) to attack this challenge. At a high level, \PCQEC expresses the
geometry using algebraic parity probes, called cuts, and translates a supplied
cut-invariant certificate alongside the extraction program. For the rotated Surface
code, which arranges qubits on a grid~\cite{fowler2012surface}, code algebra
shows that every logical residual crosses $d$ cuts in some collection.
The program invariant shows that $\ell$ errors cross at most $\ell$ cuts in
each collection:
\begin{equation}\label{eq:cut-bridge}
\pccfitmath{
\underbrace{\text{logical error: $d$ cuts in some collection}}_{\text{code algebra (Lemma~\ref{lemma:surface-main-algebra})}}
 \land\underbrace{\text{$\ell$ errors: $\le\ell$ cuts per collection}}_{\text{program invariant (Thms.~\ref{thm:qhl-sound}, \ref{thm:standard-compile-preserve})}}
 \ \Longrightarrow\ d^{\mathrm{circ}}\ge d.
}
\end{equation}

Figure~\ref{fig:cut-bridge}(a,b) illustrates why parity is a good idea: a logical Z row
and a hook error both affect three data qubits, but cross three columns and
one column, respectively. Row probes supply the corresponding measure for X
errors. A bulk stabilizer generator couples an ancilla qubit to four nearby
data qubits~\cite{beverland2022surface}. N-shaped and Z-shaped coupling orders
direct hook errors across logical error
strings~\cite{tomita2014low,kishony2026offthehook}, enabling the one-cut bound.
Cuts require no additional measurements. The bound excludes logical data
errors for every record, hence also for detector-silent executions under
Definition~\ref{def:dcirc}. Equality additionally needs a matching execution
(\S~\ref{sec:standard-exactness}).

As shown in Figure~\ref{fig:pcc}, the \PCQEC pipeline contains three different layers with different purposes.  At the top layer, CodeML is a language that allows users to implement QEC codes parametrically at the code level without worrying about the compilation. It supplies the ordered family and code-algebra premises. In the middle layer, we introduce an intermediate representation language called QStab. QStab expresses
the invariant at stabilizer-measurement boundaries, and each backend justifies
the circuit error effects needed to compile it. Following \emph{Fault
Tolerance by Construction}~\cite{rodatz2025construction}, this separates the
high-level bound from its implementation. Finally, at the bottom layer, the language QClifford works at the standard Clifford circuit level. \PCQEC translates proofs from the QStab layer to the QClifford layer under the
declared error and readout contracts for the Standard Scheme ($\mathsf{Std}$),
Flag Scheme ($\mathsf{Flag}$), Shor Scheme ($\mathsf{Shor}$) and Knill Scheme
($\mathsf{Knill}$). This specialized translation reuses supplied source proofs
and separately proved backend lemmas (Section~\ref{sec:compiler}). Following proof-carrying
code~\cite{necula1997proof,appel2001foundational}, a recipient (verifier) checks the
certificate against the requested circuit.\pcctight

\paragraph{Putting things all together, the result is a parametric derivation.}
For the Surface code, \PCQEC produces a finite proof with symbolic $d=2s+1$
under $\Gamma_{\rm surf}=\{s\ge1,d=2s+1\}$.
Write $C_{\rm surf}$ for the ordered CodeML program and
$\mathsf{Generate}\llbracket C_{\rm surf}\rrbracket(d;d,d-1)$ for its
generation into QStab at distance $d$, with $d$ extraction rounds and
an error budget of $d-1$ (Section~\ref{sec:codeml-elaboration}).
Both stages of compilation appear in the program below:
\begin{equation}\label{eq:headline-family-triple}
\pccfitmath{
\mathcal H_{\mathsf{Std}}:\quad
 \Gamma_{\rm surf}\vdash_F
 \{\tgtstate=\cleanstate(d)\}\;
 \underbrace{\circuitcompile{\mathsf{Std}}{
   \underbrace{\mathsf{Generate}\llbracket C_{\rm surf}\rrbracket(d;d,d-1)}
              _{S_d:\ \text{QStab}}
 }}_{\bar c_d:\ \text{QClifford}}\;
 \{\mathsf{Accept}_d(\tgtstate)\land\mathsf{LogFail}_d(\tgtstate)
                 \Rightarrow d\le\ell(\tgtstate)\}.
}
\end{equation}
Here $S_d$ is the QStab program generated by the ordered CodeML
family, $\bar c_d$ is its compiled QClifford circuit, and
$\cleanstate(d)$ is the clean initial error state. The error-aware Hoare
judgment $\vdash_F$ quantifies over every error choice.
$\mathsf{Accept}_d$ specifies the declared record condition, including
detector silence, while $\mathsf{LogFail}_d$ identifies a logical data
residual. The counter $\ell$ records the number of injected
errors. For the
Standard Scheme, the proof establishes the stronger postcondition
$\mathsf{LogFail}_d\Rightarrow d\le\ell$ for every record.

Induction gives $\mathcal H_{\mathsf{Std}}$ a finite representation, including
all code-algebra and compilation premises, independent of the value of $d$.
Expanded circuits and gate-by-gate derivations still grow with $d$
(\S~\ref{sec:family-proof}). Figure~\ref{fig:cut-bridge} uses QStab
states $\srcstate,\srcstate'$ and QClifford states $\tgtstate,\tgtstate'$, defined
in Figure~\ref{fig:ordered-check-circuit}.

\setlength{\textfloatsep}{6pt plus 2pt minus 2pt}
\setlength{\floatsep}{4pt plus 2pt minus 1pt}
\setlength{\intextsep}{4pt plus 2pt minus 1pt}

\begin{figure}[!htbp]
\captionsetup{skip=3pt}
\begin{figureframe}
\centering
\begingroup
\hypersetup{hidelinks}
\definecolor{overviewX}{RGB}{63,91,176}
\definecolor{overviewZ}{RGB}{191,35,45}
\definecolor{overviewCut}{RGB}{110,65,145}
\definecolor{overviewError}{RGB}{191,35,45}
\definecolor{pipeCode}{RGB}{46,91,158}
\definecolor{pipeSource}{RGB}{119,68,157}
\definecolor{pipeCompiler}{RGB}{178,103,20}
\definecolor{pipeTarget}{RGB}{22,119,130}
\definecolor{pipeChecker}{RGB}{45,118,65}
\resizebox{.955\linewidth}{!}{%
\begin{tikzpicture}[x=1cm,y=1cm,
 check/.style={font=\scriptsize,text=black!75,inner sep=1pt},
 cut/.style={draw=overviewCut,densely dashed,line width=.65pt,rounded corners=3pt},
 error/.style={draw=overviewError,line width=.9pt,fill=white},
 note/.style={font=\footnotesize,align=center,inner sep=1pt},
 heading/.style={font=\footnotesize\bfseries,align=center,inner sep=1pt},
 stage/.style={rounded corners=2pt,minimum width=2.18cm,
   minimum height=1.35cm,inner sep=0pt,line width=.7pt},
 piperef/.style={font=\scriptsize,inner sep=0pt,align=center},
 chip/.style={draw,rounded corners=1pt,fill=white,minimum width=.72cm,
   minimum height=.23cm,inner sep=0pt,font=\scriptsize},
 flow/.style={-{Latex[length=2mm]},line width=.75pt},
  proof flow/.style={flow,densely dashed},
 gate/.style={draw,fill=white,minimum width=12pt,minimum height=8pt,
   inner sep=0pt,font=\scriptsize},
 fault box/.style={minimum width=12pt,minimum height=8pt},
 fault width=12pt,fault height=8pt,fault font=5]
\path[use as bounding box] (0,-2.65) rectangle (13.45,8.20);
\begin{scope}[shift={(0,-2.55)}]
\node[heading] at (2.0,10.58) {(a) \hyperref[eq:code-predicates]{Logical Z error}};
\node[heading] at (5.85,10.58) {(b) \hyperref[fig:qstab-state-invariant]{Hook from one error}};
\foreach \offset in {1.32,5.17} {
 \begin{scope}[shift={(\offset,9.95)},x=.65cm,y=.65cm]
  \filldraw[fill=overviewZ!15,draw=overviewZ!70] (0,0) rectangle (1,-1);
  \filldraw[fill=overviewX!15,draw=overviewX!70] (1,0) rectangle (2,-1);
  \filldraw[fill=overviewX!15,draw=overviewX!70] (0,-1) rectangle (1,-2);
  \filldraw[fill=overviewZ!15,draw=overviewZ!70] (1,-1) rectangle (2,-2);
  \filldraw[fill=overviewX!15,draw=overviewX!70]
    (0,0) arc[start angle=180,end angle=0,radius=.5] -- cycle;
  \filldraw[fill=overviewX!15,draw=overviewX!70]
    (1,-2) arc[start angle=180,end angle=360,radius=.5] -- cycle;
  \filldraw[fill=overviewZ!15,draw=overviewZ!70]
    (2,0) arc[start angle=90,end angle=-90,radius=.5] -- cycle;
  \filldraw[fill=overviewZ!15,draw=overviewZ!70]
    (0,-1) arc[start angle=90,end angle=270,radius=.5] -- cycle;
  \foreach \x/\y/\p in {.5/-.5/Z,1.5/-.5/X,.5/-1.5/X,1.5/-1.5/Z,
                          .5/.26/X,1.5/-2.26/X,2.26/-.5/Z,-.26/-1.5/Z}
    \node[check] at (\x,\y) {$\p$};
  \foreach \i in {0,1,2} {
   \draw[cut] (\i-.16,.12) rectangle (\i+.16,-2.12);
   \node[check,text=overviewCut] at (\i,.48) {$\kappa^X_{\i}$};
  }
  \foreach \x/\y/\q in {0/0/0,1/0/1,2/0/2,0/-1/3,1/-1/4,2/-1/5,
                         0/-2/6,1/-2/7,2/-2/8} {
   \fill (\x,\y) circle[radius=1.7pt];
   \ifnum\q=0\relax
    \node[font=\scriptsize,anchor=south east,inner sep=1pt] at (\x-.025,\y+.025) {$q_{\q}$};
   \else\ifnum\q=3\relax
    \node[font=\scriptsize,anchor=south east,inner sep=1pt] at (\x-.025,\y+.025) {$q_{\q}$};
   \else\ifnum\q=5\relax
    \node[font=\scriptsize,anchor=north west,inner sep=1pt] at (\x+.025,\y-.025) {$q_{\q}$};
   \else\ifnum\q>5\relax
    \ifnum\q=8\relax
     \node[font=\scriptsize,anchor=north west,inner sep=1pt] at (\x+.025,\y-.025) {$q_{\q}$};
    \else
     \node[font=\scriptsize,anchor=north east,inner sep=1pt] at (\x-.025,\y-.025) {$q_{\q}$};
    \fi
   \else
    \node[font=\scriptsize,anchor=south west,inner sep=1pt] at (\x+.025,\y+.025) {$q_{\q}$};
   \fi\fi\fi\fi
  }
 \end{scope}
}
\draw[overviewError,line width=1.3pt] (1.32,9.30) -- (2.62,9.30);
\foreach \x in {1.32,1.97,2.62} {
 \draw[error] (\x,9.30) circle[radius=3.8pt];
 \node[font=\scriptsize,text=overviewError] at (\x,9.30) {$Z$};
}
\foreach \x/\y in {5.17/9.30,5.82/9.95,5.82/9.30} {
 \draw[error] (\x,\y) circle[radius=3.8pt];
 \node[font=\scriptsize,text=overviewError] at (\x,\y) {$Z$};
}
\node[font=\scriptsize,align=left,anchor=west] at (6.96,9.30)
 {parities\\cancel};
\draw[black!45,-{Latex},thin] (6.94,9.51) to[out=155,in=0] (5.99,9.62);
\node[note] at (1.97,8.00)
 {$\Nref{eq:surface-cut-values}{\chi_x}(Z_3Z_4Z_5)=(1,1,1)$\\\textbf{three cuts}};
\node[note] at (5.82,8.00)
 {$\Nref{eq:surface-cut-values}{\chi_x}(Z_3Z_1Z_4)=(1,0,0)$\\\textbf{one cut}};
\end{scope}

\foreach \y/\q in {4.72/0,4.39/3,4.06/1,3.73/4} {
 \node[piperef,anchor=east] at (.42,\y) {$q_{\q}$};
 \draw (.60,\y) -- (3.65,\y);
 \node[piperef,fill=white,inner sep=1pt] at (.82,\y) {$I$};
}
\draw[overviewError,dashed,rounded corners=2pt] (1.10,3.55) rectangle (1.60,4.57);
\foreach \y/\q in {4.39/3,4.06/1,3.73/4} {
 \draw[overviewError,line width=.9pt,-{Latex[length=1.2mm]}] (1.35,\y) -- (3.65,\y);
 \node[gate,draw=overviewError,text=overviewError] at (1.35,\y) {$Z$};
 \node[piperef,fill=white,text=overviewError,inner sep=1pt] at (3.32,\y) {$Z_{\q}$};
}
\node[piperef,fill=white,inner sep=1pt] at (3.32,4.72) {$I$};
\node[piperef,text=pcqedFaultZ,fill=white,inner sep=.4pt] at (3.15,4.57) {Hook error};
\draw[pcqedFaultZ,-{Latex[length=1mm]},thin] (3.64,4.48) -- (3.61,4.39);
\node[piperef,text=pipeSource!85!black] at (.82,5.00) {$\srcstate$};
\node[piperef,text=pipeSource!85!black] at (3.32,5.00) {$\srcstate'$};
\draw[draw=pipeSource!85!black,fill=pipeSource!5,rounded corners=2pt] (2.05,3.56) rectangle (2.55,4.89);
\foreach \y/\q in {4.72/0,4.39/3,4.06/1,3.73/4}
 \node[piperef,text=pipeSource!85!black] at (2.30,\y) {$\prog{Z_{\q}}$};
\node[heading,text=pipeSource!85!black] at (6.02,4.65) {\hyperref[sec:qstab]{QStab} $S_3$: three rounds};
\node[note] at (6.02,4.30) {\hyperref[sec:qstab-transitions]{QS-Hook}: $(E,\ell)=(I^9,0)$};
\node[note] at (6.02,3.95) {$\longrightarrow (\textcolor{overviewError}{Z_3Z_1Z_4},1),\quad G_0=0$};
\draw[pipeCompiler!85!black,line width=.65pt] (2.30,3.56) -- (2.30,3.44);
\draw[pipeCompiler!85!black,-{Latex[length=1.4mm]},line width=.65pt] (2.30,3.44) -- (2.30,3.38);
\draw[pipeCompiler!85!black,-{Latex[length=1.4mm]},line width=.65pt]
 (2.30,3.44) -- (8.28,3.44) -- (8.28,.80) -- (7.45,.80);

\node[heading,text=pipeTarget!85!black] at (4.10,3.18)
 {\hyperref[sec:compiler-translation]{(c) Circuit compiled with the Standard Scheme}};
\foreach \y/\q in {2.60/0,2.25/3,1.90/1,1.55/4} {
 \node[piperef,anchor=east] at (.48,\y) {$q_{\q}$};\draw (.60,\y) -- (5.35,\y);
}
\node[piperef,anchor=east] at (.48,1.20) {$a:\ket{0}$};\draw (.60,1.20) -- (5.45,1.20);
\node[piperef,text=pipeTarget!85!black] at (.65,2.91) {$\tgtstate$};
\node[piperef,text=pipeTarget!85!black] at (5.55,2.91) {$\tgtstate'$};
\node[piperef,anchor=west] at (5.55,2.60) {$I$};
\foreach \y/\q in {2.25/3,1.90/1,1.55/4}
 \node[piperef,text=overviewError,anchor=west] at (5.55,\y) {$Z_{\q}$};
\draw[overviewError,line width=.9pt] (2.00,1.20) -- (5.45,1.20);
\foreach \x/\y in {2.50/2.25,3.55/1.90,4.60/1.55}
 \draw[overviewError,line width=.9pt] (\x,1.20) -- (\x,\y) -- (5.35,\y);
\foreach \x/\y in {1.45/2.60,2.50/2.25,3.55/1.90,4.60/1.55} {
 \faultmark{\x-.50}{\y}
 \ifdim\x pt=2.50pt\else\faultmark{\x-.50}{1.20}\fi
 \draw (\x,\y) -- (\x,1.20);\fill (\x,\y) circle[radius=1.4pt];
 \draw[fill=white] (\x,1.20) circle[radius=.075];
 \draw (\x-.075,1.20) -- (\x+.075,1.20);\draw (\x,1.125) -- (\x,1.275);
}
\node[piperef,text=pcqedFaultZ,fill=white,inner sep=.4pt] at (4.45,2.43) {Hook error};
\draw[pcqedFaultZ,-{Latex[length=1mm]},thin] (4.98,2.36) -- (5.28,2.25);
\node[gate,draw=pcqedFaultZ,text=pcqedFaultZ,dashed] at (2.00,1.20) {$Z$};
\faultmark{4.95}{1.20}\node[gate] at (5.45,1.20) {$\prog{M_Z}$};
\node[piperef] at (7.25,1.90) {$\begin{array}{ll}
 \faultX_a&(I^9,1)\\[2pt]
 \faultY_a&(Z_3Z_1Z_4,1)\\[2pt]
 \faultZ_a&(Z_3Z_1Z_4,0)
 \end{array}$};

\node[heading,text=pipeTarget!85!black] at (4.10,.80)
 {\hyperref[sec:trust]{(d) Circuit compiled with the Flag Scheme}};
\foreach \y/\q in {.20/0,-.15/3,-.50/1,-.85/4} {
 \node[piperef,anchor=east] at (.48,\y) {$q_{\q}$};\draw (.60,\y) -- (5.35,\y);
}
\node[piperef,anchor=east] at (.48,-1.20) {$a:\ket{0}$};\draw (.60,-1.20) -- (5.35,-1.20);
\node[piperef,anchor=east] at (.48,-1.55) {$f:\ket{+}$};\draw (.60,-1.55) -- (6.20,-1.55);
\node[piperef,text=pipeTarget!85!black] at (.65,.51) {$\tgtstate$};
\node[piperef,text=pipeTarget!85!black] at (5.55,.51) {$\tgtstate'$};
\node[piperef,anchor=west] at (5.55,.20) {$I$};
\foreach \y/\q in {-.15/3,-.50/1,-.85/4}
 \node[piperef,text=overviewError,anchor=west] at (5.55,\y) {$Z_{\q}$};
\draw[overviewError,line width=.9pt] (2.18,-1.20) -- (5.35,-1.20);
\foreach \x/\y in {2.50/-.15,3.15/-.50,3.80/-.85}
 \draw[overviewError,line width=.9pt] (\x,-1.20) -- (\x,\y) -- (5.35,\y);
\draw[overviewError,line width=.9pt] (4.45,-1.20) -- (4.45,-1.55) -- (5.25,-1.55);
\draw[pcqedFaultX,line width=.9pt] (5.25,-1.55) -- (6.20,-1.55);
\foreach \x/\y in {1.20/-1.55,1.85/.20,2.50/-.15,3.15/-.50,3.80/-.85,4.45/-1.55} {
 \faultmark{\x-.32}{\y}
 \ifdim\x pt=2.50pt\else\faultmark{\x-.32}{-1.20}\fi
 \draw (\x,\y) -- (\x,-1.20);\fill (\x,\y) circle[radius=1.4pt];
 \draw[fill=white] (\x,-1.20) circle[radius=.075];
 \draw (\x-.075,-1.20) -- (\x+.075,-1.20);\draw (\x,-1.275) -- (\x,-1.125);
}
\node[piperef,text=pcqedFaultZ,fill=white,inner sep=.4pt] at (4.45,.025) {Hook error};
\draw[pcqedFaultZ,-{Latex[length=1mm]},thin] (4.98,-.045) -- (5.28,-.15);
\node[gate,draw=pcqedFaultZ,text=pcqedFaultZ,dashed] at (2.18,-1.20) {$Z$};
\faultmark{4.90}{-1.20}\node[gate] at (5.35,-1.20) {$\prog{M_Z}$};
\faultmark{4.78}{-1.55}\node[gate] at (5.25,-1.55) {$\prog{H}$};
\faultmark{5.72}{-1.55}\node[gate] at (6.20,-1.55) {$\prog{M_Z}$};
\node[piperef] at (7.25,-.50) {$\begin{array}{ll}
 \faultX_a&(I^9,1,0)\\[2pt]
 \faultY_a&(Z_3Z_1Z_4,1,1)\\[2pt]
 \faultZ_a&(Z_3Z_1Z_4,0,1)
 \end{array}$};
\node[piperef,text=black!75] at (4.10,-1.97)
 {\faultpanel{12pt}{8pt}{5}\ each site: $I,\faultchoices$. Red path: selected $Z_a$ injection.};

\filldraw[fill=overviewX!15,draw=overviewX!70] (.35,-2.53) rectangle (.55,-2.33);
\node[piperef,anchor=west] at (.65,-2.43) {X stabilizer};
\filldraw[fill=overviewZ!15,draw=overviewZ!70] (2.12,-2.53) rectangle (2.32,-2.33);
\node[piperef,anchor=west] at (2.42,-2.43) {Z stabilizer};
\draw[cut] (3.95,-2.53) rectangle (4.15,-2.33);
\node[piperef,anchor=west] at (4.25,-2.43) {proof cut};
\draw[error] (6.05,-2.43) circle[radius=3.8pt];
\node[font=\scriptsize,text=overviewError] at (6.05,-2.43) {$Z$};
\node[piperef,anchor=west] at (6.25,-2.43) {Pauli error};

\draw[black!18] (8.50,-1.65) -- (8.50,7.78);
\node[heading] at (10.95,8.03) {(e) \PCQEC pipeline};
\node[piperef,text=black!75] at (9.68,7.65) {Program / circuit};
\node[piperef,text=black!75] at (12.17,7.65) {Proof / certificate};
\foreach \name/\x/\y/\height in
 {code/9.68/6.95/1.05,algebra/12.17/6.95/1.05,
  source/9.68/4.85/1.35,sourceproof/12.17/4.85/1.35,
  target/9.68/1.20/1.35,targetproof/12.17/1.20/1.35}
 \node[stage,draw=black!55,fill=white,minimum height=\height cm] (\name) at (\x,\y) {};
\draw[flow,black!60] (code.east) -- (algebra.west);
\draw[flow,black!60] (source.east) -- (sourceproof.west);
\node[heading,text=pipeCode!85!black] at (9.68,7.26) {\hyperref[sec:codeml]{CodeML}\ \S~\ref{sec:codeml}};
\node[note] at (9.68,6.95) {$C(d,k)$: stabilizer};
\node[piperef] at (9.68,6.64) {\hyperref[sec:codeml-elaboration]{$t\mapsto P[q]$}};
\node[heading,text=pipeCode!85!black] at (12.17,7.26) {\hyperref[sec:assertion-logic]{Code algebra}};
\node[piperef] at (12.17,6.95) {$V_d$: \hyperref[eq:valid-complete]{Well formed}};
\node[piperef] at (12.17,6.64) {$W_d$: \hyperref[eq:valid-complete]{Complete}};
\draw[flow,black!60] (8.77,6.425) -- (8.77,5.525);
\draw[proof flow,black!60] (11.26,6.425) -- (11.26,5.525);
\node[piperef,fill=white,inner sep=1pt] at (9.85,5.98) {\hyperref[sec:codeml-elaboration]{\shortstack{Evaluate stabilizers\\at distance $d$}}};
\node[piperef,fill=white,inner sep=1pt] at (12.32,5.98) {\hyperref[def:hook-bound]{\shortstack{Supply cut bounds\\for hook errors}}};
\node[heading,text=pipeSource!85!black] at (9.68,5.24) {\hyperref[sec:qstab]{QStab}\ \S~\ref{sec:qstab}};
\node[note] at (9.68,4.90) {$S_d$: $R$ rounds};
\node[piperef] at (9.68,4.52) {\hyperref[sec:qstab-transitions]{\shortstack{Ordered error\\effects}}};
\node[heading,text=pipeSource!85!black] at (12.17,5.24) {\hyperref[sec:qhl-assertions]{Invariant proof}};
\node[note] at (12.17,4.90) {$\mathcal D_d$};
\node[piperef] at (12.17,4.50) {\hyperref[rule:Src-Cut]{Cut bounds}\\\hyperref[rule:Src-Dist]{and distance}};
\node[note,text=pipeCompiler!85!black] at (10.95,3.70) {\hyperref[sec:compiler]{Choose scheme $\xi$}\quad\S~\ref{sec:compiler}};
\foreach \x/\scheme/\dest in
 {9.30/Standard/sec:compiler-translation,10.40/Flag/sec:trust,11.50/Shor/sec:trust,12.60/Knill/sec:trust}
 \node[chip,draw=black!55,align=center,minimum height=.40cm] at (\x,3.20) {\hyperref[\dest]{\shortstack{\scheme\\Scheme}}};
\draw[flow,black!60] (source.south) -- (9.68,3.97) -- (8.59,3.97) -- (8.59,2.43) -- (9.68,2.43) -- (target.north);
\draw[proof flow,black!60] (sourceproof.south) -- (12.17,3.97) -- (13.39,3.97) -- (13.39,2.43) -- (12.17,2.43) -- (targetproof.north);
\node[piperef,fill=white,inner sep=1pt,text=pipeCompiler!85!black] at (9.68,2.69) {\hyperref[sec:compiler-translation]{Compile circuit}};
\node[piperef,fill=white,inner sep=1pt,text=pipeCompiler!85!black] at (12.17,2.69) {\hyperref[sec:target-cut-rules]{Translate proof}};
\draw[flow,black!60] (10.18,2.22) -- (11.67,2.22);
\node[piperef,fill=white,inner sep=1pt] at (10.95,2.05) {\hyperref[eq:fault-images-main]{error images}};
\node[heading,text=pipeTarget!85!black] at (9.68,1.58) {\hyperref[sec:qclifford]{QClifford}\ \S~\ref{sec:qclifford}};
\node[note] at (9.68,1.20) {$\bar c_{d,\xi}$};
\node[piperef] at (9.68,.80) {\hyperref[sec:qclifford]{\shortstack{Gates and\\injection sites}}};
\node[heading,text=pipeTarget!85!black] at (12.17,1.58) {\hyperref[sec:target-cut-rules]{Target proof}};
\node[note] at (12.17,1.20) {$\mathcal D'_d$};
\node[piperef] at (12.17,.80) {\hyperref[def:hoare-triple-c]{Hoare proof}\\\hyperref[sec:target-cut-rules]{and distance}};
\draw[flow,black!60] (8.77,.525) -- (8.77,-.475);
\draw[proof flow,black!60] (13.08,.525) -- (13.08,-.475);
\node[piperef,fill=white,inner sep=1pt] at (10.95,.02) {\hyperref[eq:certificate]{Certificate $\mathcal C$}};
\node[stage,draw=black!55,fill=white,minimum width=4.67cm,minimum height=.95cm] (checker) at (10.925,-.95) {};
\node[heading,text=pipeChecker!85!black] at (10.95,-.70) {\hyperref[sec:standard-checking]{Consumer checker}\ \S~\ref{sec:standard-checking}};
\node[piperef] at (10.95,-1.10) {\hyperref[sec:verifier]{$K_{d,\xi}$}: accept $\Rightarrow$ \hyperref[def:dcirc]{$d^{\mathrm{circ}}_{K_{d,\xi}}\ge d$}};
\end{tikzpicture}}

\endgroup
\end{figureframe}
\caption{Framework of \PCQEC.}
\label{fig:pcc}\label{fig:cut-bridge}
\Description{Two distance-three Surface code patches compare a logical Z row
crossing three X-column proof cuts and a hook error crossing one. Blue and red faces
are X and Z stabilizers. The QStab circuit has four data wires and a narrow measurement box whose
upright factors are Z zero, Z three, Z one and Z four from top to bottom.
A source hook error adds Z to data sites 3, 1 and 4 at cost one, with zero
readout flip. Measurement preserves that residual. Branching lines connect the
source measurement to both implementations. Below it, fully instrumented
Standard Scheme and Flag Scheme circuits produce the same propagated data
error from a selected Z error on a helper qubit. The Flag Scheme also flips a flag
readout. Tables enumerate the X, Y and Z alternatives at that site.
The right panel has paired program and proof columns across CodeML, QStab and
QClifford. Schedule well-formedness and supplied cut bounds for hook errors connect CodeML and
QStab. The chosen scheme compiles programs and structurally translates supplied
cut certificates using proved backend lemmas and emitted error images. Both outputs
form the certificate checked against the requested contract. Solid arrows
carry programs and circuits. Dashed arrows carry proofs.}
\end{figure}

\emph{Comparison with prior work.} Table~\ref{tab:prior-qec} compares
demonstrated QEC verification capabilities. QuantumSE symbolically executes
programs with specified Pauli errors~\cite{fang2024symbolic}; Veri-QEC verifies
specified error conditions with primitive stabilizer measurements and also
supports Coq proofs of scalable codes~\cite[\S~6]{huang2025efficient}.
\citet{chen2025verifying} analyze
errors inside small implementing circuits (gadgets) and detect bad extraction orders for a given circuit.
AutoQ-Para proves input-output correctness and equivalence for parameterized
circuits~\cite{abdulla2026parameterized}.
\PCQEC accounts for hook errors, proves a circuit-level distance
bound for every legal family size, and reuses the code-algebra argument and
invariant-proof structure across four schemes. Each backend justifies
its own error effects through scheme-specific propagation lemmas.\pcctight

\begin{table}[!htbp]
\captionsetup{skip=4pt}
\caption{Demonstrated QEC verification capabilities.}
\label{tab:prior-qec}
\centering\small
\setlength{\tabcolsep}{3pt}
\renewcommand{\arraystretch}{1.0}
\newcommand{\PriorYes}{\textcolor{green!45!black}{\cmark}}
\newcommand{\PriorNo}{\textcolor{red!70!black}{\xmark}}
\begin{tabular}{@{}>{\raggedright\arraybackslash}m{0.34\linewidth}
 >{\centering\arraybackslash}m{0.15\linewidth}
 >{\centering\arraybackslash}m{0.24\linewidth}
 >{\centering\arraybackslash}m{0.20\linewidth}@{}}
\toprule
Approach & Hook error analysis & All-size circuit-level code distance proof
 & Proof reuse across schemes \\
\midrule
QuantumSE~\cite{fang2024symbolic} & \PriorNo & \PriorNo & \PriorNo \\
Veri-QEC~\cite{huang2025efficient} & \PriorNo & \PriorNo & \PriorNo \\
\citet{chen2025verifying} & \PriorYes & \PriorNo & \PriorNo \\
AutoQ-Para~\cite{abdulla2026parameterized} & \PriorNo & \PriorNo & \PriorNo \\
\PCQEC (this work) & \PriorYes & \PriorYes & \PriorYes \\
\bottomrule
\end{tabular}
\end{table}

In the following sections, we use the parametric Surface code family as our running example. Specifically, we prove circuit-level code distance
$d$ for the odd-distance rotated Surface code compiled with
the Standard Scheme (Corollary~\ref{cor:parametric-standard}). Lean~4 mechanizes the metatheory and the exact circuit-level distances of the canonical rotated Surface, unrotated Surface and Toric code programs. Our approach can also be generalized to other infinite code families such as
hypergraph-product (HGP) codes and Toric codes, as well as a Cartesian lifted-product (LP) code family (\S~\ref{sec:eval}).
Figure~\ref{fig:family-geometries} shows the five code families and the
cut collections or reduced-weight measure used in their proofs.\pcctight

\section{Syndrome extraction and circuit-level code distance proofs}
\label{sec:overview}\label{sec:background}
\subsection{Background}
\label{sec:qec-primer}\label{sec:bg-pauli}\label{sec:bg-stab}

Here we briefly go through the concepts and notation we use in the paper,
which are all standard and well known in quantum error correction. For
interested readers who want to learn more about these concepts, we recommend
Nielsen and Chuang's textbook~\cite[Chapter~10]{nielsen_chuang_2010}.
In a quantum circuit with $n$ qubits (wires), we represent errors by Pauli strings. Write $[n]=\{0,\ldots,n-1\}$. A Pauli
string $E:[n]\to\{I,X,Y,Z\}$ assigns an error to each wire. Here I is the
identity, X is a bit flip, Z is a phase flip, and Y combines both (while still counting as one error). The weight
$\wt E$ counts nonidentity entries, and $P_q$
denotes P on wire $q$ and I elsewhere. For example, in Figure~\ref{fig:pcc}(c),
one $Z$ error injected on the helper qubit propagates to the data qubits,
causing the hook error $IZZZ$ on the four displayed wires, with weight
$\wt{IZZZ}=3$. The wire order is $0,3,1,4$, so this is $Z_3Z_1Z_4$,
with I on every other data qubit. We omit global phase
while retaining commutation relations.

We say two operators commute if their order does not matter. As an example, the Pauli string operator $XZ$ commutes with $ZX$. Anticommuting operators
differ by a sign when their order is reversed. Distinct nonidentity
single-qubit Paulis anticommute. The commutation parity $\langle E,F\rangle_n$
combines these single-wire bits by exclusive OR (XOR), written
$\oplus$. For example, the dashed column through qubits $1,4,7$ in
Figure~\ref{fig:pcc}(b) is the Pauli string $X_1X_4X_7$. We call this
parity probe a \emph{cut}. Against the hook error $E=Z_3Z_1Z_4$, X and Z
anticommute at qubits $1$ and $4$, giving two contributions that cancel.
The column through $0,3,6$ instead has only one anticommuting position:
\begin{equation}\label{eq:cut-commutation-example}
\pccfitmath{
\langle X_1X_4X_7,Z_3Z_1Z_4\rangle_9=1\oplus1=0,\qquad
 \langle X_0X_3X_6,Z_3Z_1Z_4\rangle_9=1.
}
\end{equation}
Thus the hook error crosses the cut in column zero but not the cut in
column one. In particular, $\langle EF,G\rangle_n=\langle E,G\rangle_n\oplus\langle F,G\rangle_n$. An error that
anticommutes with a stabilizer flips its syndrome bit.

A stabilizer QEC code is determined by a set of mutually commuting Pauli strings, also known as stabilizer generators. These generators and their products form the stabilizer
group $\mathcal S$. Its elements act trivially on encoded information.
The normalizer $\mathcal N(\mathcal S)$ contains all strings that commute with
every stabilizer. A logical error has zero syndrome but changes encoded
information, so it belongs to $\mathcal N(\mathcal S)\setminus\mathcal S$. The minimum weight of such
an error is the code distance. Section~\ref{sec:assertion-logic} makes this
criterion executable in CodeML.\pcctight

A Pauli frame records errors relative to an error-free run. We call its
restriction to the data qubits the residual. For example, in Figure~\ref{fig:pcc}(c), the residual error caused by the injected $Z$ error is $IZZZ$ on the four displayed data wires. Record bits likewise describe
readout deviations: zero denotes an unchanged readout and one denotes a
flip relative to the error-free outcome. A detector XORs selected record bits. For
example, comparing consecutive measurements of the same stabilizer detects a
change, but two flipped outcomes can cancel in this parity.

We follow the standard way to model circuit-level errors, where Pauli
errors can be injected anywhere in the circuit~\cite{gottesman2009QECIntro}. Our error model counts one injected
error for each pre-operation X, Y or Z injection on a qubit (wire) at a
selected location. A serious problem in circuit-level noise modeling is
error propagation~\cite{gottesman2009QECIntro,kishony2026offthehook}:
gates propagate errors and can amplify their weight without adding to the
injected error count, which can be harmful and decrease the number of
errors the actual circuit can protect against.\pcctight

Now we explain the definition of circuit-level code distance in a compiled
QEC circuit, which is what we want to formally verify. Circuit-level code
distance measures the minimum number of injected Pauli errors that cause
a logical residual error without triggering any detector in the circuit.
We call the absence of detector triggers the detector-silence condition.
From this definition and the complicated error propagation mechanism, we
can see that verifying circuit-level code distance is a nontrivial problem.
A direct approach would enumerate all possible error injection patterns
in the entire circuit with fewer than $d$ injected errors, simulate the
error propagation for each of them, and decide whether the residual error
is a logical error of the code. Previous work such as
QuantumSE~\cite{fang2024symbolic} uses symbolic execution to model and
analyze QEC programs more efficiently. However, what we want to achieve
is a formal proof without simulating these error injection patterns. In
the next section, we will give an overview of the idea and explain how
this is possible.\pcctight

\subsection{Overview of our approach: Turning parametric code knowledge into a parametric program invariant}
\label{sec:cut-intuition}

The key idea of \PCQEC is that we should fully utilize the parametric
code-distance knowledge provided by the design of QEC codes, and carry
this knowledge across the entire compilation pipeline. Despite the
complexity of modeling circuit-level noise and error propagation, all
errors, including hook errors, measurement errors and data qubit errors,
either have no effect on the residual data error or cause effects that we
can geometrically understand and reason about when we know how the
lower-level compilation works. Then, the question we ask first is:
\emph{What measure separates a one-injection hook error from a logical error when
both affect the same number of qubits?}

Figure~\ref{fig:pcc}(a) and (b) visualize the difference. In Figure~\ref{fig:cut-bridge}, the logical row $Z_3Z_4Z_5$ crosses all three
columns, whereas the hook error $Z_3Z_1Z_4$ crosses only column zero
(the two contributions in column one cancel). We extract this geometric information and call it \emph{cut parities}.

So the answer to the above research question is that we can use cut
parities to distinguish the geometric effects a hook error can cause
given an error budget from what it takes to form a logical error under
that budget. Combining both inspires us to reason about a program
invariant at a lower level, which inherits the CodeML layer's geometric
relationship between cause and effect.

\emph{Next we focus on using the concrete example in Figure~\ref{fig:pcc} to further elaborate the idea in detail.}

The X strings \phantomsection\label{eq:surface-cut-values}%
$(\kappa^X_0,\kappa^X_1,\kappa^X_2)=(X_0X_3X_6,X_1X_4X_7,X_2X_5X_8)$ probe the Z/Y parity in each column by anticommutation. Similarly, the Z strings $(\kappa^Z_0,\kappa^Z_1,\kappa^Z_2)=(Z_0Z_1Z_2,Z_3Z_4Z_5,Z_6Z_7Z_8)$ probe the X/Y parity in each row. The cut signatures $\Nref{eq:surface-cut-values}{\chi_x}(E)=(\langle\kappa^X_i,E\rangle_9)_{i<3}$ and $\Nref{eq:surface-cut-values}{\chi_z}(E)$ collect these column and row bits. For the two errors in Figure~\ref{fig:cut-bridge}, \phantomsection\label{eq:cut-parity-examples}%
$\Nref{eq:surface-cut-values}{\chi_x}(Z_3Z_4Z_5)=(1,1,1)$ and $\Nref{eq:surface-cut-values}{\chi_x}(Z_3Z_1Z_4)=(1,0,0)$.

We say that an error crosses a cut when the corresponding signature bit is
one. The signature's Hamming weight therefore counts the crossed cuts. These
probes are independent of the measured syndrome. A logical error, for example,
has zero syndrome even though its cut signature can be $(1,1,1)$.

First, we establish a property of the code: every logical error crosses all three cuts in at least one collection. This can be formalized by the following assertion:
\begin{equation}\label{eq:surface-cut-completeness}
\pccfitmath{
\Nref{eq:code-predicates}{\mathsf{Logical}}(E)\ \Longrightarrow\
 \wt{\Nref{eq:surface-cut-values}{\chi_x}(E)}=3\ \lor\ \wt{\Nref{eq:surface-cut-values}{\chi_z}(E)}=3.
}
\end{equation}
The column probes differ by stabilizer products. A zero-syndrome error
therefore has the same parity against every column probe. The same argument
applies to the row probes. If both signatures are zero, code algebra shows
that the error is a stabilizer. Lemma~\ref{lemma:surface-main-algebra}
establishes this property for every odd distance $d$ and every
logical representative.

Second, we establish a property of the program: each error crosses at most one
cut in each collection. \emph{Why are we sure that this is correct when errors can be injected anywhere?}

If the error is a data qubit error, clearly it affects only one row and one
column. If the error is injected on a syndrome qubit, then either it has
no effect on the residual data error or it produces a hook error. We have
already shown that a hook error also has a limited effect on crossing
these cuts. Combining the above, we prove a \emph{program invariant} that is preserved
at stabilizer-measurement boundaries regardless of what kind of error is
injected at each step.
More explicitly, we can construct the program invariant that relates the data residual $E$ to the number $\ell$ of injected errors:
\begin{equation}\label{eq:cut-overview-invariant}
\pccfitmath{
\prog{\wt{\Nref{eq:surface-cut-values}{\chi_x}(E)}\le\ell\quad\land\quad
 \wt{\Nref{eq:surface-cut-values}{\chi_z}(E)}\le\ell}.
}
\end{equation}

Combining this with Equation~\eqref{eq:surface-cut-completeness}, we now
know that the program invariant in Equation~\eqref{eq:cut-overview-invariant},
together with $\ell<3$, implies $\neg\mathsf{Logical}(E)$.
This is sufficient to prove that the circuit-level code distance is at
least $3$. Equality additionally requires a matching silent error pattern
(\S~\ref{sec:standard-exactness}). Thus, the lower-bound proof obligation is
converted to proving this program invariant across the compilation stack
in \PCQEC, as shown in Figure~\ref{fig:pcc}. To obtain one finite proof
parametric over all code distances $d$, consider the invariant's representation. If we explicitly
write down the invariant in Equation~\eqref{eq:cut-overview-invariant} for
the nine data qubits in Figure~\ref{fig:pcc}, it looks like:
\begin{equation}\label{eq:cut-overview-invariant-explicit}
\pccfitmath{
\wt{(\langle X_0X_3X_6,E\rangle_9,
       \langle X_1X_4X_7,E\rangle_9,
       \langle X_2X_5X_8,E\rangle_9)}\le\ell\quad\land
 \wt{(\langle Z_0Z_1Z_2,E\rangle_9,
       \langle Z_3Z_4Z_5,E\rangle_9,
       \langle Z_6Z_7Z_8,E\rangle_9)}\le\ell.
}
\end{equation}
Clearly, the size of the above explicitly expanded assertion keeps increasing with code distance $d$. We ask the following question: \emph{How can one finite proof cover circuits whose numbers of qubits,
stabilizers and rounds grow with $d$?}
To make our proof framework parametric over code distance, we need to
represent the program invariant parametrically. One way to achieve this
is to have a parametric language that can express the Surface code family
and its X/Z cuts symbolically with a finite-size program. This is why we
define a new parametric language in a functional style. Inspired by OCaml,
we call this language CodeML. CodeML supplies the family-level algebraic
premises and makes it possible to reason about the invariant in
Equation~\eqref{eq:cut-overview-invariant-explicit} with a proof whose size
is independent of $d$ for the entire family.

Another key insight we learn from Figure~\ref{fig:pcc} and the above
reasoning is that the source invariant depends on each error injection's
\emph{cost} and \emph{effect}. The compiler determines both by tracking
where the error is injected and how it propagates under a fixed
syndrome extraction scheme. This justifies our design of the second
abstraction layer, QStab. QStab sets the boundary between expressing the
invariant we need to prove circuit-level code distance and the detailed
gate implementation. Thus, the design of this IR also helps us reason
about circuit-level code distance parametrically over different syndrome
extraction schemes.

\emph{How does proving this invariant establish the Hoare triple in the
Introduction?} The program in the triple follows the three layers in
Figure~\ref{fig:pcc}. The CodeML program $C_{\rm surf}$ is evaluated at
distance $d$ by $\mathsf{Generate}$ to produce the ordered QStab program
$S_d$. The outer $\mathsf{Compile}_{\mathsf{Std}}$ then implements those
measurements as the QClifford circuit $\bar c_d$. For the Standard
Scheme, the intermediate goal puts the cut invariant in the
postcondition of this composition:
\begin{equation}\label{eq:overview-invariant-triple}
\pccfitmath{
\Gamma_{\rm surf}\vdash_F
 \{\tgtstate=\cleanstate(d)\}\;
 \underbrace{\circuitcompile{\mathsf{Std}}{
   \underbrace{\mathsf{Generate}\llbracket C_{\rm surf}\rrbracket(d;d,d-1)}
              _{S_d:\ \text{QStab}}
 }}_{\bar c_d:\ \text{QClifford}}\;
 \left\{\underbrace{
   \wt{\chi_z(E_{\rm raw})}\le\ell\ \land\
   \wt{\chi_x(E_{\rm raw})}\le\ell
 }_{\Phi_d(\tgtstate):\ \text{the cut invariant on target data}}\right\}.
}
\end{equation}
Here $E_{\rm raw}$ is the data residual in target state $\tgtstate$,
and $\ell=\ell(\tgtstate)$ is its injection count. At the end of the
measurement in Figure~\ref{fig:pcc}(c), these are the same
$E=Z_3Z_1Z_4$ and $\ell=1$ used above. For arbitrary $d$, the two
signatures have $d$ entries. The proof starts with zero cut parities,
shows that every permitted error effect preserves the bound, and uses
induction to cover all measurements and rounds. Compilation must also
justify this step through the actual gates, including errors still on
helper qubits. Separately proved backend lemmas establish this connection
and are reused by certificate translation.\pcctight

The invariant is one necessary piece of the argument. We also need
successful generation of the ordered program, well-formed stabilizers,
and the code-algebra fact that every logical error crosses all $d$ cuts
in some collection. With these premises, a logical residual must satisfy
$d=\wt{\chi_x(E_{\rm raw})}\le\ell$ or
$d=\wt{\chi_z(E_{\rm raw})}\le\ell$.
Thus the invariant triple yields exactly the goal
$\mathcal H_{\mathsf{Std}}$ from
Equation~\eqref{eq:headline-family-triple}.\phantomsection\label{eq:overview-family-triple}
The Standard Scheme establishes this implication for every record, so
it also holds when $\mathsf{Accept}_d$ requires detector silence. By the
definition of circuit-level distance, no accepted silent logical failure
can then cost fewer than $d$ injections. Section~\ref{ex:surface-source}
assembles the source premises, and Section~\ref{sec:family-proof} shows
the complete derivation through the compiled circuit.

\section{CodeML: A parametric language for families of QEC codes}
\label{sec:codeml}

In CodeML, we turn the static stabilizer formalism of a QEC code family
into a function that evaluates the stabilizer operator and its Pauli
scheduling order, given a code distance and a stabilizer index.
Section~\ref{sec:cut-intuition} explained why we need such a language:
both the stabilizers and the cuts used in our invariant must have finite
descriptions as $d$ grows. Here we give these objects a syntax, starting
with the stabilizer measurement already shown in Figure~\ref{fig:pcc}(c).

Given a code distance $d$ and stabilizer index $k$, a CodeML program
$C(d,k)$ returns a bounded function. Its argument $t$ is a coupling
position. For the first stabilizer in Figure~\ref{fig:pcc}(c), $d=3$,
$k=0$, and positions $t=0,1,2,3$ return
$\prog{Z[0]},\prog{Z[3]},\prog{Z[1]},\prog{Z[4]}$, respectively.
Here $\prog{P[q]}$ is the addressed Pauli written $P_q$ in
Section~\ref{sec:qec-primer}. Multiplying the four factors gives the
stabilizer $Z_0Z_3Z_1Z_4$; visiting them in order gives the four CNOT
couplings in the figure. This order determines which factors remain
after an ancilla error and therefore which hook error reaches the data.
Increasing $k$ orders the stabilizer measurements, while increasing $t$
orders the couplings within one measurement.

\paragraph{Syntax for stabilizers and Pauli strings.}\label{sec:codeml-semantics}
The types come from the objects introduced in Section~\ref{sec:qec-primer}.
The type $\prog{nat}$ represents nonnegative sizes and indices, such as
the number of qubits $n$ and a qubit index $q\in[n]$.
The type $\prog{pauli}$ contains the four letters I, X, Y and Z, and
$\prog{bool}$ represents tests used to select a stabilizer or a qubit.
A Pauli string, such as the residual error $E$ or a cut $\kappa_j^X$,
has type $\prog{str}:=\prog{nat}\to\prog{pauli}$. Its string lambda
$\prog{\lambda u.p}$ specifies the Pauli acting at data qubit $u$.
Observing the first $n$ entries recovers the strings on $[n]$ used in
Section~\ref{sec:qec-primer}; the same type serves every code size.\pcctight

To retain the coupling order, an addressed Pauli $\prog{P[q]}$ is the
pair $(P,q)$, of type $\prog{pauli}\times\prog{nat}$. A stabilizer has
type $\prog{bounded}(\prog{pauli}\times\prog{nat})$:
$\prog{\lambda t<L.e}$ gives its $L$ ordered factors. Thus the bounded
lambda for Figure~\ref{fig:pcc}(c) has four positions, whereas a string
lambda for its data error has one Pauli entry per data qubit. The former
tells the compiler which qubit to couple next; the latter describes the
error on the data qubits, as in Section~\ref{sec:qec-primer}.
Appendix~\ref{app:codeml} gives the complete CodeML syntax, typing and evaluation rules.\pcctight

\begin{definition}[Family declaration]\label{def:family}
A declaration $(C,\mathsf{Legal},n,m,f)$ supplies the program, legal code
distances, data width, stabilizer count and a bound $f$ on nested family
calls, called \emph{fuel}~\cite{amin2017interpreters}.
For every legal $d$ and $k<m(d)$, evaluating $C$ must return a stabilizer
with a defined bound and a defined addressed Pauli at every position.
\end{definition}

Here we use the simplest repetition code as an example. Against bit-flip
(Pauli X) errors, it has code distance $d$, with $d-1$ adjacent Z stabilizers
on $d\ge2$ data qubits. We can program this family in CodeML as
$\prog{\kw{family}\ C_{\rm rep}(d,k)=\lambda t<2.\ Z[k+t]}$.
The bounded lambda returns Z on qubit $k+t$ at each coupling
position $t=0,1$. Stabilizer $k$ therefore couples
to qubit $k$ before qubit $k+1$. At
$d=3$, the program returns $Z_0Z_1$ and
$Z_1Z_2$. At $d=5$, the same program returns four
ordered stabilizers. The bound two counts the couplings in one
measurement, just as the bound four does in Figure~\ref{fig:pcc}(c).
The repetition code uses $(n,m,f)=(d,d-1,1)$: its body makes no further
family calls. Successful evaluation fixes the coupling order. Address
bounds, distinct qubits and commutation are additional proof obligations.\pcctight

\paragraph{Programming the Surface code in Figure~\ref{fig:pcc}.}
We now program every stabilizer of the running Surface code example.
Figure~\ref{fig:surface-code-and-cuts} shows the same patch at $d=3$ and
$d=5$. Write $q_d(a,c)=da+c$ for the qubit in row $a$ and column $c$.
This is the numbering already used in Figure~\ref{fig:pcc}:
the first Z stabilizer visits coordinates $(0,0),(1,0),(0,1),(1,1)$,
which give qubits $0,3,1,4$ when $d=3$.
At $d=5$, those same coordinates give qubits $0,5,1,6$.
We use $s=(d-1)/2$ to count pairs of rows or columns:
\begin{equation}\label{eq:surface-main-spec}
\pccfitmath{
\Gamma_{\rm surf}=\{s\ge1,d=2s+1\},\qquad
 n(d)=d^2,\quad m(d)=4s^2+4s=d^2-1,\quad f(d)=1.
}
\end{equation}
Here $\mathsf{Legal}(d):=3\le d\land d\bmod2=1$. Each square face, or
plaquette, supports a weight-four bulk stabilizer. The curved boundary
faces support weight-two stabilizers~\cite{horsman2012surface,beverland2022surface}.
There are $(d-1)^2=4s^2$ bulk stabilizers and $2(d-1)=4s$ boundary
stabilizers, giving the count $m(d)$ above. The direct program below
makes no nested family calls, so fuel one suffices.

We number bulk stabilizers in $2\times2$ blocks; the remaining indices
select the top, bottom, right or left boundary.
For example, at $d=3$, indices $k=0,1,2,3$ visit the four
bulk faces of Figure~\ref{fig:surface-code-and-cuts}; $k=4$ selects the
top boundary $X_0X_1$. Write $L(d,k)$ for the coupling count and
$F(d,k,t)$ for the addressed Pauli at position $t<L(d,k)$.
Figure~\ref{fig:surface-code-and-cuts} illustrates the bulk and boundary
cases; the supplement gives the index decoding and their complete definitions.
Each case is expressed by an ordinary $\kw{if}$ expression. Generation evaluates the declaration only for legal $d$ and $k<m(d)$.

The Z branch visits each plaquette column by column, and the X branch
visits it row by row. For $d=3,k=0$, the Z branch has $a=c=0$ and
returns $Z[3(t\bmod2)+\lfloor t/2\rfloor]$: evaluating $t=0,1,2,3$
gives exactly the order $0,3,1,4$ in Figure~\ref{fig:pcc}(c).
As Section~\ref{sec:cut-intuition} explained, this order is what limits
the number of column cuts crossed by a hook error.

The expression $\prog{\kw{for}\ t<a\ \kw{yield}\ e}$ computes one result
for each position and abbreviates $\prog{\lambda t<a.e}$. We therefore
write the complete direct declaration as
\begin{equation}\label{eq:surface-generator-main}
\pccfitmath{
\prog{\kw{family}\ C_{\rm surf}(d,k)=
   \kw{for}\ t<L(d,k)\ \kw{yield}\ F(d,k,t)}.
}
\end{equation}
Generation evaluates this declaration for $k<m(d)$ before QStab
execution. The case distinction selects the stabilizer, and the bounded loop
returns its factors in coupling order. The program text has a fixed size
for every $d$.\pcctight

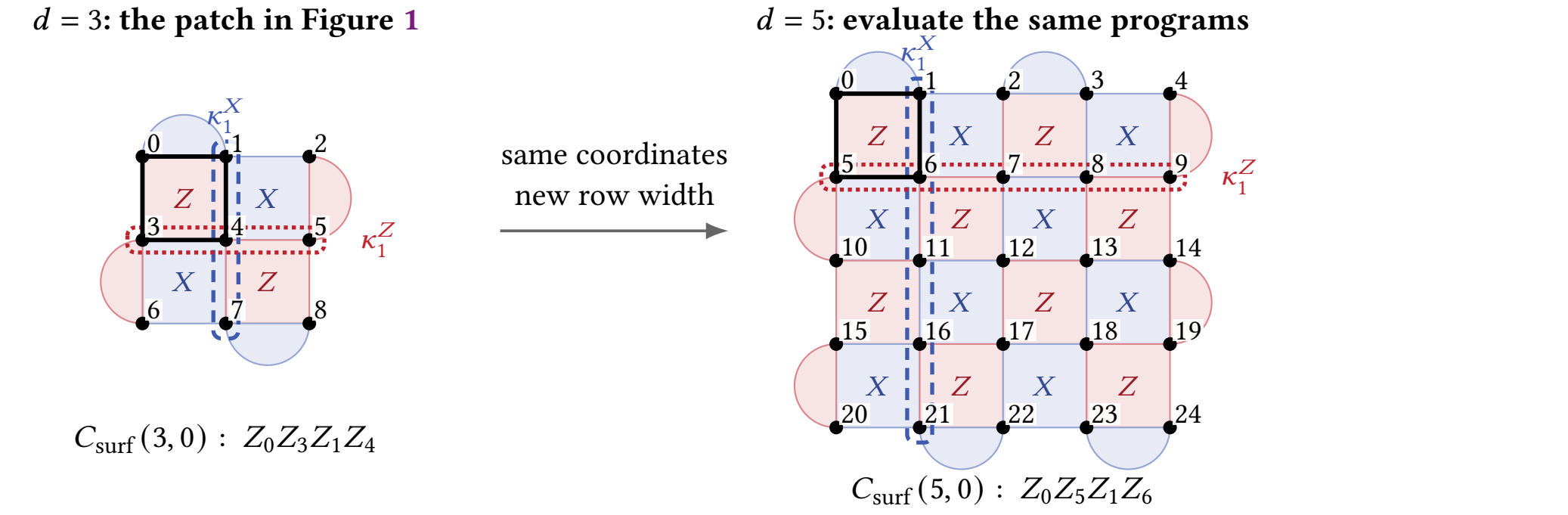
Figure~\ref{fig:surface-code-and-cuts} also shows how the same coordinate
convention programs the cuts from Section~\ref{sec:cut-intuition}.
The column containing qubit $q$ is $q\bmod d$, and its row is
$\lfloor q/d\rfloor$. A string lambda can therefore select a whole
column or row, giving finite descriptions of both cut collections.
Section~\ref{sec:assertion-logic} uses these strings to express the
invariant from Equation~\eqref{eq:cut-overview-invariant}.

\begin{figure}[!htbp]
\captionsetup{skip=3pt}
\centering
\begin{figureframe}
\begin{tikzpicture}[x=1cm,y=1cm,
 every node/.style={font=\footnotesize,inner sep=1pt},
 familylabel/.style={align=center,font=\footnotesize},
 familycutX/.style={draw=pcqedSigX,densely dashed,line width=1pt,rounded corners=2pt},
 familycutZ/.style={draw=pcqedSigZ,densely dotted,line width=1pt,rounded corners=2pt}]
\path[use as bounding box] (0,0) rectangle (13.3,4.55);
\node[familylabel,font=\footnotesize\bfseries] at (2.15,4.28)
 {$d=3$: the patch in Figure~\ref{fig:pcc}};
\node[familylabel,font=\footnotesize\bfseries] at (8.95,4.28)
 {$d=5$: evaluate the same programs};
\foreach \dd/\ox/\oy in {3/1.42/3.10,5/7.49/3.65} {
 \begin{scope}[shift={(\ox,\oy)},x=.73cm,y=.73cm]
  \pgfmathtruncatemacro{\lastvertex}{\dd-1}
  \pgfmathtruncatemacro{\lastface}{\dd-2}
  \pgfmathtruncatemacro{\lastpair}{(\dd-3)/2}
  \foreach \a in {0,...,\lastface} {
   \foreach \c in {0,...,\lastface} {
    \pgfmathtruncatemacro{\faceparity}{mod(\a+\c,2)}
    \ifnum\faceparity=0
     \def\familyfacecolor{pcqedSigZ}\def\familyletter{Z}
    \else
     \def\familyfacecolor{pcqedSigX}\def\familyletter{X}
    \fi
    \filldraw[fill=\familyfacecolor!12,draw=\familyfacecolor!55]
      (\c,-\a) rectangle (\c+1,-\a-1);
    \node[font=\scriptsize,text=\familyfacecolor!80!black]
      at (\c+.5,-\a-.5) {$\familyletter$};
   }
  }
  \foreach \j in {0,...,\lastpair} {
   \filldraw[fill=pcqedSigX!12,draw=pcqedSigX!55]
    (2*\j,0) arc[start angle=180,end angle=0,radius=.5] -- cycle;
   \filldraw[fill=pcqedSigX!12,draw=pcqedSigX!55]
    (2*\j+1,-\lastvertex) arc[start angle=180,end angle=360,radius=.5] -- cycle;
   \filldraw[fill=pcqedSigZ!12,draw=pcqedSigZ!55]
    (\lastvertex,-2*\j) arc[start angle=90,end angle=-90,radius=.5] -- cycle;
   \filldraw[fill=pcqedSigZ!12,draw=pcqedSigZ!55]
    (0,-2*\j-1) arc[start angle=90,end angle=270,radius=.5] -- cycle;
  }
  \draw[familycutX] (.85,.18) rectangle (1.15,-\lastvertex-.18);
  \draw[familycutZ] (-.18,-.85) rectangle (\lastvertex+.18,-1.15);
  \draw[black,line width=1.1pt] (0,0) rectangle (1,-1);
  \foreach \a in {0,...,\lastvertex} {
   \foreach \c in {0,...,\lastvertex} {
    \pgfmathtruncatemacro{\address}{\dd*\a+\c}
    \fill (\c,-\a) circle[radius=1.7pt];
    \node[font=\scriptsize,fill=white,fill opacity=.88,text opacity=1,
      anchor=south west,inner sep=.4pt] at (\c+.025,-\a+.025)
      {$\address$};
   }
  }
  \node[font=\scriptsize,text=pcqedSigX,anchor=south]
    at (1,.22) {$\kappa^X_1$};
  \node[font=\scriptsize,text=pcqedSigZ,anchor=west]
    at (\lastvertex+.56,-1) {$\kappa^Z_1$};
 \end{scope}
}
\draw[-{Latex[length=2mm]},black!60,line width=.7pt] (4.55,2.45) -- (6.55,2.45);
\node[familylabel] at (5.55,2.95) {same coordinates\\new row width};
\node[familylabel] at (2.15,.63) {$C_{\rm surf}(3,0):\ Z_0Z_3Z_1Z_4$};
\node[familylabel] at (8.95,.18) {$C_{\rm surf}(5,0):\ Z_0Z_5Z_1Z_6$};
\end{tikzpicture}
\par\smallskip
\footnotesize
\begin{tabular}{@{}l@{\quad}l@{}}
\textbf{First bulk stabilizer} &
 $C_{\rm surf}(d,0)=\prog{\lambda t<4.\ Z[d(t\bmod2)+\lfloor t/2\rfloor]}$\\[.5pt]
\textbf{Top boundary} &
 $C_{\rm surf}(d,(d-1)^2)=\prog{\lambda t<2.\ X[t]}$\\[.5pt]
\textbf{Column cuts} &
 $\kappa_j^X=\prog{\lambda q.\ \kw{if}\ q\bmod d=j\ \kw{then}\ X\ \kw{else}\ I}$\\[.5pt]
\textbf{Row cuts} &
 $\kappa_i^Z=\prog{\lambda q.\ \kw{if}\ \lfloor q/d\rfloor=i\ \kw{then}\ Z\ \kw{else}\ I}$\\[.5pt]
\textbf{Cut signatures} &
 $\chi_x(E)_j=\langle\kappa_j^X,E\rangle_{d^2},\quad
  \chi_z(E)_i=\langle\kappa_i^Z,E\rangle_{d^2}\quad(i,j<d)$\\[.5pt]
\textbf{Invariant} &
 $\wt{\chi_x(E)}\le\ell\ \land\ \wt{\chi_z(E)}\le\ell$
\end{tabular}
\par\smallskip
The hook errors $Z_3Z_1Z_4$ and $Z_5Z_1Z_6$ have column signatures
$(1,0,0)$ and $(1,0,0,0,0)$, respectively.
\end{figureframe}
\caption{Programming the Surface code and its cut invariant with CodeML.
Red and blue faces are Z and X stabilizers. Black outlines select the first
bulk stabilizer. Qubit addresses are $q=da+c$ for row $a$ and column $c$.}
\label{fig:surface-code-and-cuts}
\Description{Complete distance-three and distance-five rotated Surface code
patches. The first bulk Z stabilizer is outlined in black. A dashed blue
column and dotted red row show the X and Z cuts through column and row one.
The programs below the patches define the first ordered stabilizer, a top
boundary stabilizer, both cut families and the invariant. Evaluating the
same programs changes the addresses and the number of cut parities.}
\end{figure}

\paragraph{From scheduled factors to a Pauli operator.}
The bounded value $C_{\rm surf}(3,0)$ lists four couplings, but the
commutation calculation in Section~\ref{sec:qec-primer} needs a Pauli
string on all nine data qubits. We recover that string by putting each
factor on its addressed qubit and multiplying the results. For a
stabilizer value $S$, this is
\begin{equation}\label{eq:pauli-word-expansion}
\pccfitmath{
\mathsf{operator}(S):=\prog{\prod_{P[q]\ \kw{in}\ S}
  \lambda u.\ \kw{if}\ u=q\ \kw{then}\ P\ \kw{else}\ I}.
}
\end{equation}
The product visits each coupling position once and multiplies the factors
addressed to $u$. It therefore specifies the Pauli acting at
data address $u$, whereas $S[t]$ identifies the
qubit coupled at position $t$. We write $T_k:=\prog{\mathsf{operator}(C(d,k))}$
and use $\denote C_f(d,k)$ for this derived string interpretation. For a
concrete operand list $r$, $\mathsf{pauli}(r)$ denotes the
same product. In the running example, $T_0=Z_0Z_3Z_1Z_4$, with I
on the other five data qubits. The ordered factors and this string thus
describe the same stabilizer at the two levels needed for compilation
and the commutation tests from Section~\ref{sec:qec-primer}.

\subsection{From stabilizer families to fixed QStab programs}
\label{sec:codeml-elaboration}
We write $\mathsf{Generate}\llbracket C\rrbracket(d;R,B)$ for the
CodeML-to-QStab translation used in the Introduction's triple. For a
family declaration $(C,\mathsf{Legal},n,m,f)$, it evaluates $C(d,k)$
with fuel $f(d)$ for every $k<m(d)$. Let $r_{d,k}$ be the resulting
list of addressed Pauli factors, in increasing coupling order $t$.
When generation succeeds, it returns
\begin{equation}\label{eq:codeml-generation}
\pccfitmath{
\mathsf{Generate}\llbracket C\rrbracket(d;R,B)
   =\langle n(d),[r_{d,k}]_{k<m(d)},R,B\rangle.
}
\end{equation}
The declaration supplies $n,m,f$; the explicit arguments $R,B$ supply
the number of extraction rounds and the error budget. Each list
$r_{d,k}$ becomes a $\prog{\kw{MeasStab}}$ instruction with numeric
operands in that order. Generation is defined only for legal $d$ with
successful evaluation and the width and operand checks stated below.

For $d=3$, $C_{\rm surf}(3,0)$ gives $Z_0Z_3Z_1Z_4$,
the measurement already shown at the QStab level in
Figure~\ref{fig:pcc}(c). Figure~\ref{fig:ordered-check-circuit} expands
that same example: the four ordered factors determine the CNOTs, and
the selected ancilla Z error leaves the residual $Z_3Z_1Z_4$ used in
Section~\ref{sec:cut-intuition}. The generated QStab program contains
numeric operands in this order, with no unevaluated CodeML expressions.

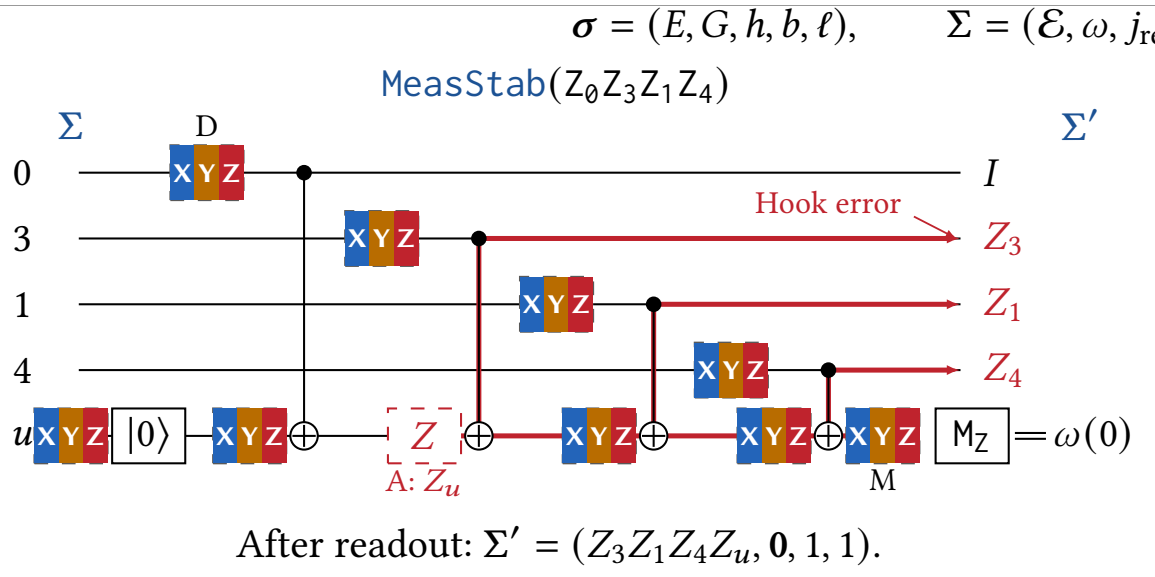
\begin{figure}[!htbp]
\captionsetup{skip=3pt}
\small
\centering
\resizebox{.86\linewidth}{!}{%
\begin{minipage}{\linewidth}
\centering
\begin{figureframe}
\small\noindent
\centering
$\srcstate=(E,G,h,b,\ell),\qquad
 \tgtstate=(\mathcal E,\omega,j_{\rm rec},\ell)$.
\par\smallskip
\begin{minipage}[t]{245pt}\centering
$\prog{\kw{MeasStab}(Z_0Z_3Z_1Z_4)}$\par\smallskip
\begin{tikzpicture}[x=.85pt,y=.8pt,line width=.45pt,
 every node/.style={font=\small,inner sep=0pt,outer sep=0pt},
 cg/.style={draw,fill=white,minimum width=16pt,minimum height=12pt,inner sep=1pt},
 fault width=16pt,fault height=12pt,fault font=6,
 fault box/.style={minimum width=16pt,minimum height=12pt}]
\path[use as bounding box] (0,-17) rectangle (280,87);
\node[text=pcqedCodeKeyword] at (15,85) {$\tgtstate$};
\node[text=pcqedCodeKeyword] at (275,85) {$\tgtstate'$};
\foreach \name/\yy in {0/72,3/54,1/36,4/18} {
 \node[anchor=west] at (0,\yy) {$\name$};
 \draw (17,\yy) -- (244,\yy);
}
\node[anchor=west] at (0,0) {$u$};
\draw (10,0) -- (230,0);
\draw[pcqedFaultZ,line width=1.1pt] (106,0) -- (230,0);
\foreach \xx/\yy in {120/54,165/36,210/18}
 \draw[pcqedFaultZ,line width=1.1pt,-{Latex[length=3pt]}]
   (\xx,0) -- (\xx,\yy) -- (244,\yy);
\node[anchor=west] at (250,72) {$I$};
\foreach \yy/\q in {54/3,36/1,18/4}
 \node[anchor=west,text=pcqedFaultZ] at (250,\yy) {$Z_{\q}$};
\faultmark{15}{0}
\node[cg] at (35,0) {$\lvert0\rangle$};
\foreach \xx/\yy in {75/72,120/54,165/36,210/18} {
 \faultmark{\xx-25}{\yy}
 \ifnum\xx=120\relax\else\faultmark{\xx-14}{0}\fi
 \draw (\xx,\yy) -- (\xx,0);
 \fill (\xx,\yy) circle[radius=1.7pt];
 \draw[fill=white] (\xx,0) circle[radius=3pt];
 \draw (\xx-3,0) -- (\xx+3,0);
 \draw (\xx,-3) -- (\xx,3);
}
\node[cg,draw=pcqedFaultZ,dashed,text=pcqedFaultZ] at (106,0) {$Z$};
\faultmark{224}{0}
\node[cg] at (247,0) {$\prog{M_Z}$};
\draw (258,.8) -- (266,.8);\draw (258,-.8) -- (266,-.8);
\node[anchor=west] at (268,0) {$\omega(0)$};
\node[font=\scriptsize,text=pcqedFaultZ,fill=white,inner sep=.4pt] at (210,64) {Hook error};
\draw[pcqedFaultZ,-{Latex[length=2.4pt]},thin] (232,60) -- (243,54);
\node[font=\scriptsize,anchor=south] at (50,82) {D};
\node[font=\scriptsize,anchor=north,text=pcqedFaultZ] at (106,-9) {A: $Z_u$};
\node[font=\scriptsize,anchor=north] at (224,-9) {M};
\end{tikzpicture}
\par\smallskip
After readout: $\tgtstate'=(Z_3Z_1Z_4Z_u,\mathbf0,1,1)$.
\end{minipage}\hfill
\begin{minipage}[t]{145pt}\centering
\textbf{Each row selects one error}\par\smallskip
\setlength{\tabcolsep}{3pt}
\begin{tabular}{@{}lcc@{}}
Injection & Final data error $E$ & $\omega(0)$\\\hline
D: $\faultZ_0$ & $Z_0$ & $0$\\
D: $\faultX_0$ & $X_0$ & $1$\\
D: $\faultY_0$ & $Y_0$ & $1$\\[2pt]
A: $\faultX_u$ & $I^9$ & $1$\\
\textcolor{pcqedFaultZ}{A: $Z_u$} & $\textcolor{pcqedFaultZ}{Z_3Z_1Z_4}$ & $\textcolor{pcqedFaultZ}{0}$\\
A: $\faultY_u$ & $Z_3Z_1Z_4$ & $1$\\[2pt]
M: $\faultX_u$ & $I^9$ & $1$\\
M: $\faultY_u$ & $I^9$ & $1$\\
M: $\faultZ_u$ & $I^9$ & $0$
\end{tabular}
\end{minipage}\par\smallskip
\end{figureframe}
\end{minipage}}
\caption{The Standard Scheme circuit and single-injection outcomes.
Each table row starts clean and idles elsewhere. Striped sites allow $I,\faultchoices$.
The red path selects $Z_u$ at A. Final $E$ excludes the helper qubit.}
\label{fig:ordered-check-circuit}
\Description{Four data wires control CNOTs targeting a fresh helper qubit u in
the order zero, three, one, four. The helper qubit is prepared in zero and
measured in Z. Ten injection sites independently allow idle, blue X,
ochre Y or red Z. The selected red Z error at A occurs after the first
CNOT. Red propagation paths reach only data wires three, one and four.
The displayed full-frame calculation retains Z on the helper qubit, while
the final data residual is explicitly Z three Z one Z four with readout
flip zero. The adjacent table writes all Pauli products explicitly and
lists separate X, Y and Z injections at D, A and M.}
\end{figure}

\phantomsection\label{def:surface-schedule}
The evaluated stabilizers $\vec r_d=[r_{d,k}]_{k<m(d)}$ form the fixed
program $S_d=\langle n(d),\vec r_d,R(d),B(d)\rangle$.
Thus $S_d=\mathsf{Generate}\llbracket C_{\rm surf}\rrbracket(d;R(d),B(d))$.
For the Surface code, each round contains
$d^2-1$ stabilizers, the round count is $R(d)=d$, and
the error budget is $B(d)=d-1$. For the nine-qubit patch in
Figure~\ref{fig:pcc}, this gives
$S_3=\mathsf{Generate}\llbracket C_{\rm surf}\rrbracket(3;3,2)
    =\langle9,\vec r_3,3,2\rangle$:
eight stabilizers measured in each of three rounds, with a budget of two
injected errors. This is the budget $\ell<3$ used in
Section~\ref{sec:cut-intuition}. The list $\vec r_d$ stores one round;
QStab repeats it $R(d)$ times.

Generation checks sizes and distinct in-range operands
(Appendix~\ref{app:codeml-evaluation}). The proof additionally establishes
stabilizer commutation and the
cut bound for hook errors. The latter depends on the coupling order: for example,
reordering the first stabilizer as $Z_0Z_1Z_3Z_4$ preserves its operator
but violates that bound (\S~\ref{sec:cut-proof-rules}).
Figure~\ref{fig:ordered-check-circuit} shows how the coupling order determines
which couplings remain after an ancilla-qubit error.\pcctight

\subsection{Assertions, cuts, and code-algebra proofs}
\label{sec:assertion-logic}\label{def:codeml-vcgen}

In this subsection, we introduce an assertion language for stating the
proof obligations of a CodeML program and the invariant introduced in
Section~\ref{sec:cut-intuition}. An \emph{assertion} is a logical
condition on the code parameters and error state. For example,
Equation~\eqref{eq:cut-overview-invariant} asserts that the number of
crossed cuts in each collection is at most the number $\ell$ of injected
errors. We need a finite expression for this condition so that the same
invariant can be stated and proved for every legal $d$.

Appendix~\ref{app:assertions} gives the complete assertion language and its rules~\cite{hoare1969axiomatic,leino2010dafny}.
String equality compares entries up to the stated width; commutation
and weight use that same width. For the hook error in
Figure~\ref{fig:pcc}(b), the test
$\prog{\langle X_1X_4X_7,E\rangle_9=0}$ is precisely the two cancelling
contributions calculated in Equation~\eqref{eq:cut-commutation-example}.
Conjunction combines the row and column bounds, while bounded
quantification ranges over stabilizers or cuts. Their bounds may depend
on $d$ without changing the assertion's syntax.\pcctight

\paragraph{Expressing the logical-error condition.}
Section~\ref{sec:qec-primer} characterized a logical error as a string
in $\mathcal N(\mathcal S)\setminus\mathcal S$. We give the two membership
tests program names. The predicate $\mathsf{Norm}(E)$ checks whether $E$
commutes with every stabilizer, and $\mathsf{Gen}(E)$ checks whether $E$
is a product of stabilizers. Using the strings $T_k$
derived in Section~\ref{sec:codeml}, an $m$-bit mask
$a<2^m$ selects this subset through $a_k:=\prog{\nrem{\nquot{a}{2^k}}{2}}$. Each
stabilizer need be selected at most once because the stabilizers commute and
square to the identity. A logical error satisfies Norm and fails Gen:

\pccwrapdisplay{4}{.60\linewidth}{eq:code-predicates}{1.25\baselineskip}{%
\setlength{\jot}{1pt}
\begin{aligned}[t]
 \mathsf{Norm}(E)&:=\prog{\forall k<m.\ \langle T_k,E\rangle_n=0},\\
 \mathsf{Gen}(E)&:=\prog{\exists a<2^m.\ E\equiv_n\prod\nolimits_{k<m}(\kw{if}\ a_k=0\ \kw{then}\ \lambda q.I\ \kw{else}\ T_k)},\\
 \mathsf{Logical}(E)&:=\prog{\mathsf{Norm}(E)\land\neg\mathsf{Gen}(E)}.
\end{aligned}
}
Thus $\mathsf{Logical}(E)$ implements the criterion
$E\in\mathcal N(\mathcal S)\setminus\mathcal S$~\cite{gottesman1997stabilizer}.
For example, the mask $a=1$ selects only $T_0=Z_0Z_3Z_1Z_4$,
so $\mathsf{Gen}(T_0)$ holds.
For the logical row $Z_3Z_4Z_5$ in Figure~\ref{fig:pcc}(a), Norm
holds and Gen fails. For the hook error $Z_3Z_1Z_4$ in
Figure~\ref{fig:pcc}(b), Norm fails: it anticommutes with the top-boundary
stabilizer $X_0X_1$. These are the same two errors distinguished by the
cut parities in Section~\ref{sec:cut-intuition}.

\paragraph{Expressing the cut invariant.}
We can now replace the explicit lists of three cuts in
Equation~\eqref{eq:cut-overview-invariant-explicit} by programs that work
for every $d$. Recall that $q\bmod d$ and $\lfloor q/d\rfloor$ give the
column and row of data qubit $q$. We name these functions $L_x$ and $L_z$;
they select the dashed column and dotted row in
Figure~\ref{fig:surface-code-and-cuts}. The X and Z string lambdas below
are therefore the parametric versions of the probes $\kappa_j^X$ and
$\kappa_i^Z$ already used in Section~\ref{sec:cut-intuition}:

\pccwrapdisplay{5}{.565\linewidth}{eq:surface-main-cuts}{1.8\baselineskip}{%
\setlength{\jot}{1pt}
\begin{aligned}[t]
 L_x(d,q)&=\prog{\nrem{q}{d}},\qquad L_z(d,q)=\prog{\nquot{q}{d}},\\
 \kappa_j^X&:=\prog{\lambda q.\ \kw{if}\ L_x(d,q)=j\ \kw{then}\ X\ \kw{else}\ I},\\
 \kappa_i^Z&:=\prog{\lambda q.\ \kw{if}\ L_z(d,q)=i\ \kw{then}\ Z\ \kw{else}\ I},\\
 \chi_x(E)_j&=\prog{\langle\kappa_j^X,E\rangle_n},\qquad
 \chi_z(E)_i=\prog{\langle\kappa_i^Z,E\rangle_n},\qquad i,j<d.
\end{aligned}
}
\phantomsection\label{eq:signature-core-expansion}%
At $d=3,j=1$, the first string lambda gives $X_1X_4X_7$, exactly
the column cut in Equation~\eqref{eq:cut-commutation-example}. Its parity
against $Z_3Z_1Z_4$ is still zero. Thus these definitions compute the
same bits as the overview; only their description is now parametric.
The X/Z signature labels name the probe types.
Appendix~\ref{app:assertions} explains how the core language stores each
parity bit as a string entry, so that string weight counts the crossed
cuts. We omit the width when it is clear from context. The condition
$\prog{\wt[d]{\chi_x(E)}\le\ell\land\wt[d]{\chi_z(E)}\le\ell}$
is therefore the same invariant as Equation~\eqref{eq:cut-overview-invariant},
now expressed by one finite assertion for the whole family.

\paragraph{Proving the code properties required by the invariant.}
\phantomsection\label{eq:valid-complete}
The overview used two facts about the code: its generators define a
stabilizer code, and every logical error crosses all cuts in at least
one collection.
The predicate $\mathsf{WellFormed}$ requires definedness and pairwise
commutation of the generated stabilizer strings. For example, the
neighboring stabilizers $Z_0Z_3Z_1Z_4$ and $X_1X_2X_4X_5$ in
Figure~\ref{fig:surface-code-and-cuts} overlap at qubits $1$ and $4$.
Their two anticommuting contributions cancel, just as in the cut
calculation in Section~\ref{sec:qec-primer}. If a program omitted
the factor $X_4$ from the second stabilizer, the strings would no longer
commute, even though every factor still had the correct type.
Commutation ensures $\mathsf{Gen}(E)\Rightarrow\mathsf{Norm}(E)$,
as required by the stabilizer-group definition in
Section~\ref{sec:qec-primer}. The distance rules require this
$\mathsf{WellFormed}$ proof, and compilation preserves it unchanged.

The cut-completeness contract, $\mathsf{Complete}_{L_x,L_z}$, ensures that cut signatures do not
conceal a logical error. For every fully defined data error,
$\mathsf{Norm}(E)$ together with an uncrossed row and an uncrossed column
implies $\mathsf{Gen}(E)$. Consequently, a logical error must cross all
$d$ cuts in at least one collection. This is exactly the code property
in Equation~\eqref{eq:surface-cut-completeness}, with three replaced by $d$.

These contracts are proved symbolically. Evaluating an assertion answers a
question at a particular size. A judgment $\Gamma\vdash_C\assertion{F}$ establishes
$\assertion{F}$ for every parameter and residual permitted by
$\Gamma$. For a Surface code property $\assertion{F}(d)$, a proof
of $\assertion{F}(3)$ and a proof of $\assertion{F}(d+2)$ assuming
$\assertion{F}(d)$ for arbitrary odd $d\ge3$ establish the family
property by induction. This finite proof justifies the use of the same cut
summary at every size without enumerating all errors.

\begin{lemma}[Surface code algebra]\label{lemma:surface-main-algebra}
Under Equation~\eqref{eq:surface-main-spec},
$C_{\rm surf}$ satisfies the well-formedness and completeness contracts above
for the cuts in
Equation~\eqref{eq:surface-main-cuts}.
\end{lemma}

\begin{proof}[Proof idea]
Induct on odd $d$, starting with eight independent stabilizers at $d=3$.
The step $d\mapsto d+2$ establishes definedness, commutation and independence
for the larger collection and preserves the probe intersections. At code distance $d$,
among zero-syndrome errors, the $d^2-1$ independent stabilizers leave two independent
logical directions beyond stabilizer products.
The row and column probes distinguish these directions, so zero parity
for both leaves only a stabilizer product. As in
Section~\ref{sec:cut-intuition}, adjacent column probes differ by a
stabilizer product, and likewise for adjacent row probes. A zero-syndrome
error therefore has the same parity against every probe in a collection.
If it is logical, one collection must have parity one throughout.
This gives the cut-completeness property used by the invariant. The
supplement supplies the independence proof for the direct construction.
\end{proof}

\begin{corollary}\label{cor:surface-static-distance}
$C_{\rm surf}$ has static code distance $d$ for every legal odd $d\ge3$.
\end{corollary}
An error of weight below $d$ leaves an untouched row and column.
Completeness therefore excludes a logical error of that weight. A full Z row
attains the bound: it commutes with every stabilizer but anticommutes with an
X-column probe, so it cannot be a stabilizer product. At $d=3$, this is
the logical row shown in Figure~\ref{fig:pcc}(a).

We have now expressed the stabilizers, cuts and invariant from
Section~\ref{sec:cut-intuition} as finite programs and assertions, and
established the code property that separates logical errors from
errors crossing too few cuts. The remaining question is the program
property from that section: \emph{How do we prove that extraction steps
preserve the invariant?} QStab lets us state and prove that obligation
using the error costs and effects illustrated in Figure~\ref{fig:pcc}(c).

\section{QStab: From syndrome extraction to an invariant}
\label{sec:qstab}

Section~\ref{sec:assertion-logic} gives us a parametric expression for the
cut invariant. To prove it, we need the cost and effect of each error, as
explained in Section~\ref{sec:cut-intuition}. In Figure~\ref{fig:pcc}(c),
one helper-qubit error leaves $Z_3Z_1Z_4$ on the data. The Flag Scheme in
panel~(d) produces the same hook error and also flips a flag readout. The
circuits differ, but the data error crosses the same single column cut.

\emph{How can an IR retain these circuit-level error effects while remaining
parametric over syndrome-extraction schemes?} QStab records data errors,
readout changes and injection counts at stabilizer-measurement boundaries.
This gives the parametric invariant proof a common interface: a step has an
error cost and a data and readout effect. The ordered stabilizers come from
CodeML; the ancillas and gates depend on the extraction scheme. Each backend
must justify the error-effect bounds used by the proof under its declared
readout contract. We first prove the invariant through this interface, then
show how compilation preserves it.

\subsection{Ordered syndrome extraction and error effects}
\label{sec:qstab-prog}\label{sec:qstab-sem}

\phantomsection\label{def:QStabProgram}
Recall the program $S=\langle n,\vec r,R,B\rangle$ generated in
Section~\ref{sec:codeml-elaboration}: $n$ counts data qubits, $\vec r$ lists
one round of ordered stabilizers, $R$ counts rounds, and $B$ is the error
budget. For the patch in Figure~\ref{fig:pcc}, these are nine qubits, eight
stabilizers, three rounds and two allowed injections. A fixed program has
numeric addresses and counts; its CodeML declaration and parametric
well-formedness proof remain in the proof context.\pcctight

Let $m=|\vec r|$. The measurement index $h<mR$ identifies both the round
and the stabilizer within that round: $r_h=\vec r[h\bmod m]$, with operator
$T_h=\mathsf{pauli}(r_h)$. For the first measurement, $r_0$ lists
$Z_0,Z_3,Z_1,Z_4$ in that order. The helper error in
Figure~\ref{fig:ordered-check-circuit} occurs after the first coupling, so
its data effect is the remaining suffix $Z_3Z_1Z_4$. In general, we write
$\mathsf{Hooks}(h)=\{\prod_{j\le i<|r_h|}r_h[i]\mid0\le j\le|r_h|\}$
for these suffixes. The last suffix is $I^n$: after all couplings, a helper
error has no remaining data effect in this block.\pcctight

\phantomsection\label{def:QStabState}
The source state $\srcstate=(E,G,h,b,\ell)$ in
Figure~\ref{fig:ordered-check-circuit} keeps the data residual $E$ and
injection count $\ell$ used by the invariant. It also records the remaining
budget $b$, the next measurement $h$, and a vector $G\in\{0,1\}^{mR}$ of
readout flips. Past entries of $G$ are committed; $G_h$ holds an offset
for the current measurement, whose data-error contribution is added below.
Execution starts at $\srcstate_0=(I^n,\mathbf0,0,B,0)$.
A configuration $\statecfg$ adds a running, finished or budget-exhausted
status (Appendix~\ref{app:qstab}).

\paragraph{A nondeterministic transition relation.}
\phantomsection\label{sec:qstab-transitions}

We write $\statecfg\xrightarrow{k}_S\statecfg'$ for a step costing
$k\in\{0,1\}$ injected errors.
\emph{We observe that, no matter where errors are injected, their cost
and effect can always be categorized into four types.}
These types describe the effects at a stabilizer-measurement boundary:
Type-0 changes one data qubit and leaves the readout offset unchanged;
Type-1 changes one data qubit and may also flip the readout;
Type-2 adds a hook error and may also flip the readout; and
Type-3 flips the readout while leaving the residual data error unchanged.
Each type costs one injection.\pcctight

Figure~\ref{fig:ordered-check-circuit} gives concrete examples:
$Z_0$, $X_0$ or $Y_0$ at D has a Type-0 effect; $X_0$ and $Y_0$ flip
$\omega(0)$ only through the commutation test at the measurement. A
helper $Y$ before the last CNOT has a Type-1 effect: it adds $Z_4$ and an
additional readout flip. Both $Z_u$ and $Y_u$ at A have Type-2 effects, producing
the same hook $Z_3Z_1Z_4$; only $Y_u$ also flips the readout.
Finally, $X_u$ at M has a Type-3 effect.
The four rules below express these effects. An error step requires $b>0$ and
chooses, as needed, a data qubit $q<n$, a Pauli $P\in\{X,Y,Z\}$, an
additional readout offset $t\in\mathbb B=\{0,1\}$, or a suffix
$H\in\mathsf{Hooks}(h)$:
\[
\pccfitmath[\linewidth]{
\begin{array}{@{}l@{\;}l@{\quad}l@{\quad}l@{}}
 \text{\sffamily Type-0}&\textsc{QS-Data}:&E'=EP_q,\quad G'_h=G_h,
   &\textcolor{blue}{\text{\scriptsize Change one data qubit only}}\\
 \text{\sffamily Type-1}&\textsc{QS-DataReadout}:&E'=EP_q,\ G'_h=G_h\oplus t,
   &\textcolor{blue}{\text{\scriptsize Data error; optional readout flip}}\\
 \text{\sffamily Type-2}&\textsc{QS-Hook}:&E'=EH,\ G'_h=G_h\oplus t,
   &\textcolor{blue}{\text{\scriptsize Hook error; optional readout flip}}\\
 \text{\sffamily Type-3}&\textsc{QS-Readout}:&E'=E,\quad G'_h=G_h\oplus1.
   &\textcolor{blue}{\text{\scriptsize Flip readout; keep residual}}
\end{array}
}
\]
The choices are nondeterministic: the proof must cover every permitted
effect. In particular, \textsc{QS-DataReadout} and \textsc{QS-Hook} charge
once even when both data and readout change.

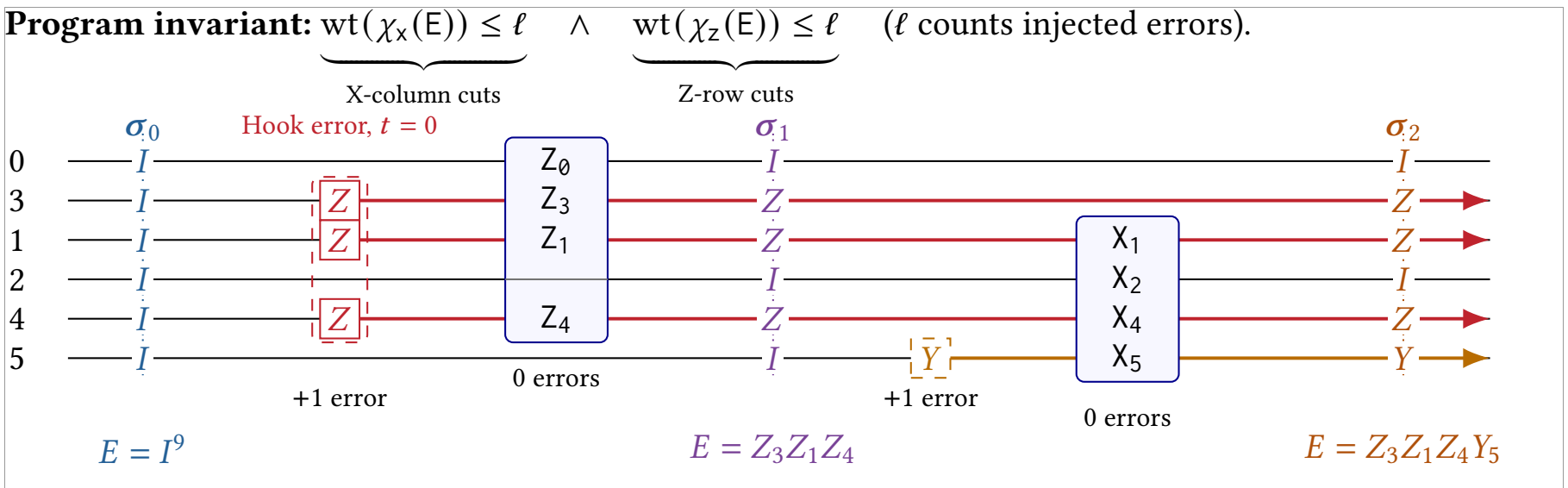
\begin{figure}[!htb]
\captionsetup{skip=3pt}
\begingroup\small\centering
\centering
\resizebox{.86\linewidth}{!}{%
\begin{minipage}{\linewidth}
\centering
\begin{figureframe}
\definecolor{invariantStateZero}{RGB}{35,95,150}
\definecolor{invariantStateOne}{RGB}{120,65,150}
\definecolor{invariantStateTwo}{RGB}{175,80,10}
\textbf{Program invariant:}
$\underbrace{\prog{\wt{\chi_x(E)}\le\ell}}_{\text{X-column cuts}}
 \quad\land\quad
 \underbrace{\prog{\wt{\chi_z(E)}\le\ell}}_{\text{Z-row cuts}}$
\quad($\ell$ counts injected errors).
\par\smallskip
\begin{tikzpicture}[x=1pt,y=1pt,line width=.45pt,
 every node/.style={font=\small,inner sep=1pt,outer sep=0pt}]
\path[use as bounding box] (0,-17) rectangle (390,71);
\foreach \addr/\yy in {0/60,3/50,1/40,2/30,4/20,5/10} {
 \node[anchor=west] at (0,\yy) {$\addr$};\draw (16,\yy) -- (377,\yy);
}
\foreach \yy in {50,40,20}
 \draw[pcqedFaultZ,line width=.85pt,->] (85,\yy) -- (377,\yy);
\draw[pcqedFaultY,line width=.85pt,->] (235,10) -- (377,10);
\draw[pcqedFaultZ,dashed,rounded corners=2pt] (78,14) rectangle (92,56);
\foreach \yy in {50,40,20}
 \node[draw=pcqedFaultZ,text=pcqedFaultZ,fill=white,minimum size=10pt] at (85,\yy) {$Z$};
\node[draw=pcqedFaultY,text=pcqedFaultY,fill=white,dashed,minimum size=10pt] at (235,10) {$Y$};
\node[font=\scriptsize,align=center,text=pcqedFaultZ] at (85,69) {Hook error, $t=0$};
\node[font=\scriptsize] at (85,0) {$+1$ error};
\node[font=\scriptsize] at (235,0) {$+1$ error};
\draw[fill=blue!3,draw=blue!55!black,rounded corners=2pt] (127,14) rectangle (153,66);
\draw[black!55] (127,30) -- (153,30);
\foreach \yy/\q in {60/0,50/3,40/1,20/4}
 \node at (140,\yy) {$\prog{Z_\q}$};
\draw[fill=blue!3,draw=blue!55!black,rounded corners=2pt] (272,4) rectangle (298,46);
\foreach \yy/\q in {40/1,30/2,20/4,10/5}
 \node at (285,\yy) {$\prog{X_\q}$};
\node[font=\scriptsize] at (140,5) {$0$ errors};
\node[font=\scriptsize] at (285,-5) {$0$ errors};
\foreach \xx/\state/\statecolor in {35/0/invariantStateZero,195/1/invariantStateOne,355/2/invariantStateTwo} {
 \draw[\statecolor,densely dotted] (\xx,67) -- (\xx,5);
 \node[text=\statecolor,font=\small\bfseries] at (\xx,68) {$\srcstate_\state$};
}
\foreach \yy in {60,50,40,30,20,10}
 \node[fill=white,text=invariantStateZero] at (35,\yy) {$I$};
\foreach \yy in {60,30,10}
 \node[fill=white,text=invariantStateOne] at (195,\yy) {$I$};
\foreach \yy in {50,40,20}
 \node[fill=white,text=invariantStateOne] at (195,\yy) {$Z$};
\foreach \yy in {60,30}
 \node[fill=white,text=invariantStateTwo] at (355,\yy) {$I$};
\foreach \yy in {50,40,20}
 \node[fill=white,text=invariantStateTwo] at (355,\yy) {$Z$};
\node[fill=white,text=invariantStateTwo] at (355,10) {$Y$};
\node[text=invariantStateZero] at (35,-13) {$E=I^9$};
\node[text=invariantStateOne] at (195,-13) {$E=Z_3Z_1Z_4$};
\node[text=invariantStateTwo] at (355,-13) {$E=Z_3Z_1Z_4Y_5$};
\end{tikzpicture}
\par
\smallskip\centering
\begin{tabular}{@{}lccc@{}}
 & $\srcstate_0$ & $\srcstate_1$ & $\srcstate_2$\\
\hline
$\ell$ & $0$ & $1$ & $2$\\
$\chi_x(E)$ & $(0,0,0):\ 0\le0$ & $(1,0,0):\ 1\le1$ & $(1,0,1):\ 2\le2$\\
$\chi_z(E)$ & $(0,0,0):\ 0\le0$ & $(0,0,0):\ 0\le1$ & $(0,1,0):\ 1\le2$
\end{tabular}
\end{figureframe}
\end{minipage}}
\caption{The first two QStab instructions of $S_3$. A hook error $Z_3Z_1Z_4$
and a data error $Y_5$ each cost one injection. Each measurement preserves
$E,\ell$. Both cut counts remain below three. At $\srcstate_2$, $G_1=1$ and all other entries are zero.}
\label{fig:qstab-state-invariant}
\Description{The first two ordered stabilizer measurements carry a hook
error and a Y5 data error. Three boundary states show the successive residuals.
A table evaluates both cut signatures and their injection-count bounds at each state.}
\endgroup
\end{figure}

Each error rule updates $(b,\ell)$ to $(b-1,\ell+1)$ and preserves every
unlisted field. Thus $b+\ell=B$, even if errors later cancel. The classes
can overlap: an ancilla Y error before the last CNOT can be viewed as a
one-qubit hook error with a readout change. Either classification assigns
it cost one.

The zero-cost \textsc{QS-Meas} step commits
$G_h\oplus\langle T_h,E\rangle_n$ and increments $h$, preserving
$E,b,\ell$~\cite[\S~6.2]{fang2024symbolic}. This is the commutation test
from Section~\ref{sec:qec-primer}: an error anticommuting with the measured
stabilizer flips its syndrome bit. Measurements can continue when $b=0$;
only an additional error exhausts the budget.\pcctight

Figure~\ref{fig:qstab-state-invariant} follows the first two measurements
of $S_3$. After the hook error, $E=Z_3Z_1Z_4$ and $\ell=1$, just as in
Figure~\ref{fig:pcc}(c). The next injection, $Y_5$, adds one crossed column
and one crossed row, giving signatures $(1,0,1)$ and $(0,1,0)$ at
$\ell=2$. It also anticommutes with the second stabilizer,
$X_1X_2X_4X_5$, so that measurement commits $G_1=1$. The two overlaps
of the earlier hook error, at qubits 1 and 4, cancel in this readout.
Appendix~\ref{app:qstab} shows how a readout error can hide such evidence.

\subsection{Assertions at every reachable state}
\label{sec:hoare}\label{sec:qhl}\label{sec:qhl-assertions}
\label{sec:qhl-triple}\label{sec:qhl-judgments}

Intermediate residuals can have nonzero syndrome: the hook error in
Figure~\ref{fig:qstab-state-invariant} is detectable, and later errors
may change its syndrome. We prove the cut invariant from
Equation~\eqref{eq:cut-overview-invariant} for every reachable QStab state.
It holds initially, and each step preserves it. Induction then covers
every measurement in every round~\cite{mannapnueli1995temporal,padon2016ivy}.
This is a parametric proof: $d$ remains arbitrary throughout the argument.\pcctight

Appendix~\ref{app:qstab} defines the source invariance judgment and the complete transition relation.
\label{sec:qstab-accept}\label{def:qstab-accept}
To prove the two cut bounds together, we use the common property of the
row and column selectors in Equation~\eqref{eq:surface-main-cuts}: each
data qubit belongs to at most one cut per collection. For example, qubit 5
belongs to column 2 and row 1, which explains the two effects of $Y_5$ in
Figure~\ref{fig:qstab-state-invariant}. We write $\nu$ for the number of
cuts and $L(\nu,q)$ for a total CodeML term selecting qubit $q$'s cut.
For the Surface code, $\nu=d$, and $L$ is the existing $L_x$ or $L_z$.
A value outside $[\nu]$ selects no cut. With Pauli kind $P$, the generic
cut and signature are
\begin{equation}\label{eq:cut-program}
\pccfitmath{
\kappa_i:=\prog{\lambda q.\ \kw{if}\ L(\nu,q)=i\ \kw{then}\ P\ \kw{else}\ I},
 \qquad \Nref{eq:cut-program}{\chi_{L,P}}(E)_i:=\prog{\langle\kappa_i,E\rangle_n}\quad(i<\nu).
}
\end{equation}
Each $\kappa_i$ is a Pauli string on $n$ data qubits, while
$\chi_{L,P}(E)$ has one bit per cut, hence length $\nu$.
The earlier signatures are exactly
$\colsig=\chi_{L_x,X}$ and $\rowsig=\chi_{L_z,Z}$, with $n=d^2$ and
$\nu=d$. We use $e_i$ for a signature with only bit $i$ set. At $d=3$,
$e_0=(1,0,0)$ is the hook signature from Figure~\ref{fig:pcc}(b).
Thus a data error $Z_q$ contributes $e_{\nrem{q}{d}}$ to $\colsig$ and
zero to $\rowsig$; $X_q$ contributes zero to $\colsig$ and
$e_{\nquot{q}{d}}$ to $\rowsig$. A Y error contributes to both and still
costs one injection.

\subsection{From hook error cut bounds to invariants}
\label{sec:qstab-hoare}\label{sec:qhl-prog}\label{sec:qhl-rules}
\label{sec:cut-proof-rules}\label{sec:qstab-dcirc}

A data error therefore crosses at most one cut per collection.
\emph{Why does the same bound hold when one injection reaches several
data qubits?} Return to the order $Z_0,Z_3,Z_1,Z_4$ in
Figure~\ref{fig:ordered-check-circuit}. Its column contributions are
$e_0,e_0,e_1,e_1$. Reading backwards, $Z_4$ crosses column 1;
adding $Z_1$ cancels that crossing; adding $Z_3$ leaves only column 0;
and adding $Z_0$ cancels that crossing too. Every suffix therefore crosses
at most one column, including the hook error $Z_3Z_1Z_4$ from Figure~\ref{fig:pcc}(b).
The whole stabilizer crosses none. We now express this calculation
parametrically so the proof covers every stabilizer and legal distance.
Appendix~\ref{app:source-logic} defines the suffix contributions and the hook-bound judgment.\pcctight
\begin{figure}[!htbp]
\captionsetup{skip=3pt}
\centering
\resizebox{.86\linewidth}{!}{%
\begin{minipage}{\linewidth}
\centering
\begin{figureframe}
\noindent
\begin{minipage}[c]{.14\linewidth}\centering
\RuleDef{Pair-Nil}\\[3pt]
$\inferrule{ }{\Gamma\vdash_{\mathrm{hook}}[]}$
\end{minipage}\hfill
\begin{minipage}[c]{.37\linewidth}\centering
\RuleDef{Pair-Step}\\[3pt]
$\inferrule
 {\Gamma\vdash_C\prog{v=v'}\quad\Gamma\vdash_C\prog{\wt{v'}\le1}\\
  \Gamma\vdash_{\mathrm{hook}}\vec v}
 {\Gamma\vdash_{\mathrm{hook}}v::v'::\vec v}$
\end{minipage}\hfill
\begin{minipage}[c]{.46\linewidth}\centering
\RuleDef{Hook-Check}\\[3pt]
$\inferrule
 {\Gamma\vdash_C\prog{\bigoplus\nolimits_{i<w}v_i=\mathbf0}\\
  \Gamma\vdash_C\prog{\forall j\le w.\ \wt{\bigoplus\nolimits_{j\le i<w}v_i}\le1}}
 {\Gamma\vdash_{\mathrm{hook}}[v_0,\ldots,v_{w-1}]}$
\end{minipage}
\end{figureframe}
\end{minipage}}
\caption{Read a rule as ``if the premises above the line hold, the conclusion
below follows.'' \Rule{Pair-Nil} starts with no couplings left.
\Rule{Pair-Step} adds two equal contributions before an already certified tail.
\Rule{Hook-Check} instead checks every coupling boundary directly.
All CodeML premises include definedness.}
\label{fig:suffix-rules}
\Description{Three rules establish the empty certificate, prepend an equal pair, or check every remaining cut contribution directly.}
\end{figure}

Figure~\ref{fig:suffix-rules} turns the cancellation into proof rules.
\Rule{Pair-Nil} starts after all couplings. \Rule{Pair-Step} prepends two
equal contributions $v,v'$ to a tail whose total contribution is zero.
Before the pair, $v\oplus v'=0$; inside it, only $v'$ remains, with weight
at most one. \Rule{Hook-Check} allows a direct proof of the same suffix
bounds when the pairing rule is not used.

Figure~\ref{fig:surface-pairing} gives the parametric version of the
four-coupling calculation, following program-annotation
proofs~\cite[Fig.~1]{zhou2024vsta}. Here
$\Gamma=\{d=2s+1,s\ge1,0\le a,b<d-1,(a+b)\bmod2=0\}$ and
$(q_0,q_1,q_2,q_3)=(da+b,d(a+1)+b,da+b+1,d(a+1)+b+1)$.
These are the addresses from the bulk Z branch of
Figure~\ref{fig:surface-code-and-cuts}, with $b$ naming the column.
At $d=3,a=b=0$, they are precisely $(0,3,1,4)$. For arbitrary $d$, the
contributions still form two equal pairs:
$v_0=v_1=e_b$ and $v_2=v_3=e_{b+1}$. The proof tree thus has the same
two \Rule{Pair-Step} applications at every distance.

\begin{figure}[!htbp]
\captionsetup{skip=3pt}
\small
\centering
\resizebox{.86\linewidth}{!}{%
\begin{minipage}{\linewidth}
\centering
\begin{figureframe}
\small\noindent
\begin{minipage}[t]{.43\linewidth}\vspace{0pt}
\textbf{Remaining cut contributions}\par\smallskip
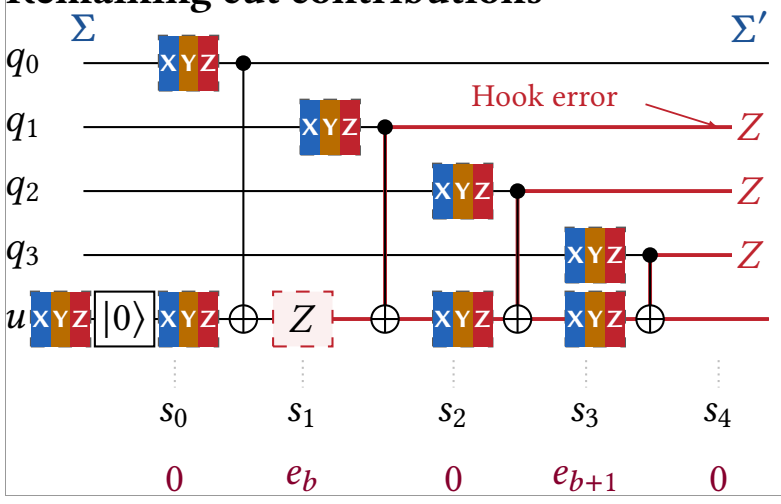
\begin{tikzpicture}[x=1pt,y=1pt,line width=.45pt,
 every node/.style={font=\small,inner sep=0pt,outer sep=0pt},
 cg/.style={draw,fill=white,minimum width=13pt,minimum height=12pt,inner sep=1pt},
 fault width=13pt,fault height=12pt,fault font=5.5,
 fault box/.style={minimum width=13pt,minimum height=12pt}]
\path[use as bounding box] (0,-38) rectangle (168,64);
\node[text=pcqedCodeKeyword] at (17,64) {$\tgtstate$};
\node[text=pcqedCodeKeyword] at (163,64) {$\tgtstate'$};
\foreach \name/\yy in {0/56,1/42,2/28,3/14} {
 \node[anchor=west] at (0,\yy) {$q_{\name}$};
 \draw (17,\yy) -- (167,\yy);
}
\node[anchor=west] at (0,0) {$u$};\draw (17,0) -- (167,0);
\draw[pcqedFaultZ,line width=.85pt] (65,0) -- (167,0);
\foreach \xx/\yy in {83/42,112/28,141/14}
 \draw[pcqedFaultZ,line width=.85pt] (\xx,0) -- (\xx,\yy) -- (167,\yy);
\node[font=\scriptsize,text=pcqedFaultZ,fill=white,inner sep=.3pt] at (118,49) {Hook error};
\draw[pcqedFaultZ,-{Latex[length=2.2pt]},thin] (140,47) -- (156,42);
\faultmark{12}{0}
\node[cg] at (26,0) {$|0\rangle$};
\foreach \xx/\yy in {52/56,83/42,112/28,141/14} {
 \draw (\xx,\yy) -- (\xx,0);
 \fill (\xx,\yy) circle[radius=1.7pt];
 \draw[fill=white] (\xx,0) circle[radius=3pt];
 \draw (\xx-3,0) -- (\xx+3,0);\draw (\xx,-3) -- (\xx,3);
 \faultmark{\xx-12}{\yy}
 \ifdim\xx pt=83pt\else\faultmark{\xx-12}{0}\fi
}
\node[cg,draw=pcqedSigZ,dashed,fill=pcqedSigZ!7] at (65,0) {$Z$};
\foreach \yy in {42,28,14}
 \node[anchor=east,text=pcqedSigZ,fill=white,inner sep=1pt] at (167,\yy) {$Z$};
\foreach \xx/\j/\val in {37/0/0,65/1/{e_b},98/2/0,127/3/{e_{b+1}},156/4/0} {
 \draw[black!25,densely dotted] (\xx,-9) -- (\xx,-15);
 \node at (\xx,-21) {$s_{\j}$};
 \node[text=purple!70!black] at (\xx,-35) {$\val$};
}
\end{tikzpicture}
\end{minipage}\hfill
\begin{minipage}[t]{.54\linewidth}\vspace{0pt}
\textbf{Certificate for every boundary}\par\smallskip
\begin{subequations}\label{eq:surface-pair-empty}\label{eq:surface-pair-last}
\renewcommand{\theequation}{\theparentequation}
\begin{equation}\label{eq:surface-pair-first}
\inferrule
 {\inferrule
   {\inferrule{ }{\Gamma\vdash_{\mathrm{hook}}[]}
    \\ \Gamma\vdash_C\prog{e_{b+1}=e_{b+1}}\quad\Gamma\vdash_C\prog{\wt{e_{b+1}}\le1}}
   {\Gamma\vdash_{\mathrm{hook}}[e_{b+1},e_{b+1}]}
  \\ \Gamma\vdash_C\prog{e_b=e_b}\quad\Gamma\vdash_C\prog{\wt{e_b}\le1}}
 {\Gamma\vdash_{\mathrm{hook}}[e_b,e_b,e_{b+1},e_{b+1}]}
\end{equation}
\end{subequations}
Start with no couplings. Add the last pair, then the first.
Each leaves either zero or one cut.
\end{minipage}
\end{figureframe}
\end{minipage}}
\caption{A paired Surface code stabilizer and its hook-error cut certificate.
The selected helper error leaves $Z_{q_1}Z_{q_2}Z_{q_3}$ on data at cost one.
At $d=3,a=b=0$, this is $Z_3Z_1Z_4$.}
\label{fig:surface-pairing}
\Description{A five-wire circuit sits beside a proof tree for the cut bound on hook errors.
A Z error on a helper qubit before the second CNOT propagates to the last three
data operands. Under the couplings, five remaining cut contributions are zero,
e b, zero, e b plus one, zero. The adjacent tree applies Pair-Nil and
two Pair-Step rules to the list e b, e b, e b plus one, e b plus one.}
\end{figure}

The bulk X branch of Figure~\ref{fig:surface-code-and-cuts} pairs
contributions along rows instead. Its boundary branches have two operands
in the same row or column, contributing $[e_i,e_i]$ to that collection
and zeros to the other. These cases establish both hook bounds for every
legal $d$.

We can now prove the program property from Section~\ref{sec:cut-intuition}.
Initially, $E=I^n$ crosses no cuts and $\ell=0$. Commutation parity gives
$\chi(EH)=\chi(E)\oplus\chi(H)$, where $\chi=\chi_{L,P}$.
The data and hook bounds show that an error step increases each cut count
by at most one, while also increasing $\ell$ by one. Readout-only errors
increase $\ell$ without changing either count; measurement steps change
neither $E$ nor $\ell$. This is exactly the progression shown in
Figure~\ref{fig:qstab-state-invariant}, now for an arbitrary step of cost
$k$ and arbitrary $d$:
\begin{equation}\label{eq:cost-preservation}
\pccfitmath{
\wt{\chi_{L,P}(E')}\le\wt{\chi_{L,P}(E)}+k\le\ell+k=\ell'.
}
\end{equation}

\subsection{Putting the Surface code proof together}
\label{ex:surface-source}

We now assemble the premises needed for the invariant triple in
Equation~\eqref{eq:overview-invariant-triple} and, eventually, the
circuit-level distance guarantee. Each premise comes from a part of the
argument already explained.

First, Section~\ref{sec:codeml-elaboration} establishes that
$\mathsf{Generate}\llbracket C_{\rm surf}\rrbracket(d;d,d-1)$ succeeds
and returns $S_d$ with defined, distinct, in-range operands, positive data
width, stabilizer count and round count, and uniform X- or Z-type stabilizers.
We call this parametric generation proof $\mathcal G$.
Second, Lemma~\ref{lemma:surface-main-algebra} supplies proofs $V$ of
$\mathsf{WellFormed}$ and $W$ of $\mathsf{Complete}_{L_x,L_z}$.
As explained in Section~\ref{sec:assertion-logic}, these establish that
the generated operators define a stabilizer code and that a logical
error must cross all $d$ cuts in some collection.\pcctight

Third, the hook proofs $\mathcal P_z,\mathcal P_x$ from
Section~\ref{sec:cut-proof-rules} establish the row and column bounds.
For example, the two equal pairs in Figure~\ref{fig:surface-pairing}
prove that every suffix of $Z_0Z_3Z_1Z_4$ crosses at most one column.
Equation~\eqref{eq:cost-preservation} combines this hook bound with the
data and readout cases to show that an arbitrary QStab step preserves
the invariant. We name the source invariant $I_d$ and use $\Phi_d$ from
the overview for the same assertion on target data:
\begin{equation}\label{eq:family-invariants}
\pccfitmath{
I_d(E,\ell):=\wt[d]{\chi_{L_z,Z}(E)}\le\ell
                   \land\wt[d]{\chi_{L_x,X}(E)}\le\ell,\qquad
 \Phi_d(\tgtstate):=I_d(E_{\rm raw}(\tgtstate),\ell(\tgtstate)).
}
\end{equation}
\begin{figure}[!htbp]
\captionsetup{skip=3pt}
\small
\centering
\resizebox{.86\linewidth}{!}{%
\begin{minipage}{\linewidth}
\centering
\begin{figureframe}
\centering\small

\setlength{\abovedisplayskip}{3pt}\setlength{\belowdisplayskip}{3pt}
\begin{mathpar}[\lineskiplimit=4pt\lineskip=4pt]
\inferrule*[right=\RuleDef{Src-Cut}]
 {\Gamma\vdash\Nref{def:hook-bound}{\mathsf{HookBound}}_{L,P}(S)}
 {\Gamma\mid S\vdash\Box(\prog{\wt{\Nref{eq:cut-program}{\chi_{L,P}}(E)}\le\ell})}
\and
\inferrule*[right=\RuleDef{Src-Conj}]
 {\Gamma\mid S\vdash\Box(\assertion{A})\\\Gamma\mid S\vdash\Box(\assertion{A}')}
 {\Gamma\mid S\vdash\Box(\prog{\assertion{A}\land\assertion{A}'})}
\and
\inferrule*[right=\RuleDef{Src-Dist}]
 {\Gamma\mid S\vdash\Box(\prog{\wt{\Nref{eq:cut-program}{\chi_{L_z,Z}}(E)}\le\ell\land\wt{\Nref{eq:cut-program}{\chi_{L_x,X}}(E)}\le\ell})\\
  \Gamma\vdash_C\Nref{eq:valid-complete}{\mathsf{WellFormed}}\quad\Gamma\vdash_C\Nref{eq:valid-complete}{\mathsf{Complete}}_{L_x,L_z}}
 {\Gamma\mid S\vdash\Box(\prog{\ell<\nu\Rightarrow\neg\Nref{eq:code-predicates}{\mathsf{Logical}}(E)})}
\end{mathpar}
\end{figureframe}
\end{minipage}}
\caption{Source proof rules: establish both cut bounds, conjoin them,
and exclude logical errors below $\nu$ errors. Both selectors have $\nu$ cuts.
All premises retain the same code body, fuel, width and static context.}
\label{fig:qhl-rules}
\Description{Cut establishes each count invariant, conjunction combines them, and distance adds code well-formedness and cut completeness.}
\end{figure}

Figure~\ref{fig:qhl-rules} expresses these steps as proof rules.
\Rule{Src-Cut} and \Rule{Src-Conj} build $\mathcal D_{\rm inv}$;
\Rule{Src-Dist} adds $V,W$ to exclude logical errors when $\ell<d$:
\begin{equation}\label{eq:source-family-tree}
\pccfitmath{
\mathcal D_{\rm inv}
  =\textsc{Src-Conj}(\textsc{Src-Cut}(\mathcal P_z),
                     \textsc{Src-Cut}(\mathcal P_x)),\qquad
 \mathcal D=\textsc{Src-Dist}(\mathcal D_{\rm inv},V,W).
}
\end{equation}
The complete proof uses two inductions for different purposes. Induction
on odd $d$ proves the code-algebra facts. With $d$ arbitrary, induction on
execution proves that errors preserve the invariant through all rounds.
Together they give one parametric QStab derivation:
\begin{equation}\label{eq:surface-source-family-conclusion}
\pccfitmath{
\mathcal D:\quad\Gamma_{\rm surf}\mid S_d\vdash
 \Box\bigl(\prog{\ell<d\Rightarrow\neg\mathsf{Logical}(E)}\bigr).
}
\end{equation}
\emph{What remains between this source derivation and the circuit-level
guarantee from the Introduction?} We must establish the target cut premises
corresponding to $\mathcal D_{\rm inv}$ for the actual circuit. The two states observe the
same data residual at measurement boundaries, but circuit errors may
first occur on helpers and propagate through several gates. The
generation proof $\mathcal G$ and the gate argument in
Section~\ref{ex:surface-gate-proof} justify that connection.
Appendix~\ref{app:surface-derivations} gives the complete source and target derivations.

\section{QClifford: Translating cut certificates by example}
\label{sec:qclifford}\label{sec:compiler}
QStab lets us prove the invariant parametrically over code distance and
syndrome-extraction scheme. Our implementation translates the supported
cut certificates: it preserves their logical structure and code-algebra
premises, and replaces source cut leaves by target checks of propagated
error effects. Users supply the family proofs; backend developers prove
the scheme-specific propagation lemmas. Circuit emission, error-image
construction and structural certificate translation are automatic for an
instantiated program.\pcctight

We explain the backend argument using the same two errors in
Figure~\ref{fig:qstab-state-invariant}: the hook $Z_3Z_1Z_4$ and the
data error $Y_5$. The key is to account for the part of an error still
on a helper before it reaches the data.\pcctight

\phantomsection\label{sec:qcliff-syntax}\label{sec:compiler-translation}
Appendix~\ref{app:qclifford} defines QClifford and Appendix~\ref{app:target-logic} gives its Hoare rules.
QClifford uses standard Clifford-circuit syntax and Pauli-frame
notation~\cite{gottesman2009QECIntro}. We refer readers to the standard
constructions for single-helper measurements
(Standard)~\cite{gottesman2009QECIntro}, flagged measurements
(Flag)~\cite{chao2020flag}, cat-state extraction
(Shor)~\cite{shor1996fault}, and teleportation-based extraction
(Knill)~\cite{knill2005quantum}. The supplement fixes our circuit
templates and readout contracts. Here we work through the Standard
Scheme circuit already shown in Figure~\ref{fig:ordered-check-circuit}.\pcctight

\subsection{Following the hook error through the gates}
\label{sec:qcliff-state}\label{sec:qcliff-sem}\label{sec:bg-prop}
Recall the target state $\tgtstate=(\mathcal E,\omega,j_{\rm rec},\ell)$
from Figure~\ref{fig:ordered-check-circuit}: $\mathcal E$ records Pauli
errors on data and helpers, $\omega$ records readout flips,
$j_{\rm rec}$ counts readouts, and $\ell$ counts injections.
Its data restriction is $E_{\rm raw}$, and the clean state is
$\cleanstate=(I^N,\mathbf0,0,0)$ for $N$ total wires. Each operand has a
pre-operation $\kw{err}(q)$ site that idles or injects X, Y or Z at cost
one, following Section~\ref{sec:qec-primer}'s error model.

The first measurement in Figure~\ref{fig:qstab-state-invariant} has
ordered operands $(0,3,1,4)$. Its compiled circuit prepares helper
$u=9$, applies four CNOTs with these data controls and target $u$, and
measures $u$ in Z. Select $Z_u$ at site A in
Figure~\ref{fig:ordered-check-circuit}, after the first CNOT, and idle
elsewhere. CNOT propagates Z from target to control, so the next three
gates successively add $Z_3$, $Z_1$ and $Z_4$ to the data.
Figure~\ref{fig:compiled-hook-proof} places the selected QStab step
above a local QClifford derivation. The source step adds the whole
hook at once. The target instead injects $Z_u$ and propagates it through
three CNOTs, with a different assertion $B_i$ at each boundary.
The $B_i$ annotations inside the circuit give the data residual,
combined cut vector and injection count. We XOR the current column cuts with
the helper's pending contribution to obtain $V_j$; the assertion $A_j$
bounds this vector and the row signature by $\ell$.
Thus the intermediate assertions can establish the same cut bound even
before the complete hook has reached the data. At $B_0$, the source
precondition is extended with a clean helper. At $B_4$, restricting the
target frame to data recovers the source postcondition; the remaining
helper Z does not flip the subsequent Z readout.\pcctight

\begin{figure}[!htbp]
\captionsetup{skip=3pt}
\centering
\begin{figureframe}
\begingroup
\definecolor{pccEightTypeZero}{RGB}{30,100,175}
\definecolor{pccEightTypeOne}{RGB}{185,92,0}
\definecolor{pccEightTypeTwo}{RGB}{133,55,157}
\definecolor{pccEightTypeThree}{RGB}{0,119,92}
\centering\small
\textbf{(a) QStab effect: one Type-2 step, $H=Z_3Z_1Z_4$, $t=0$.}\par
\begingroup
\setlength{\abovedisplayskip}{2pt}\setlength{\belowdisplayskip}{2pt}
\[
\pccfitmath[.98\linewidth]{
\underbrace{\{E=I^9,\ G_h=0,\ \ell=0\}}_{I_3:\ 0\le0}\;
\textcolor{pccEightTypeTwo}{\textsc{QS-Hook}(H,t)}\;
\underbrace{\{E=Z_3Z_1Z_4,\ G_h=0,\ \ell=1\}}_{I_3:\ 1\le1\ \land\ 0\le1}.
}
\]
\endgroup
\textcolor{pccEightTypeTwo}{$\Downarrow$\quad
 \textit{A local gate derivation of the same effect and cost.}}
\par\smallskip
\resizebox{.94\linewidth}{!}{%
\begin{tikzpicture}[x=1pt,y=1pt,line width=.45pt,
 every node/.style={font=\small,inner sep=0pt,outer sep=0pt},
 pccEightGate/.style={draw,fill=white,minimum width=13pt,
   minimum height=12pt,inner sep=1pt},
 pccEightAssertion/.style={font=\scriptsize,align=center,
   fill=white,inner sep=.7pt,text=pccEightTypeTwo},
 pccEightFlow/.style={line width=.85pt},
 fault width=13pt,fault height=12pt,fault font=5.5,
 fault box/.style={minimum width=13pt,minimum height=12pt}]
\path[use as bounding box] (0,-53) rectangle (399,159);
\node[anchor=west,font=\footnotesize\bfseries] at (3,154)
 {(b) Local QClifford derivation: $\{B_0\}$ injection; three CNOTs $\{B_4\}$};
\foreach \q/\yy in {0/110,3/80,1/50,4/20} {
 \node[anchor=east] at (12,\yy) {$\q$};
 \draw (18,\yy) -- (397,\yy);
}
\node[anchor=east] at (12,0) {$u$};
\draw (18,0) -- (397,0);

\faultmark{20}{0}
\node[pccEightGate] at (37,0) {$|0\rangle$};
\faultmark{58}{0}
\faultmark{144}{80}
\faultmark{224}{50}\faultmark{224}{0}
\faultmark{304}{20}

\draw[pccEightFlow,pccEightTypeZero,-{Latex[length=2.2pt]}]
 (52,110) -- (52,114) -- (396,114);
\draw[pccEightFlow,pccEightTypeTwo,-{Latex[length=2.2pt]}]
 (106,0) -- (106,8) -- (374,8);
\foreach \xx/\yy in {159/84,239/54,319/24}
 \draw[pccEightFlow,pccEightTypeTwo,-{Latex[length=2.2pt]}]
  (\xx,8) -- (\xx,\yy) -- (396,\yy);
\draw[pccEightFlow,pccEightTypeOne,-{Latex[length=2.2pt]}]
 (294,0) -- (294,-8) -- (374,-8);
\draw[pccEightFlow,pccEightTypeOne,-{Latex[length=2.2pt]}]
 (312,-8) -- (312,16) -- (396,16);
\draw[pccEightFlow,pccEightTypeThree,-{Latex[length=2.2pt]}]
 (362,0) -- (362,-16) -- (376,-16);

\foreach \xx/\yy in {70/110,156/80,236/50,316/20} {
 \draw (\xx,\yy) -- (\xx,0);
 \fill (\xx,\yy) circle[radius=1.7pt];
 \draw[fill=white] (\xx,0) circle[radius=3pt];
 \draw (\xx-3,0) -- (\xx+3,0);\draw (\xx,-3) -- (\xx,3);
}
\node[pccEightGate] at (384,0) {$M_Z$};
\foreach \xx/\yy/\letter/\typecolor in
 {52/110/Z/pccEightTypeZero,106/0/Z/pccEightTypeTwo,
  294/0/Y/pccEightTypeOne,362/0/X/pccEightTypeThree}
 \node[pccEightGate,draw=\typecolor,dashed,fill=\typecolor!7,
   text=\typecolor] at (\xx,\yy) {$\letter$};
\node[font=\scriptsize,text=pccEightTypeZero,anchor=east] at (53,126) {D: Type-0};
\node[font=\scriptsize,text=pccEightTypeTwo] at (106,-17) {A: Type-2};
\node[font=\scriptsize,text=pccEightTypeOne] at (282,-20) {B: Type-1};
\node[font=\scriptsize,text=pccEightTypeThree] at (357,-26) {M: Type-3};

\node[font=\scriptsize,text=pccEightTypeZero,fill=white,inner sep=.5pt] at (388,114) {$Z$};
\foreach \xx/\yy in {194/84,274/84,348/84,274/54,348/54,348/24,134/8,194/8,274/8,338/8}
 \node[font=\scriptsize,text=pccEightTypeTwo,fill=white,inner sep=.5pt]
  at (\xx,\yy) {$Z$};
\node[font=\scriptsize,text=pccEightTypeOne,fill=white,inner sep=.5pt] at (382,16) {$Z$};
\node[font=\scriptsize,text=pccEightTypeOne,fill=white,inner sep=.5pt] at (338,-8) {$Y$};

\node[pccEightAssertion] at (91,137)
 {$\{B_0:E=I^9\}$\\[-1pt]$V=0,\ \ell=0$};
\node[pccEightAssertion] at (132,96)
 {$\{B_1:E=I^9\}$\\[-1pt]$V=e_0,\ \ell=1$};
\node[pccEightAssertion] at (194,66)
 {$\{B_2:E=Z_3\}$\\[-1pt]$V=e_0,\ \ell=1$};
\node[pccEightAssertion] at (274,96)
 {$\{B_3:E=Z_3Z_1\}$\\[-1pt]$V=e_0,\ \ell=1$};
\node[pccEightAssertion] at (348,66)
 {$\{B_4:E=Z_3Z_1Z_4\}$\\[-1pt]$V=e_0,\ \ell=1$};
\draw[black!30,densely dotted] (91,125) -- (91,117);
\foreach \xx/\yy in {132/86,194/56,274/86,348/56}
 \draw[black!30,densely dotted] (\xx,\yy) -- (\xx,\yy-3);
\node[pccEightAssertion] at (314,37)
 {$\{B_3\}\;\mathsf{CNOT}[4,u]\;\{B_4\}$\\[-1pt]
  $V_3=(1,1,0)\oplus(0,1,0)=(1,0,0)=V_4$};
\node[pccEightAssertion] at (169,-18)
 {$\{B_0\}\;[\kw{err}(u)=Z]\;\{B_1\}$};

\node[anchor=west,font=\scriptsize,text=pccEightTypeZero] at (3,-37)
 {Type-0: $\textsc{QS-Data}$; $E=Z_0$, $\omega(0)=0$};
\node[anchor=west,font=\scriptsize,text=pccEightTypeOne] at (205,-37)
 {Type-1: $\textsc{QS-DataReadout}$; $E=Z_4$, $\omega(0)=1$};
\node[anchor=west,font=\scriptsize,text=pccEightTypeTwo] at (3,-50)
 {Type-2: $\textsc{QS-Hook}$; $E=Z_3Z_1Z_4$, $\omega(0)=0$};
\node[anchor=west,font=\scriptsize,text=pccEightTypeThree] at (205,-50)
 {Type-3: $\textsc{QS-Readout}$; $E=I^9$, $\omega(0)=1$};
\end{tikzpicture}}
\endgroup
\end{figureframe}
\caption{Relating one abstract effect to concrete propagation.
Colored paths are separate single-injection runs, each costing one;
other sites idle. Purple assertions describe the Type-2 branch, with
$E=E_{\rm raw}$, $e_0=(1,0,0)$, and $V=\chi_x(E)\oplus z_us_j$.
The source budget is two. The helper is clean at $B_0$ and carries Z
at $B_1,\ldots,B_4$; row cuts and readout flips remain zero.}
\label{fig:compiled-hook-proof}
\Description{A source triple sits above a local derivation on one five-wire circuit
in the style of Figure 6. The four color-coded alternatives are Z on
data 0, Z on the helper after the first coupling, Y on the helper
before the last coupling, and X on the helper before readout.
Propagation paths are slightly offset to keep them separate. Assertions
B0 through B4 sit between the wires and show the changing data residual,
combined cut vector and injection count. Two local triples inside the
circuit show the injection step and the final CNOT cancellation.
A two-row legend maps each path to its QStab rule and final output.}
\end{figure}

\emph{How do we account for the helper's pending contribution between
gates?} After the third coupling, $E_{\rm raw}=Z_3Z_1$ crosses two
columns, although $\ell=1$. The helper still carries $Z_u$, which will
add $Z_4$ and cancel the crossing in column 1. This is the same
$1\oplus1=0$ calculation on qubits 1 and 4 from
Section~\ref{sec:qec-primer}. An instruction-level derivation must include
this pending contribution.

\subsection{A local derivation of the target cut bound}
\label{ex:surface-gate-proof}\label{sec:fhoare-rules}
The source hook proof already computes what remains after each
coupling. Evaluate Figure~\ref{fig:surface-pairing}'s suffixes at
$d=3,a=b=0$, with $e_0=(1,0,0)$ and $e_1=(0,1,0)$:
\begin{equation}\label{eq:surface-target-suffixes}
\pccfitmath{
(s_0,s_1,s_2,s_3,s_4)=(0,e_0,0,e_1,0),\qquad
 s_j=e_{q_j\bmod3}\oplus s_{j+1},\quad(q_0,q_1,q_2,q_3)=(0,3,1,4).
}
\end{equation}
Let $z_u$ indicate whether the helper carries Z or Y. Only then does it
contribute $s_j$. We add that contribution to the data signature and
use the same bound as in the source invariant:
\begin{equation}\label{eq:surface-target-pending}
\pccfitmath{
V_j:=\chi_x(E_{\rm raw})\oplus z_us_j,\qquad
A_j:=\wt{V_j}\le\ell\ \land\ \wt{\chi_z(E_{\rm raw})}\le\ell,
\qquad A_0=A_4=\Phi_3.
}
\end{equation}
The last equality follows from $s_0=s_4=0$: at measurement boundaries,
we recover exactly the target invariant of
Equation~\eqref{eq:family-invariants}.

\paragraph{An injected error spends one unit of the bound.}
A helper Z or Y toggles $z_u$ and therefore adds $s_j$ to $V_j$.
Every displayed $s_j$ has weight at most one, so
$\wt{V_j\oplus s_j}\le\wt{V_j}+1\le\ell+1=\ell'$.
A helper X changes no data cut. A data injection changes at most one
bit in each cut collection, and satisfies the same inequality.
Idle changes nothing. Thus every choice at an injection site preserves
$A_j$, for any incoming state satisfying it. At site A, in particular,
the clean state becomes $V_1=e_0$, $\ell=1$, in assertion $B_1$ of
Figure~\ref{fig:compiled-hook-proof}(b).\pcctight

\paragraph{A gate transfers the pending contribution to data.}
CNOT $[q_j,u]$ adds $z_ue_{q_j\bmod3}$ to the data signature and leaves
$z_u$ and $\ell$ unchanged. Using the suffix recurrence gives
\begin{equation}\label{eq:surface-target-cnot}
\pccfitmath{
V_{j+1}'=(\chi_x(E_{\rm raw})\oplus z_ue_{q_j\bmod3})\oplus z_us_{j+1}
=\chi_x(E_{\rm raw})\oplus z_us_j=V_j.
}
\end{equation}
The row signature $\chi_z$ is unchanged too. For the last gate, the
calculation is explicitly
$(1,1,0)\oplus(0,1,0)=(1,0,0)$ before the gate and
$(1,0,0)\oplus(0,0,0)=(1,0,0)$ after it.
Preparation and readout leave data unchanged at boundaries where
the pending term is zero.

\phantomsection\label{def:hoare-triple-c}\label{def:hoare-diagram}
Consequently the instruction proofs compose into
$\vdash_F\{\Phi_3\}\,\bar c_{3,0}\,\{\Phi_3\}$ for this first
measurement block. The Hoare triple covers every incoming state
satisfying $\Phi_3$ and every error choice~\cite{hoare1969axiomatic};
the displayed execution illustrates its gate calculation.
Replacing the selected helper Z by Y gives the same hook and also a
readout flip, precisely as in Figure~\ref{fig:ordered-check-circuit}'s
table. This is how knowing the scheme makes QStab's abstract hook and
readout choices concrete.

We can now continue to the second error in
Figure~\ref{fig:qstab-state-invariant}. Injecting $Y_5$ adds one column
bit and one row bit at cost one:
\begin{equation}\label{eq:compiled-second-error}
\pccfitmath{
\chi_x(Z_3Z_1Z_4Y_5)=(1,0,0)\oplus(0,0,1)=(1,0,1),\quad
\chi_z(Z_3Z_1Z_4Y_5)=(0,1,0),\quad\ell=2.
}
\end{equation}
Thus the two cut counts are $2$ and $1$, both at most $\ell$.
The second measurement is X-type: its gate argument uses the same pending
contribution for row cuts and transports the assertion through its
Hadamard basis changes. The supplement gives these local rules.
At its boundary we recover $\Phi_3$ again, with exactly the residual
and counts in Figure~\ref{fig:qstab-state-invariant}.

\subsection{From the example to the parametric circuit guarantee}
\label{sec:compiler-preservation}\label{sec:external-compile-tree}
\label{sec:family-proof}
\emph{How does this calculation complete the parametric proof from
Section~\ref{ex:surface-source}?} The backend proof generalizes the
preceding gate argument over the ordered operands and coupling index.
For a bulk Z stabilizer, Figure~\ref{fig:surface-pairing} replaces
$e_0,e_1$ by $e_b,e_{b+1}$; the cancellation is unchanged for every
legal $d$. The X and boundary cases come from the same CodeML branches
in Figure~\ref{fig:surface-code-and-cuts}. The generation proof
$\mathcal G$ fixes their operands and order; the hook proofs
$\mathcal P_x,\mathcal P_z$ bound their pending contributions.

\phantomsection\label{eq:fault-images-main}
Propagating one injection through the remaining circuit gives its
\emph{error image}. The backend lemmas establish that each
record-corrected image crosses at most one cut per collection; for Standard
and Flag the correction is empty. The structural translator replaces each
\Rule{Src-Cut} leaf of $\mathcal D$ by a target cut node and erases that
leaf's source branch and pairing derivations. The target checker checks
the bound on the actual circuit's images. Conjunction, consequence and
the code-algebra proofs $V,W$ are reused. The source derivations justify
the backend lemmas' premises in the preservation proof. This is a structural certificate
translation~\cite{shao2002certified,barthe2006certificate};
The supplement's ``Structural proof translation and checking'' section gives its rules.

\emph{Why do these local checks cover every error combination?}
The circuit includes an injection site for every operation operand, and
image construction covers X, Y and Z at every site; idle contributes zero.
For any clean-start execution with $\ell$ injections,
linearity of Pauli propagation expresses its final data residual as the
product of their data images $E_1,\ldots,E_\ell$. For either cut collection,
\begin{equation}\label{eq:compiled-image-composition}
\pccfitmath{
 E_{\rm raw}=\prod_{i=1}^{\ell}E_i,\qquad
 \chi(E_{\rm raw})=\bigoplus_{i=1}^{\ell}\chi(E_i),\qquad
 \wt{\chi(E_{\rm raw})}\le\sum_{i=1}^{\ell}\wt{\chi(E_i)}\le\ell.
}
\end{equation}
This execution-composition theorem covers cancellations and multiple
injections. The purple path in Figure~\ref{fig:compiled-hook-proof}
illustrates the image of the helper Z injection. The supplement proves the corresponding
composition for record-corrected observations in the other schemes.
The block-composition and final consequence steps appear in Appendix~\ref{app:surface-derivations}.

\phantomsection\label{def:standard-lift}
The invariant excludes logical failure for every record when $\ell<d$;
when $\ell\ge d$, the postcondition already holds. Thus a certificate
with budget $d-1$ establishes the final triple for every execution. It uses
$A_d:=\ell<d\Rightarrow\neg\mathsf{Logical}(E)$, interpreted on
$E=E_{\rm raw}$ and $b=d-1-\ell$.

\begin{theorem}[Standard Scheme cut-certificate translation]\label{thm:standard-compile-preserve}
With the Standard Scheme propagation lemmas above, the supplied symbolic
Surface code source derivation $\mathcal D$ and proof
$\mathcal G$ of $S_d=\mathsf{Generate}\llbracket C_{\rm surf}\rrbracket(d;d,d-1)$
yield one finite derivation schema
$\mathcal H_{\mathsf{Std}}$ of Equation~\eqref{eq:headline-family-triple}.
For every valuation satisfying $\Gamma_{\rm surf}$, it instantiates to
an error-aware Hoare derivation for the corresponding emitted circuit.
Its target certificate $\mathcal D'$ proves $A_d$ at budget $d-1$.
\end{theorem}
The derivation in Appendix~\ref{app:surface-derivations} proves the theorem.
Its size is independent of $d$: it stores the finite code-algebra proofs,
the symbolic bulk and boundary gate cases, and the induction step,
including their proof dependencies. Expanded circuits still grow with $d$.
Appendix~\ref{app:distance} defines the record and detector predicates used in the circuit-level distance.

\paragraph{A matching silent failure.}
\label{sec:endtoend}\label{sec:standard-exactness}
The full Z row in Figure~\ref{fig:pcc}(a) suggests the matching upper
bound. On each top-row qubit, inject X before its last H, or Z before
its last data-controlled CNOT. Each becomes a final Z without reaching
a helper. These $d$ injections leave the logical row
$\prod_{b<d}Z_{q_d(0,b)}$ and zero readout flips, so every detector is
silent. At $d=3$ the row is $Z_0Z_1Z_2$.
\begin{corollary}[Surface code circuits compiled with the Standard Scheme]
\label{cor:parametric-standard}
For $S_d$ compiled by the Standard Scheme translation defined above and any declared
record-parity detector family $\mathcal J_d$,
$\forall d\in\mathbb N.\ (d\ge3\land d\bmod2=1)
\Rightarrow d^{\rm circ}_{K_{d,\mathsf{Std}}}
=d^{\rm circ}_{K_{d,\mathsf{Std}},\mathcal J_d}=d$.
\end{corollary}

\paragraph{Changing the extraction scheme.}\label{sec:trust}\label{sec:limitations}
The same method is parametric over schemes through QStab's error
interface: code algebra is reused, while each backend must justify its
own propagated errors and readout effects. Standard and Flag bound the
raw data residual for every record. Shor and Knill use virtual Pauli
corrections determined by ancilla-check and Bell-measurement readout
flips, respectively. Each backend's
$\mathsf{RecordOK}_K$ ensures the corrected residual agrees with
$E_{\rm raw}$ when claiming a raw-residual distance. The supplement
gives the translations and obligations, and
Section~\ref{sec:verifier} explains how checking binds the proof to the
emitted circuit and its contract.

\section{Evaluation}
\label{sec:eval}\label{sec:surface}
The evaluation demonstrates family-wide certification across five code
families, structural source-certificate translation for a finite Shor code program, and
certificate checking for eight finite target circuits
(Table~\ref{tab:intro-results}). It measures checking time and shows how
extraction choices affect the certified distance.\pcctight

\paragraph{Certificate checking.}
\phantomsection
\label{sec:verifier}\label{sec:verifier-function}\label{def:standard-contract}
\label{sec:standard-checking}\label{sec:qcliff-pcc}\label{sec:target-cut-rules}
\label{eq:certificate}\label{eq:standard-certificate}\label{eq:target-check}
The certificate binds the proof to the ordered program, error model and
record contract: changing the coupling order in Figure~\ref{fig:pcc}(c)
can change its hook errors. For a parametric family, the checker verifies
the generation equation, source derivation $\mathcal D$, proof translation
and every cited CodeML premise, including each induction's base case and
arbitrary step, with $d$ symbolic. At a fixed distance, the certificate
$\mathcal C=(\bar c,\mathcal I,\mathcal D')$ contains the circuit, its
single-injection images and the translated proof. The checker regenerates
the instructions, recomputes the images by a shared backward pass, and
checks the cut bounds and remaining proof premises
(Appendix~\ref{app:evaluation-results}).

\subsection{Experimental setup}
We develop the verifier in OCaml. It is the trusted executable checker for symbolic family
certificates and finite target certificates. Lean mechanizes the metatheory and exactness for canonical
rotated Surface, unrotated Surface and Toric code programs.
Table~\ref{tab:intro-results} reports $d^{\rm circ}_K$ from
Definition~\ref{def:dcirc}. Its lower bounds transfer to detector silence
because $d^{\rm circ}_{K,\mathcal J}\ge d^{\rm circ}_K$.
We sum verifier wall times from process launch to completion, including
loading, on an Intel i5-12400 Windows host with 64\,GiB RAM.
Appendix~\ref{app:evaluation-results} gives the timed stages and measurement counts.

\subsection{Family-wide distance guarantees}
\emph{Can one certificate establish a useful distance guarantee for every
size of a code family?} For rotated Surface code, the verifier checks one
supplied Standard Scheme certificate proving a circuit-level code distance
lower bound $d$ for every odd $d\ge3$ in \textbf{214\,s on average}.
The matching silent execution in \S~\ref{sec:standard-exactness} establishes
equality for the same ordered schedule.

The other family checks take 23.9\,s for unrotated Surface code, 3692\,s for
Cartesian LP code, 9.48\,s for Toric code, and 5511\,s for triangular color
code. The interleaved Shor source certificate takes 105\,s; the eight finite
target certificates take 0.209--0.536\,s each.
Appendix~\ref{app:evaluation-results} gives the full distance and timing table.\pcctight

\subsection{Comparison with numerical distance search}
We compare the cost of checking one supplied lower-bound proof for a
family with the cost of finding the exact distance of one instance
(Figure~\ref{fig:distance-scaling}).
Each search instance comes from our own pipeline: we instantiate the
family's source program at distance $d$ and compile it with our Standard Scheme
compiler, so the search receives the QClifford circuit that the family proof
covers. We translate this circuit gate by gate into
Stim~\cite{gidney2021stim}, which propagates every single-wire X, Y and Z
injection to its final data error under the same error model and unrestricted
readouts. The encoder of codeDistance~\cite{webster2026distancefinding} turns
the resulting syndrome and logical-parity matrices into a MaxSAT problem, which
we solve with two PySAT algorithms, its default linear search (LSU) and
core-guided RC2, both on the Glucose 4 SAT solver.
Every optimum agrees with our proofs: it equals the proved distance for
the Surface and LP families, and for the color code it is $(d+1)/2$, above
the certified bound $\lceil d/3\rceil$.
Search time grows steeply with $d$: within the limit, the better configuration
reaches $d=11,\ 15,\ 4$ and $17$ in the four panels,
whereas each family proof is checked once for every $d$.
Appendix~\ref{app:evaluation-results} gives the timing boundaries and run counts.

\begin{figure}[!htbp]
\captionsetup{skip=3pt}
\centering
\begin{tikzpicture}
\begin{groupplot}[
 group style={group size=4 by 1,horizontal sep=.035\linewidth},
 scale only axis,width=.198\linewidth,height=56pt,
 ymode=log,ymin=.5,ymax=20000,
 xlabel={Code distance $d$},xlabel style={yshift=4pt},
 tick label style={font=\scriptsize},label style={font=\scriptsize},
 title style={font=\scriptsize,yshift=-6pt},
 axis lines=left,ymajorgrids,grid style={black!12},
 clip=false,ytick={1,10,100,1000,10000},
 xticklabel style={/pgf/number format/fixed},
 legend style={font=\scriptsize,draw=none,legend columns=4,
               /tikz/every even column/.append style={column sep=4pt}},
]
\nextgroupplot[title={\strut Unrotated Surface},xmin=2.4,xmax=13.6,xtick={3,7,11},ylabel={Wall time (s)},legend to name=pccTimingLegend]
\addplot[forget plot,black!45,densely dotted,line width=.5pt] coordinates {(2.4,7200) (13.6,7200)};
\addplot[forget plot,teal!70!black,densely dotted,mark=square*,mark size=1.2pt,line width=.6pt,mark options={solid}] coordinates {(3,0.8159) (5,1.78) (7,6.855) (9,81.04) (11,3970)};
\addplot[forget plot,only marks,teal!70!black,mark=triangle,mark size=2.1pt,mark options={solid,fill=white,xshift=-2.2pt}] coordinates {(13,7200)};
\addplot[forget plot,blue!70!black,mark=*,mark size=1.5pt,line width=.65pt] coordinates {(3,1.207) (5,1.634) (7,6.182) (9,31.32) (11,493)};
\addplot[forget plot,only marks,blue!70!black,mark=triangle,mark size=2.1pt,mark options={solid,fill=white}] coordinates {(13,7200)};
\addplot[forget plot,orange!75!black,dashed,line width=.8pt] coordinates {(2.4,27.2948) (13.6,27.2948)};
\node[anchor=south west,font=\scriptsize,text=orange!75!black,inner sep=1.5pt] at (axis cs:2.4,27.2948) {27.3\,s};
\node[anchor=south west,font=\scriptsize,text=black!60,inner sep=1pt,xshift=3pt] at (axis cs:2.4,7200) {2\,h limit};
\addlegendimage{blue!70!black,mark=*,mark size=1.5pt,line width=.65pt}
\addlegendentry{MaxSAT search (LSU)}
\addlegendimage{teal!70!black,densely dotted,mark=square*,mark size=1.2pt,line width=.6pt,mark options={solid}}
\addlegendentry{MaxSAT search (RC2)}
\addlegendimage{only marks,black!70,mark=triangle,mark size=2.1pt}
\addlegendentry{search timeout}
\addlegendimage{orange!75!black,dashed,line width=.8pt}
\addlegendentry{one proof for every $d$}
\nextgroupplot[title={\strut Rotated Surface},xmin=2.4,xmax=17.6,xtick={3,7,11,15},yticklabels={}]
\addplot[forget plot,black!45,densely dotted,line width=.5pt] coordinates {(2.4,7200) (17.6,7200)};
\addplot[forget plot,teal!70!black,densely dotted,mark=square*,mark size=1.2pt,line width=.6pt,mark options={solid}] coordinates {(3,0.8751) (5,1.394) (7,3.102) (9,22.66) (11,364) (13,6759)};
\addplot[forget plot,blue!70!black,mark=*,mark size=1.5pt,line width=.65pt] coordinates {(3,0.8856) (5,1.808) (7,2.935) (9,8.004) (11,30.34) (13,174) (15,1989)};
\addplot[forget plot,only marks,blue!70!black,mark=triangle,mark size=2.1pt,mark options={solid,fill=white}] coordinates {(17,7200)};
\addplot[forget plot,orange!75!black,dashed,line width=.8pt] coordinates {(2.4,219.529) (17.6,219.529)};
\node[anchor=south west,font=\scriptsize,text=orange!75!black,inner sep=1.5pt] at (axis cs:2.4,219.529) {219.5\,s};
\nextgroupplot[title={\strut Cartesian LP},xmin=2.8,xmax=9.2,xtick={4,8},yticklabels={}]
\addplot[forget plot,black!45,densely dotted,line width=.5pt] coordinates {(2.8,7200) (9.2,7200)};
\addplot[forget plot,teal!70!black,densely dotted,mark=square*,mark size=1.2pt,line width=.6pt,mark options={solid}] coordinates {(4,7.695)};
\addplot[forget plot,only marks,teal!70!black,mark=triangle,mark size=2.1pt,mark options={solid,fill=white,xshift=-2.2pt}] coordinates {(8,7200)};
\addplot[forget plot,blue!70!black,mark=*,mark size=1.5pt,line width=.65pt] coordinates {(4,5.744)};
\addplot[forget plot,only marks,blue!70!black,mark=triangle,mark size=2.1pt,mark options={solid,fill=white}] coordinates {(8,7200)};
\addplot[forget plot,orange!75!black,dashed,line width=.8pt] coordinates {(2.8,3715.77) (9.2,3715.77)};
\node[anchor=north west,font=\scriptsize,text=orange!75!black,inner sep=1.5pt] at (axis cs:2.8,3715.77) {3716\,s};
\nextgroupplot[title={\strut Triangular color},xmin=2.4,xmax=19.6,xtick={3,7,11,15,19},yticklabels={}]
\addplot[forget plot,black!45,densely dotted,line width=.5pt] coordinates {(2.4,7200) (19.6,7200)};
\addplot[forget plot,teal!70!black,densely dotted,mark=square*,mark size=1.2pt,line width=.6pt,mark options={solid}] coordinates {(3,0.8734) (5,1.004) (7,2.196) (9,5.348) (11,15.75) (13,83.56) (15,2036)};
\addplot[forget plot,only marks,teal!70!black,mark=triangle,mark size=2.1pt,mark options={solid,fill=white}] coordinates {(17,7200)};
\addplot[forget plot,blue!70!black,mark=*,mark size=1.5pt,line width=.65pt] coordinates {(3,0.8043) (5,1.044) (7,2.156) (9,4.932) (11,13.07) (13,51.28) (15,162.2) (17,1051)};
\addplot[forget plot,only marks,blue!70!black,mark=triangle,mark size=2.1pt,mark options={solid,fill=white}] coordinates {(19,7200)};
\addplot[forget plot,orange!75!black,dashed,line width=.8pt] coordinates {(2.4,5510.58) (19.6,5510.58)};
\node[anchor=north west,font=\scriptsize,text=orange!75!black,inner sep=1.5pt] at (axis cs:2.4,5510.58) {5511\,s$^\dagger$};
\end{groupplot}
\end{tikzpicture}
\par\vspace{1pt}\ref{pccTimingLegend}
\caption{Finding the exact circuit-level distance of one instance by MaxSAT
search (one run, 7,200\,s limit; a series ends at its first unsolved instance), compared
with compiling and checking one Standard Scheme proof for the entire family.
$^\dagger$One check of the color-code lower bound $\lceil d/3\rceil$.}
\label{fig:distance-scaling}
\Description{Four logarithmic-time panels show unrotated Surface, rotated
Surface, Cartesian LP and triangular color code. In every panel the search
time of both MaxSAT configurations grows with the code distance until the
7,200 second limit, while the family proof cost is a horizontal line at
27.3, 219.5, 3715.8 and 5511 seconds, respectively.}
\end{figure}


\section{Conclusion}
\label{sec:conclusion}
\PCQEC connects supplied code-algebra facts and QStab invariants to circuit-error obligations, then checks the resulting certificate against the requested artifact. For the Surface code, it establishes circuit-level code distance $d$ for every odd $d\ge3$, together with a matching silent logical failure.

\paragraph{AI disclosure.}
OpenAI Codex assisted with proof development, implementation, experiment scripts, and manuscript editing.
\label{page:body-end}
\endgroup

\section*{Data-Availability Statement}
The artifact provides sources, measurements, reproduction scripts
and a claim-to-evidence map at
\url{https://anonymous.4open.science/r/PLDI2027-artifact-6F50/};
its family certificates are distributed with it as a separate proof-input archive.
\label{page:main-end}
\clearpage
\bibliography{main-pcqec-references}
\label{page:references-end}
\clearpage
\raggedbottom
\appendix
\label{page:appendix-start}
\section{Formal definitions}
\label{app:formal}
We first fix the notation for qubit indices, Pauli errors, commutation and weight.
\[
 [n]=\{0,\ldots,n-1\},\qquad \mathbb B=\{0,1\},\qquad
 \mathbb P=\{I,X,Y,Z\}\cong\mathbb B^2,
\]
\[
\pccfitmath[\linewidth]{
 I=(0,0),\ X=(1,0),\ Z=(0,1),\ Y=(1,1),\quad
 (x,z)(x',z')=(x\oplus x',z\oplus z'),\quad
 \mathsf{anti}((x,z),(x',z'))=xz'\oplus zx'.
}
\]
\[
 P_q(u)=\begin{cases}P&u=q,\\I&u\ne q,\end{cases}\qquad
 \langle E,F\rangle_n=\bigoplus_{q<n}\mathsf{anti}(E(q),F(q)),\qquad
 \wt[n]{E}=|\{q<n:E(q)\ne I\}|.
\]

\subsection{CodeML syntax and typing}
\label{app:codeml}
The following definitions give the complete CodeML syntax and its types.
\[
 x\in\mathsf{Var},\quad v\in\mathbb N\uplus\mathbb B\uplus\mathbb P,\quad
 a,a':\kw{nat},\quad p:\kw{pauli},\quad T,U:\kw{str},
\]
\[
\pccfitmath[\linewidth]{
 \kw{str}:=\kw{nat}\to\kw{pauli},\quad
 \kw{factor}:=\kw{pauli}\times\kw{nat},\quad
 \kw{stabilizer}:=\mathsf{bounded}(\kw{factor}),\quad P[q]:=(P,q).
}
\]

\begin{equation}\label{eq:codeml-syntax}
\pccfitmath{
\begin{aligned}
 \tau&::=\kw{nat}\mid\kw{bool}\mid\kw{pauli}\mid\tau\times\tau
          \mid\mathsf{bounded}(\tau)\mid\kw{str},\\
 e&::=x\mid v\mid o(e,e)\mid(e,e)\mid\kw{let}\ (x,y)=e\ \kw{in}\ e\mid\kw{if}\ e\ \kw{then}\ e\ \kw{else}\ e
       \mid\lambda x<a.e\mid e[a]\mid\lambda x.p\mid\prod_{x<a}T\mid C(a,a),\\
 o&::=+\mid-\mid\cdot_{\mathbb N}\mid\mathsf{div}\mid\mathsf{pow}
       \mid=_{\mathbb N}\mid\cdot_{\mathbb P}\mid\mathsf{anti},\\
 C&::=\kw{family}\ C(d,k)=S_C.
\end{aligned}
}
\end{equation}

The operator types and typing rules are as follows.
\[
 \begin{array}{c|c}
 o&\mathsf{type}(o)\\\hline
 +,-,\cdot_{\mathbb N},\mathsf{div},\mathsf{pow}&\kw{nat}\times\kw{nat}\to\kw{nat}\\
 =_{\mathbb N}&\kw{nat}\times\kw{nat}\to\kw{bool}\\
 \cdot_{\mathbb P}&\kw{pauli}\times\kw{pauli}\to\kw{pauli}\\
 \mathsf{anti}&\kw{pauli}\times\kw{pauli}\to\kw{bool}
 \end{array}
\]
\begin{mathpar}
 \inferrule{x:\tau\in\Delta}{\Delta\vdash x:\tau}
 \and \inferrule{v\in V_\tau\quad\tau\in\{\kw{nat},\kw{bool},\kw{pauli}\}}{\Delta\vdash v:\tau}
 \and \inferrule{\mathsf{type}(o)=\tau_1\times\tau_2\to\tau\quad
 \Delta\vdash e_1:\tau_1\quad\Delta\vdash e_2:\tau_2}{\Delta\vdash o(e_1,e_2):\tau}
 \and \inferrule{\Delta\vdash b:\kw{bool}\quad\Delta\vdash e_1:\tau\quad\Delta\vdash e_2:\tau}
 {\Delta\vdash\kw{if}\ b\ \kw{then}\ e_1\ \kw{else}\ e_2:\tau}
\end{mathpar}

\begin{mathpar}
 \inferrule{\Delta\vdash e_1:\tau_1\quad\Delta\vdash e_2:\tau_2}
 {\Delta\vdash\prog{(e_1,e_2)}:\tau_1\times\tau_2}
 \and
 \inferrule{\Delta\vdash e:\tau_1\times\tau_2\quad
 \Delta,x:\tau_1,y:\tau_2\vdash e':\tau}
 {\Delta\vdash\prog{\kw{let}\ (x,y)=e\ \kw{in}\ e'}:\tau}
 \and
 \inferrule{\Delta\vdash a:\kw{nat}\quad\Delta,x:\kw{nat}\vdash e:\tau}
 {\Delta\vdash\prog{\lambda x<a.e}:\mathsf{bounded}(\tau)}
 \and
 \inferrule{\Delta\vdash e:\mathsf{bounded}(\tau)\quad\Delta\vdash a:\kw{nat}}
 {\Delta\vdash\prog{e[a]}:\tau}
 \and
 \inferrule{\Delta,x:\kw{nat}\vdash p:\kw{pauli}}
 {\Delta\vdash\prog{\lambda x.p}:\kw{str}}
 \and
 \inferrule{\Delta\vdash T:\kw{str}\quad\Delta\vdash a:\kw{nat}}
 {\Delta\vdash\prog{T[a]}:\kw{pauli}}
 \and
 \inferrule{\Delta\vdash a:\kw{nat}\quad\Delta,x:\kw{nat}\vdash T:\kw{str}}
 {\Delta\vdash\prog{\prod_{x<a}T}:\kw{str}}
 \and
 \inferrule{\Delta\vdash a,a':\kw{nat}}
 {\Delta\vdash\prog{C(a,a')}:\kw{stabilizer}}
 \and
 \inferrule{d:\kw{nat},k:\kw{nat}\vdash S_C:\kw{stabilizer}}
 {\vdash\prog{\kw{family}\ C(d,k)=S_C}\ \mathsf{ok}}
\end{mathpar}

\[
 x\ne y\quad\text{in }\kw{let}\ (x,y)=e\ \kw{in}\ e',\qquad
 \llbracket u-v\rrbracket=\max(u-v,0),\qquad
 \llbracket\mathsf{div}(u,v)\rrbracket=
 \begin{cases}\lfloor u/v\rfloor&v>0,\\0&v=0,\end{cases}
\]
\[
 a\le b\equiv(a-b=0),\quad a<b\equiv(a+1-b=0),\quad
 a\bmod b\equiv a-\mathsf{div}(a,b)\cdot_{\mathbb N}b.
\]

\subsection{CodeML evaluation and generation}
\label{app:codeml-evaluation}
We define CodeML values and evaluation with fuel $f$, followed by the generation of a QStab program.
\[
 V_{\kw{nat}}=\mathbb N,\quad V_{\kw{bool}}=\mathbb B,\quad V_{\kw{pauli}}=\mathbb P,\quad
 V_{\tau_1\times\tau_2}=V_{\tau_1}\times V_{\tau_2},
\]
\[
 V_{\mathsf{bounded}(\tau)}=\{(L,F):L\in\mathbb N,\ F:[L]\to V_\tau\},\quad
 V_{\kw{str}}=\mathbb N\to\mathbb P_\bot,\quad \mathbb P_\bot=\mathbb P\uplus\{\bot\},
\]
\[
 \denote e^C_{f,\rho}=
 \begin{cases}v&\rho\vdash_f^C e\Downarrow v,\\\bot&\text{otherwise}.\end{cases}
\]
\begin{mathpar}
 \inferrule{\rho(x)=v}{\rho\vdash_f^C x\Downarrow v}
 \and \inferrule{}{\rho\vdash_f^C v\Downarrow v}
 \and \inferrule{\rho\vdash_f^C e_1\Downarrow v_1\quad\rho\vdash_f^C e_2\Downarrow v_2}
 {\rho\vdash_f^C o(e_1,e_2)\Downarrow\llbracket o\rrbracket(v_1,v_2)}
 \and \inferrule{\rho\vdash_f^C b\Downarrow1\quad\rho\vdash_f^C e_1\Downarrow v}
 {\rho\vdash_f^C\kw{if}\ b\ \kw{then}\ e_1\ \kw{else}\ e_2\Downarrow v}
 \and \inferrule{\rho\vdash_f^C b\Downarrow0\quad\rho\vdash_f^C e_2\Downarrow v}
 {\rho\vdash_f^C\kw{if}\ b\ \kw{then}\ e_1\ \kw{else}\ e_2\Downarrow v}
 \and \inferrule{\rho\vdash_f^C e_1\Downarrow v_1\quad\rho\vdash_f^C e_2\Downarrow v_2}
 {\rho\vdash_f^C(e_1,e_2)\Downarrow(v_1,v_2)}
 \and \inferrule{\rho\vdash_f^C e\Downarrow(v_1,v_2)\quad
 \rho[x\mapsto v_1,y\mapsto v_2]\vdash_f^C e'\Downarrow v}
 {\rho\vdash_f^C\kw{let}\ (x,y)=e\ \kw{in}\ e'\Downarrow v}
\end{mathpar}

\begin{mathpar}
 \inferrule*[right=\textsc{Bounded}]
 {\rho\vdash_f^C a\Downarrow L\quad
  \forall t<L.\ \rho[x\mapsto t]\vdash_f^C e\Downarrow F(t)}
 {\rho\vdash_f^C\prog{\lambda x<a.e}\Downarrow(L,F)}
 \and
 \inferrule*[right=\textsc{Apply}]
 {\rho\vdash_f^C e\Downarrow(L,F)\quad\rho\vdash_f^C a\Downarrow t\quad t<L}
 {\rho\vdash_f^C\prog{e[a]}\Downarrow F(t)}
 \and
 \inferrule*[right=\textsc{String}]
 {V(q)=\denote p^C_{f,\rho[x\mapsto q]}\quad(q\in\mathbb N)}
 {\rho\vdash_f^C\prog{\lambda x.p}\Downarrow V}
 \and
 \inferrule*[right=\textsc{Entry}]
 {\rho\vdash_f^C T\Downarrow V\quad\rho\vdash_f^C a\Downarrow q\quad V(q)=P\in\mathbb P}
 {\rho\vdash_f^C\prog{T[a]}\Downarrow P}
 \and
 \inferrule*[right=\textsc{Call}]
 {\rho\vdash_f^C a\Downarrow d_0\quad\rho\vdash_f^C a'\Downarrow k_0\quad
  [d\mapsto d_0,k\mapsto k_0]\vdash_f^C S_C\Downarrow W}
 {\rho\vdash_{f+1}^C\prog{C(a,a')}\Downarrow W}
\end{mathpar}

\[
 \rho\not\vdash_0^C C(a,a')\Downarrow v,
\]
\[
 W_L(q):=\begin{cases}
 \prod_{i<L}V_i(q)&\forall i<L.\ \rho[x\mapsto i]\vdash_f^C T\Downarrow V_i\ \land\ V_i(q)\in\mathbb P,\\
 \bot&\text{otherwise}
 \end{cases}
\]
\begin{mathpar}
 \inferrule{\rho\vdash_f^C a\Downarrow L}
 {\rho\vdash_f^C\prod_{x<a}T\Downarrow W_L}
\end{mathpar}
\[
 \prod_{i<0}V_i(q)=I,\qquad
 \mathcal F=(C,\mathsf{Legal},n,m,f),\quad n,m,f:\mathbb N\to\mathbb N,
\]
\[
 \mathsf{GenOK}(d,R,B):=\mathsf{Legal}(d)\land n(d)>0\land m(d)>0\land R>0\land B\ge0
\]
\[
 {}\land\ \forall k<m(d).\ \varnothing\vdash^C_{f(d)}C(d,k)\Downarrow(L_k,F_k)
 \ \land\ \forall t<L_k.\ F_k(t)=(P_{k,t},q_{k,t})\in\mathbb P\times[n(d)]
\]
\[
 {}\land\ \forall k<m(d).\ \forall t,t'<L_k.\ q_{k,t}=q_{k,t'}\Rightarrow t=t',
\]
\[
 r_k=[(P_{k,t})_{q_{k,t}}]_{t<L_k},\qquad
 T_k(q)=\prod_{t<L_k,\ q_{k,t}=q}P_{k,t},
\]
\[
 \mathsf{Generate}\llbracket C\rrbracket(d;R,B)=
 \begin{cases}\langle n(d),[r_k]_{k<m(d)},R,B\rangle&\mathsf{GenOK}(d,R,B),\\
 \bot&\text{otherwise}.\end{cases}
\]

\subsection{Assertions and code-algebra judgments}
\label{app:assertions}
The following grammar defines assertions over CodeML terms and the residual data error $E$.
\[
 e_A::=e\mid E\mid\kw{let}\ x=a\ \kw{in}\ e_A,\qquad
 E:\kw{str},\quad b,\ell:\kw{nat}.
\]

\begin{equation}\label{eq:assertion-core}
\pccfitmath{
A::=a=a'\mid T\equiv_a U\mid\langle T,U\rangle_a=0\mid\wt[a]{T}\le a'\mid A\land A\mid\neg A\mid\kw{let}\ x=a\ \kw{in}\ A
       \mid\forall x<a.\ A.
}
\end{equation}

The following evaluator defines satisfaction, including the propagation of undefined values.
\[
 \mathsf{test}_{C,f,\rho,V}(A)\in\mathbb B_\bot,\quad
 \rho(E)=V,\quad \mathbb B_\bot=\mathbb B\uplus\{\bot\},\qquad
 (\rho,V)\models A\iff\mathsf{test}_{C,f,\rho,V}(A)=1.
\]
\[
 \mathsf{bind}(u,F)=\begin{cases}F(u)&u\ne\bot,\\\bot&u=\bot,\end{cases}\qquad
 \mathsf{and}(u,v)=\begin{cases}0&u=0,\\v&u=1,\\\bot&u=\bot,\end{cases}\quad
 \mathsf{neg}(u)=\begin{cases}1-u&u\in\mathbb B,\\\bot&u=\bot.\end{cases}
\]
\[
 \mathsf{scan}(L,F)=\mathsf{go}(0),\qquad
 \mathsf{go}(q)=\begin{cases}1&q=L,\\\mathsf{and}(F(q),\mathsf{go}(q+1))&q<L.\end{cases}
\]
\[
 \mathsf{eq}_{\mathbb P}(u,v)=
 \begin{cases}[u=v]&u,v\in\mathbb P,\\\bot&\text{otherwise},\end{cases}\qquad
 \mathsf{eqstr}_n(T,U)=\mathsf{scan}(n,\lambda q.\mathsf{eq}_{\mathbb P}(T(q),U(q))).
\]
\[
 p_0=0,\quad p_{q+1}=p_q\oplus\mathsf{anti}(T(q),U(q)),\quad
 w_0=0,\quad w_{q+1}=w_q+[T(q)\ne I],
\]
\[
 \mathsf{anti}(\bot,u)=\mathsf{anti}(u,\bot)=\bot,\quad
 [\bot\ne I]=\bot,\quad u\oplus\bot=u+\bot=\bot.
\]
\[
 \begin{aligned}
 \mathsf{test}(a=a')&=\mathsf{bind}(\denote a,\lambda u.\mathsf{bind}(\denote{a'},\lambda v.[u=v])),\\
 \mathsf{test}(T\equiv_a U)&=\mathsf{bind}(\denote a,\lambda n.\mathsf{bind}(\denote T,\lambda T_0.
                   \mathsf{bind}(\denote U,\lambda U_0.\mathsf{eqstr}_n(T_0,U_0)))),\\
 \mathsf{test}(\langle T,U\rangle_a=0)&=\mathsf{bind}(\denote a,\lambda n.
            \mathsf{bind}(\denote T,\lambda T_0.\mathsf{bind}(\denote U,\lambda U_0.
            \mathsf{bind}(p_n(T_0,U_0),\lambda p.[p=0])))),\\
 \mathsf{test}(\wt[a]{T}\le a')&=\mathsf{bind}(\denote a,\lambda n.
       \mathsf{bind}(\denote{a'},\lambda k.\mathsf{bind}(\denote T,\lambda T_0.
       \mathsf{bind}(w_n(T_0),\lambda w.[w\le k])))),\\
 \mathsf{test}(A\land B)&=\mathsf{and}(\mathsf{test}(A),\mathsf{test}(B)),\\
 \mathsf{test}(\neg A)&=\mathsf{neg}(\mathsf{test}(A)),\\
 \mathsf{test}(\kw{let}\ x=a\ \kw{in}\ A)&=\mathsf{bind}(\denote a,\lambda u.
                            \mathsf{test}_{\rho[x\mapsto u],V}(A)),\\
 \mathsf{test}(\forall x<a.\ A)&=\mathsf{bind}(\denote a,\lambda L.
                            \mathsf{scan}(L,\lambda i.\mathsf{test}_{\rho[x\mapsto i],V}(A))).
 \end{aligned}
\]

The remaining logical connectives are defined from conjunction, negation and bounded universal quantification.
\begin{equation}
\begin{gathered}
 \top:=(0=0),\quad\bot_A:=\neg\top,\quad
 A\lor B:=\neg(\neg A\land\neg B),\\
 A\Rightarrow B:=\neg(A\land\neg B),\quad
 \exists x<a.A:=\neg\forall x<a.\neg A.
\end{gathered}
\end{equation}

The following judgments and inference rules define reasoning about assertions under assumptions $\Gamma$.
\[
 \mathsf{Total}(n,V):=\forall q<n.\ V(q)\ne\bot,\qquad
 \mathsf{Def}(A):=A\lor\neg A,
\]
\[
 \Gamma\models_{C,f,n}A\iff
 \forall\rho,V.\ \mathsf{Total}(n,V)\land(\rho,V)\models\Gamma
 \Rightarrow(\rho,V)\models A.
\]
\begin{mathpar}
 \inferrule{A\in\Gamma}{\Gamma\vdash A}
 \and \inferrule{\Gamma\vdash A}{\Gamma,B\vdash A}
 \and \inferrule{\Gamma\vdash\bot_A}{\Gamma\vdash A}
 \and \inferrule{\Gamma\vdash A\quad\Gamma\vdash B}{\Gamma\vdash A\land B}
 \and \inferrule{\Gamma\vdash A\land B}{\Gamma\vdash A}
 \and \inferrule{\Gamma\vdash A\land B}{\Gamma\vdash B}
 \and \inferrule{\Gamma\vdash A\lor B\quad\Gamma,A\vdash D\quad\Gamma,B\vdash D}{\Gamma\vdash D}
 \and \inferrule{\Gamma\vdash A\quad\Gamma\vdash A\Rightarrow B}{\Gamma\vdash B}
\end{mathpar}

\begin{mathpar}
 \inferrule{\Gamma\vdash A}{\Gamma\vdash A\lor B}
 \and
 \inferrule{\Gamma\vdash\mathsf{Def}(A)\quad\Gamma\vdash B}{\Gamma\vdash A\lor B}
 \and
 \inferrule{\Gamma\vdash\mathsf{Def}(A)\quad\Gamma,A\vdash B}
 {\Gamma\vdash A\Rightarrow B}
 \and
 \inferrule{\Gamma\vdash\mathsf{Def}(A)\quad\Gamma,A\vdash\bot_A}
 {\Gamma\vdash\neg A}
 \and
 \inferrule{\Gamma,x<h\vdash A}{\Gamma\vdash\forall x<h.A}
 \and
 \inferrule{\Gamma\vdash\forall x<h.A\quad\Gamma\vdash t<h}
 {\Gamma\vdash A[t/x]}
 \and
 \inferrule{\Gamma\vdash\forall x<h.\mathsf{Def}(A)\quad
  \Gamma\vdash t<h\quad\Gamma\vdash A[t/x]}
 {\Gamma\vdash\exists x<h.A}
 \and
 \inferrule{\Gamma\vdash\exists x<h.A\quad\Gamma,x<h,A\vdash B}
 {\Gamma\vdash B}
\end{mathpar}

\[
 \vdash\ =\ \vdash_{C,f,n},\qquad
 \mathsf{TotalNat}(h)\land\mathsf{TotalNat}(t),\qquad
 x\notin\mathsf{FV}(\Gamma,h),\qquad
 x\notin\mathsf{FV}(B)\quad(\exists\text{-elimination}).
\]
Here $\mathsf{TotalNat}(h)$ states that $h$ evaluates to a natural number under
every valuation satisfying $\Gamma$. The outer judgment $\Gamma\vdash_C F$
fixes the family $C$ and quantifies over natural parameters and total
residuals. It uses ordinary first-order deduction over these assertions,
including evaluation of closed atomic assertions, together with natural
induction on a variable that is fresh for $\Gamma$:
\begin{mathpar}
 \inferrule{\Gamma\vdash_C F(0)\quad\Gamma,k\in\mathbb N,F(k)\vdash_C F(k+1)\quad k\notin\mathsf{FV}(\Gamma)}
 {\Gamma\vdash_C\forall k\in\mathbb N.\ F(k)}
\end{mathpar}
The Surface code induction on odd $d$ instantiates $F(k)$ at $d=2k+3$.
The code-algebra predicates below express normalizer membership, stabilizer generation, logical errors and well-formedness.
\[
 a_k=\lfloor a/2^k\rfloor\bmod2,\qquad T^0=I^n,\qquad T^1=T.
\]

\begin{equation}
\begin{aligned}
 \mathsf{Norm}(E)&:=\forall k<m.\ \langle T_k,E\rangle_n=0,\\
 \mathsf{Gen}(E)&:=\exists a<2^m.\ E\equiv_n\prod_{k<m}T_k^{a_k},\\
 \mathsf{Logical}(E)&:=\mathsf{Norm}(E)\land\neg\mathsf{Gen}(E),\\
 \mathsf{WellFormed}&:=\bigl(\forall k<m.\ T_k\equiv_nT_k\bigr)
       \land\bigl(\forall i,j<m.\langle T_i,T_j\rangle_n=0\bigr).
\end{aligned}
\end{equation}

\paragraph{Cut signatures as strings.}\label{app:signature-strings}
To use the string-weight test in Equation~\eqref{eq:assertion-core} on
the cut signatures of Equation~\eqref{eq:surface-main-cuts}, we store each
parity bit as I for zero or X for one. Since
$X\cdot X=I$, string multiplication computes XOR. For each cut, a
product over the $n$ data wires contributes X at anticommuting
sites and I elsewhere. Thus $U\oplus V:=\prog{\lambda j.U[j]\cdot V[j]}$, equality uses
$\equiv_d$, and string weight counts the crossed cuts. Bounded products and string lambdas
express these operations in the core language. For example, the
signature $(1,0,0)$ of the hook error becomes the string $XII$, whose
weight is one. The logical row's signature $(1,1,1)$ becomes $XXX$,
whose weight is three. The signature $e_i$ with only bit $i$ set has the
CodeML representation
$e_i:=\prog{\lambda j.\ \kw{if}\ j=i\ \kw{then}\ X\ \kw{else}\ I}$.

\subsection{QStab syntax and transitions}
\label{app:qstab}
We define QStab programs and states, then give the error and measurement transitions.
\[
 r::=[(P_j)_{q_j}]_{j<w},\quad S=\langle n,\vec r,R,B\rangle,\quad
 n>0,\ m=|\vec r|>0,\ R>0,\ M=mR,\ B\in\mathbb N,
\]
\[
 r_h=\vec r[h\bmod m],\quad
 T_h=\prod_{i<|r_h|}r_h[i],\quad
 \mathsf{Hooks}(h)=\left\{\prod_{j\le i<|r_h|}r_h[i]:0\le j\le|r_h|\right\},
\]
\[
 \srcstate=(E,G,h,b,\ell)\in\mathbb P^n\times\mathbb B^M\times[M]\times\mathbb N^2,
\]
\[
 \statecfg::=\kw{run}(\srcstate)\mid\kw{done}(\srcstate)\mid\kw{error}(\srcstate),\quad
 \statecfg_0=\kw{run}(I^n,\mathbf0,0,B,0),\quad v=G_h\oplus\langle T_h,E\rangle_n.
\]
\begin{mathpar}
 \inferrule*[right=\textsc{QS-Data}]
 {b>0\quad q<n\quad P\in\{X,Y,Z\}}
 {\kw{run}(E,G,h,b,\ell)\xrightarrow{1}_S\kw{run}(EP_q,G,h,b-1,\ell+1)}
 \and \inferrule*[right=\textsc{QS-DataReadout}]
 {b>0\quad q<n\quad P\in\{X,Y,Z\}\quad t\in\mathbb B}
 {\kw{run}(E,G,h,b,\ell)\xrightarrow{1}_S\kw{run}(EP_q,G[h\mapsto G_h\oplus t],h,b-1,\ell+1)}
 \and \inferrule*[right=\textsc{QS-Hook}]
 {b>0\quad H\in\mathsf{Hooks}(h)\quad t\in\mathbb B}
 {\kw{run}(E,G,h,b,\ell)\xrightarrow{1}_S\kw{run}(EH,G[h\mapsto G_h\oplus t],h,b-1,\ell+1)}
 \and \inferrule*[right=\textsc{QS-Readout}]
 {b>0}
 {\kw{run}(E,G,h,b,\ell)\xrightarrow{1}_S\kw{run}(E,G[h\mapsto G_h\oplus1],h,b-1,\ell+1)}
 \and \inferrule*[right=\textsc{QS-Meas}]
 {h+1<M}{\kw{run}(\srcstate)\xrightarrow{0}_S\kw{run}(E,G[h\mapsto v],h+1,b,\ell)}
 \and \inferrule*[right=\textsc{QS-Halt}]
 {h+1=M}{\kw{run}(\srcstate)\xrightarrow{0}_S\kw{done}(E,G[h\mapsto v],h,b,\ell)}
 \and \inferrule*[right=\textsc{QS-Exhaust}]
 {b=0}{\kw{run}(\srcstate)\xrightarrow{0}_S\kw{error}(\srcstate)}
\end{mathpar}
A readout error can hide the evidence of a data error. For the first stabilizer
$T_0=Z_0Z_3Z_1Z_4$ of $S_3$, a data error $X_3$ followed by a readout error gives
$E=X_3$, $G_0=1$ and $\ell=2$. Measurement commits
$1\oplus\langle T_0,X_3\rangle_9=1\oplus1=0$.
This hides one syndrome bit. The full failure test checks all detectors
and whether the residual is logical.

Source invariance requires an assertion to hold at every reachable QStab state.
A configuration $\kw{run}(\srcstate)$, $\kw{done}(\srcstate)$ or
$\kw{error}(\srcstate)$ satisfies $A$ when $A$ holds for the residual $E$
and the values $b,\ell$ of $\srcstate$.
\begin{definition}[Source invariance]\label{def:hoare-triple-qstab}
\[
 S\models\Box A\iff\forall\statecfg.\ \statecfg_0\xrightarrow{*}_S\statecfg\Rightarrow\statecfg\models A,
\]
\[
 \Gamma\mid S\models\Box A\iff\forall\rho\models\Gamma.\ S_\rho\models\Box A_\rho.
\]
\end{definition}

\subsection{QStab invariant rules}
\label{app:source-logic}
The following definitions describe cut signatures and the bounds on hook errors used to prove QStab invariants.
\[
\pccfitmath[\linewidth]{
 L:\mathbb N\times[n]\to\mathbb N,\quad \nu>0,\quad P\in\{X,Z\},\quad
 \kappa_i^{L,P}(q)=\begin{cases}P&L(\nu,q)=i,\\I&L(\nu,q)\ne i,\end{cases}\quad
 \chi_{L,P}(E)_i=\langle\kappa_i^{L,P},E\rangle_n\quad(i<\nu).
}
\]
\begin{definition}[Cut bounds for hook errors]\label{def:hook-bound}\label{def:pair-trees}
\[
 w=|r|,\qquad\mathsf{sig}_{L,P}(r)=[v_j]_{j<w},\qquad v_j=\chi_{L,P}(r[j])\in\mathbb B^\nu.
\]

\begin{equation}\label{eq:suffix-recurrence}
\pccfitmath{
s_w=\mathbf0,\quad s_j=v_j\oplus s_{j+1}\ (0\le j<w),
}
\end{equation}

\begin{equation}\label{eq:hook-bound}
\pccfitmath{
\prog{s_0=\mathbf0\ \land\ \forall j\le w.\ \wt{s_j}\le1}.
}
\end{equation}

\[
 \Gamma\vdash\mathsf{HookBound}_{L,P}(S)\iff
 \forall h<M.\ \Gamma\vdash_{\rm hook}\mathsf{sig}_{L,P}(r_h)
 \ \land\ \mathsf{DistinctInRange}_n(r_h).
\]
Here $\mathsf{DistinctInRange}_n([(P_j)_{q_j}]_{j<w})$ requires $q_j<n$ for every
$j<w$ and $q_j\ne q_{j'}$ for $j\ne j'$.
\end{definition}

The following rules establish hook bounds from the sequence of cut signatures.
\begin{mathpar}
 \inferrule*[right=\textsc{Pair-Nil}]{}{\Gamma\vdash_{\rm hook}[]}
 \and
 \inferrule*[right=\textsc{Pair-Zero}]
 {\Gamma\vdash_Cv=\mathbf0\quad\Gamma\vdash_{\rm hook}\vec v}
 {\Gamma\vdash_{\rm hook}v::\vec v}
 \and
 \inferrule*[right=\textsc{Pair-Step}]
 {\Gamma\vdash_Cv=v'\quad\Gamma\vdash_C\wt{v'}\le1\quad
  \Gamma\vdash_{\rm hook}\vec v}
 {\Gamma\vdash_{\rm hook}v::v'::\vec v}
 \and
 \inferrule*[right=\textsc{Hook-Check}]
 {\Gamma\vdash_Cs_0=\mathbf0\quad\Gamma\vdash_C\forall j\le w.\ \wt{s_j}\le1}
 {\Gamma\vdash_{\rm hook}[v_j]_{j<w}}
\end{mathpar}

Completeness connects the two cut invariants to the distance bound in the rules below.
\begin{equation}
 \mathsf{Complete}_{L_x,L_z}:\quad
 \forall E\in\mathbb P^n.\ \forall i,j<\nu.\
 \mathsf{Norm}(E)\land\chi_{L_z,Z}(E)_i=0\land\chi_{L_x,X}(E)_j=0
 \Rightarrow\mathsf{Gen}(E).
\end{equation}

\[
 I_X:=\wt[\nu]{\chi_{L_x,X}(E)}\le\ell,\qquad
 I_Z:=\wt[\nu]{\chi_{L_z,Z}(E)}\le\ell.
\]

\begin{mathpar}
 \inferrule*[right=\textsc{Src-Cut}]
 {\Gamma\vdash\mathsf{HookBound}_{L,P}(S)}
 {\Gamma\mid S\vdash\Box(\wt[\nu]{\chi_{L,P}(E)}\le\ell)}
 \and
 \inferrule*[right=\textsc{Src-Conj}]
 {\Gamma\mid S\vdash\Box A\quad\Gamma\mid S\vdash\Box A'}
 {\Gamma\mid S\vdash\Box(A\land A')}
 \and
 \inferrule*[right=\textsc{Src-Conseq}]
 {\Gamma\mid S\vdash\Box A\quad\Gamma,A\vdash_{C,f,n}A'}
 {\Gamma\mid S\vdash\Box A'}
 \and
 \inferrule*[right=\textsc{Src-Dist}]
 {\Gamma\mid S\vdash\Box(I_Z\land I_X)\quad
  \Gamma\vdash_C\mathsf{WellFormed}\quad\Gamma\vdash_C\mathsf{Complete}_{L_x,L_z}}
 {\Gamma\mid S\vdash\Box(\ell<\nu\Rightarrow\neg\mathsf{Logical}(E))}
\end{mathpar}

The following soundness statement connects derivable invariants to reachable-state semantics.
\begin{theorem}[Checked source invariance]\label{thm:qhl-sound}\label{sec:qhl-soundness}
\[
 \frac{\forall\rho\models\Gamma.\ \mathsf{GenOK}_\rho\ \land\ \mathsf{WellFormedRound}(S_\rho)
       \qquad\Gamma\mid S\vdash\Box A}
      {\Gamma\mid S\models\Box A}.
\]
\end{theorem}
Here $\mathsf{WellFormedRound}(S_\rho)$ requires
$\mathsf{DistinctInRange}_n(r)$ for every stabilizer $r$ of the round.

\subsection{QClifford syntax and execution}
\label{app:qclifford}
We define QClifford instructions and states, followed by their error-propagation and execution rules.
\[
 N\ge n>0,\quad q,u\in[N],\quad q\ne u,\quad
 \beta\in\{|0\rangle,|+\rangle\},
\]
\[
 g::=\kw H[q]\mid\kw{CNOT}[q,u]\mid q:=\beta\mid\kw{Meas}_Z(q),\quad
 i::=\kw{gate}(g)\mid\kw{err}(q),\quad\bar c::=\epsilon\mid i\pseq\bar c,
\]
\[
 \tgtstate=(\mathcal E,\omega,j_{\rm rec},\ell)\in\mathbb P^N\times\mathbb B^{\mathbb N}\times\mathbb N^2,
 \quad\cleanstate=(I^N,\mathbf0,0,0),\quad E_{\rm raw}=\mathcal E|_{[n]},\quad\mathcal E(q)=X^{x_q}Z^{z_q}.
\]

The state updates below track Pauli errors and measurement records through each instruction.
\begin{equation}
\begin{aligned}
 \mathsf{prop}_{\kw H[q]}(\tgtstate)&=\tgtstate[x_q\mapsto z_q,z_q\mapsto x_q],\\
 \mathsf{prop}_{\kw{CNOT}[q,u]}(\tgtstate)&=\tgtstate[x_u\mapsto x_u\oplus x_q,z_q\mapsto z_q\oplus z_u],\\
 \mathsf{prop}_{q:=\beta}(\tgtstate)&=\tgtstate[x_q\mapsto0,z_q\mapsto0]
       \quad(\beta\in\{|0\rangle,|+\rangle\}),\\
 \mathsf{prop}_{\kw{Meas}_Z(q)}(\tgtstate)&=
 \tgtstate[\omega\mapsto\omega[j_{\rm rec}\mapsto x_q],j_{\rm rec}\mapsto j_{\rm rec}+1],\\
 \mathsf{inj}_{q,X^aZ^b}(\tgtstate)&=
 \tgtstate[x_q\mapsto x_q\oplus a,z_q\mapsto z_q\oplus b,\ell\mapsto\ell+1]
       \quad((a,b)\ne(0,0)).
\end{aligned}
\end{equation}

The execution rules compose these updates, allowing each error instruction to inject one Pauli error or none.
\begin{mathpar}
 \inferrule{}{\tgtstate\xrightarrow{\kw{gate}(g)}\mathsf{prop}_g(\tgtstate)}
 \and
 \inferrule{}{\tgtstate\xrightarrow{\kw{err}(q)}\tgtstate}
 \and
 \inferrule{P\in\{X,Y,Z\}}{\tgtstate\xrightarrow{\kw{err}(q)}\mathsf{inj}_{q,P}(\tgtstate)}
 \and
 \inferrule{}{\langle\epsilon,\tgtstate\rangle\Downarrow\tgtstate}
 \and
 \inferrule{\tgtstate\xrightarrow{i}\tgtstate_1\quad
 \langle\bar c,\tgtstate_1\rangle\Downarrow\tgtstate_2}
 {\langle i\pseq\bar c,\tgtstate\rangle\Downarrow\tgtstate_2}
\end{mathpar}

\subsection{QClifford Hoare logic}
\label{app:target-logic}
The following definitions give the meaning of a QClifford Hoare triple and the rules for deriving it.
\[
 \varphi,\psi,\theta:\mathsf{State}_{\rm QC}\to\mathbb B,\qquad
 (\psi\circ F)(\tgtstate)=\psi(F(\tgtstate)),
\]
\[
 \models\{\varphi\}\bar c\{\psi\}\iff
 \forall\tgtstate,\tgtstate'.\ \varphi(\tgtstate)\land
 \langle\bar c,\tgtstate\rangle\Downarrow\tgtstate'\Rightarrow\psi(\tgtstate').
\]

\begin{mathpar}
 \inferrule*[right=\textsc{H-Skip}]{}{\vdash\{\varphi\}\epsilon\{\varphi\}}
 \and
 \inferrule*[right=\textsc{H-Gate}]{}
 {\vdash\{\psi\circ\mathsf{prop}_g\}\kw{gate}(g)\{\psi\}}
 \and
 \inferrule*[right=\textsc{H-Err}]{}
 {\vdash\{\psi\land\bigwedge_{P\in\{X,Y,Z\}}\psi\circ\mathsf{inj}_{q,P}\}
  \kw{err}(q)\{\psi\}}
 \and
 \inferrule*[right=\textsc{H-Seq}]
 {\vdash\{\varphi\}\bar c\{\theta\}\quad\vdash\{\theta\}\bar c'\{\psi\}}
 {\vdash\{\varphi\}\bar c\pseq\bar c'\{\psi\}}
 \and
 \inferrule*[right=\textsc{H-Conj}]
 {\vdash\{\varphi_1\}\bar c\{\psi_1\}\quad\vdash\{\varphi_2\}\bar c\{\psi_2\}}
 {\vdash\{\varphi_1\land\varphi_2\}\bar c\{\psi_1\land\psi_2\}}
 \and
 \inferrule*[right=\textsc{H-Conseq}]
 {\varphi\Rightarrow\varphi'\quad\vdash\{\varphi'\}\bar c\{\psi'\}\quad\psi'\Rightarrow\psi}
 {\vdash\{\varphi\}\bar c\{\psi\}}
\end{mathpar}

\[
 \vdash_F\{\varphi\}\bar c\{\psi\}\equiv\vdash\{\varphi\}\bar c\{\psi\}.
\]

The repetition rule assembles a finite sequence of indexed circuit blocks.
\begin{equation}
\inferrule*[right=\textsc{H-Repeat}]
 {\Gamma\vdash k\in\mathbb N\quad
  \Gamma,i\in\mathbb N,i<k\vdash_F
       \{\mathcal I(i)\}c(i)\{\mathcal I(i+1)\}}
 {\Gamma\vdash_F\{\mathcal I(0)\}
       \mathop{\mathsf{concat}}_{i<k}c(i)\{\mathcal I(k)\}}.
\end{equation}

\subsection{Circuit-level distance}
\label{app:distance}\label{sec:qcliff-ft}\label{sec:compiler-observations}
We define circuit-level distance using the backend's record predicate and, when required, detector silence.
\begin{definition}[Circuit-level code distance]\label{def:qcliff-ft}\label{def:dcirc}
\[
 K=(C,\mathsf{Legal},n,m,f,d,N,\bar c_K,\mathsf{RecordOK}_K),\quad
 \mathcal J=(J_a)_{a<v},\quad J_a\subseteq[M_K],
\]
where $\mathsf{Legal}(d)$ holds, $N\ge n(d)$, and $M_K$ counts the
$\kw{Meas}_Z$ instructions of $\bar c_K$. The predicate $\mathsf{Logical}$ and
the restriction $E_{\rm raw}$ are taken at $n(d)$, $m(d)$ and the stabilizers
$T_k$ of $C(d,k)$.
\[
 \mathsf{Silent}_{\mathcal J}(\omega):=\forall a<v.\ \bigoplus_{j\in J_a}\omega(j)=0,\qquad
 \mathsf{Accept}_{K,\mathcal J}(\tgtstate):=\mathsf{RecordOK}_K(\omega)\land\mathsf{Silent}_{\mathcal J}(\omega).
\]

\begin{equation}\label{eq:detector-failure}
\pccfitmath{
d^{\rm circ}_{K,\mathcal J}:=
\min\{\ell(\tgtstate'):\
\langle\bar c_K,\cleanstate\rangle\Downarrow\tgtstate',\
\mathsf{Accept}_{K,\mathcal J}(\tgtstate')\land\mathsf{Logical}(E'_{\rm raw})\},
\qquad\min\varnothing=\infty.
}
\end{equation}

\[
 d^{\rm circ}_K:=\min\{\ell(\tgtstate'):
  \langle\bar c_K,\cleanstate\rangle\Downarrow\tgtstate',\
  \mathsf{RecordOK}_K(\omega')\land\mathsf{Logical}(E'_{\rm raw})\},
\]
\[
 \mathsf{Accept}_d=\mathsf{Accept}_{K_d,\mathcal J_d},\qquad
 \mathsf{LogFail}_d=\mathsf{Logical}(E_{\rm raw}).
\]
\end{definition}

\section{Surface code proof derivations}
\label{app:surface-derivations}
The following derivations assemble the Surface code cut invariant and its compiled circuit-distance guarantee.
\[
 \Gamma_{\rm surf}=\{s\ge1,d=2s+1\},\quad
 n(d)=d^2,\quad m(d)=d^2-1,\quad f(d)=1,\quad R(d)=d,\quad B(d)=d-1.
\]
\[
 \mathcal G:\ S_d=\mathsf{Generate}\llbracket C_{\rm surf}\rrbracket(d;d,d-1),\qquad
 \bar c_d=\circuitcompile{\mathsf{Std}}{S_d}
            =\mathop{\mathsf{concat}}_{h<m(d)R(d)}\bar c_{d,h}.
\]
\[
 \mathcal P_z:\Gamma_{\rm surf}\vdash\mathsf{HookBound}_{L_z,Z}(S_d),\qquad
 \mathcal P_x:\Gamma_{\rm surf}\vdash\mathsf{HookBound}_{L_x,X}(S_d),
\]
\[
 V:\Gamma_{\rm surf}\vdash_C\mathsf{WellFormed},\qquad
 W:\Gamma_{\rm surf}\vdash_C\mathsf{Complete}_{L_x,L_z}.
\]

First, the hook proofs establish the QStab cut invariant by the source rules of
Figure~\ref{fig:qhl-rules}.
\begin{equation}\label{eq:source-invariant-induction}
\pccfitmath{
\inferrule*[right=\textsc{Src-Conj}]
 {\inferrule*[right=\textsc{Src-Cut}]
   {\mathcal P_z:\ \Gamma_{\rm surf}\vdash\mathsf{HookBound}_{L_z,Z}(S_d)}
   {\Gamma_{\rm surf}\mid S_d\vdash\Box(\wt[d]{\chi_{L_z,Z}(E)}\le\ell)}
  \qquad
  \inferrule*[right=\textsc{Src-Cut}]
   {\mathcal P_x:\ \Gamma_{\rm surf}\vdash\mathsf{HookBound}_{L_x,X}(S_d)}
   {\Gamma_{\rm surf}\mid S_d\vdash\Box(\wt[d]{\chi_{L_x,X}(E)}\le\ell)}}
 {\mathcal D_{\rm inv}:\quad
  \Gamma_{\rm surf}\mid S_d\vdash\Box I_d(E,\ell)}
}
\end{equation}
\Rule{Src-Cut} is sound by induction on execution: $I_d(I^{n(d)},0)$ holds
because both cut signatures are zero, and
Equation~\eqref{eq:cost-preservation} shows that every step
$\statecfg\xrightarrow{k}_{S_d}\statecfg'$ preserves $I_d$.
\Rule{Src-Dist} then adds well-formedness and completeness.
\begin{equation}\label{eq:source-to-headline}
\pccfitmath{
\mathcal D=\textsc{Src-Dist}(\mathcal D_{\rm inv},V,W):\quad
\Gamma_{\rm surf}\mid S_d\vdash
  \Box\bigl(\ell<d\Rightarrow\neg\mathsf{Logical}(E)\bigr).
}
\end{equation}

Second, the Standard Scheme propagation lemmas carry the same hook proofs
to every compiled block. Section~\ref{ex:surface-gate-proof} derives the
first block at $d=3$; the supplement's pending-contribution argument proves
the triple for an arbitrary ordered stabilizer whose suffix signatures
satisfy the hook bound. The generation proof $\mathcal G$ supplies the
ordered operands of block $h$.
\begin{equation}\label{eq:compiled-block}
\pccfitmath{
\frac{\mathcal G\qquad\mathcal P_z\qquad\mathcal P_x}
 {\Gamma_{\rm surf},\ h<m(d)R(d)\vdash_F
       \{\Phi_d\}\,\bar c_{d,h}\,\{\Phi_d\}}.
}
\end{equation}

The following induction composes the compiled syndrome-extraction blocks
of Equation~\eqref{eq:compiled-block} by \textsc{H-Repeat}.
\begin{equation}\label{eq:compiled-invariant-induction}
\pccfitmath{
\frac{\Phi_d(\cleanstate(d))\qquad
 \Gamma_{\rm surf},\ h<m(d)R(d)\vdash_F
       \{\Phi_d\}\,\bar c_{d,h}\,\{\Phi_d\}}
 {\Gamma_{\rm surf}\vdash_F
       \{\tgtstate=\cleanstate(d)\}\,\bar c_d\,\{\Phi_d\}}.
}
\end{equation}

Finally, cut completeness turns the compiled invariant into the circuit-distance guarantee.
\begin{equation}\label{eq:invariant-implies-distance}
\pccfitmath{
\Phi_d\land\mathsf{Logical}(E_{\rm raw})
\ \Longrightarrow\
d=\wt{\chi_x(E_{\rm raw})}\le\ell
\ \lor\ d=\wt{\chi_z(E_{\rm raw})}\le\ell
\ \Longrightarrow\ d\le\ell.
}
\end{equation}
This implication is the target reading of $\mathcal D$: it reuses $V$ and
$W$ and gives the target certificate $\mathcal D'$ of
Theorem~\ref{thm:standard-compile-preserve}, which proves
$A_d$ at budget $d-1$. \textsc{H-Conseq} weakens the postcondition.

\begin{equation}\label{eq:complete-family-derivation}
\pccfitmath{
\frac{
 \underbrace{\Gamma_{\rm surf}\vdash_F
  \{\tgtstate=\cleanstate(d)\}\,\bar c_d\,\{\Phi_d\}}
  _{\text{initialization and induction: Equation~\eqref{eq:compiled-invariant-induction}}}
 \qquad V:\ \Gamma_{\rm surf}\vdash_C\mathsf{WellFormed}
 \qquad W:\ \Gamma_{\rm surf}\vdash_C\mathsf{Complete}_{L_x,L_z}
}{
 \mathcal H_{\mathsf{Std}}:\quad\Gamma_{\rm surf}\vdash_F
 \{\tgtstate=\cleanstate(d)\}\;
 \underbrace{\circuitcompile{\mathsf{Std}}{
   \underbrace{\mathsf{Generate}\llbracket C_{\rm surf}\rrbracket(d;d,d-1)}
              _{S_d:\ \text{QStab}}
 }}_{\bar c_d:\ \text{QClifford}}\;
 \{\mathsf{Accept}_d(\tgtstate)\land\mathsf{LogFail}_d(\tgtstate)
                 \Rightarrow d\le\ell(\tgtstate)\}
}.
}
\end{equation}

\clearpage
\section{Code-family geometries}
\label{app:family-geometries}

\begin{figure}[!p]
\centering
\begin{figureframe}

\end{figureframe}
\caption{Code geometries and proof measures. Panel (b) uses the standard
edge-qubit Surface code layout; blue marks an X star and red shades a Z
plaquette. The colored loops in (c) are one X cut and one Z cut on the
torus; the two indexed grids show all four collections. In (d), a cut
is the Pauli product on the selected dots, not the connecting guide line.
The enlarged cut uses seed $s^0=1001011$ in column 2 of block 0. Larger
$t$ repeats each seven-site pattern. Panel (e) uses no cut collection:
red dots are Z errors, and multiplication by the face stabilizer reduces
their weight. All bounds retain the orders and contracts of
Table~\ref{tab:intro-results}.}
\label{fig:family-geometries}
\Description{Five stacked panels show a rotated Surface patch, an unrotated
Surface lattice, a three-dimensional torus, two Cartesian product blocks,
and a triangular honeycomb patch. The first four panels visualize two,
two, four, and twelve cut collections, respectively. The color-code panel
shows reduction of a five-qubit hook to one qubit modulo a face stabilizer.}
\end{figure}

\emph{What must we prove again when the code family has a different geometry
or more logical qubits?} Figure~\ref{fig:family-geometries} shows how the
Surface code argument extends. Panel (a) retains the columns and rows of
Figure~\ref{fig:pcc}(a). Panel (b) places the unrotated Surface code's
data qubits on lattice edges, with X stars and Z plaquettes~\cite{horsman2012surface}.
Its cuts use rows and columns of vertical-edge qubits (block $A$);
horizontal-edge qubits (block $B$) belong to no counted cut.
Panel (c) unwraps the Toric code's two periodic edge blocks to show
four collections. Panel (d) replaces complete rows and columns by
seed-selected subsets, giving twelve algebraic cut collections for LP code.
Each of these four families supplies a new proof that every logical residual crosses enough
cuts in some collection, and each ordered extraction program must establish
its one-error bounds. The invariant induction and structural certificate-translation rules
are reused. Under the declared contracts, unrotated Surface code and
Toric code attain $d$, and Cartesian LP code attains $4t$. Toric code uses four
cut collections for two logical qubits, and Cartesian LP code uses twelve for
eighteen. Code algebra identifies logical residuals, while extraction proofs
bound error effects.

\paragraph{What is a cut in Cartesian LP code?}
As in Figure~\ref{fig:pcc}(a), a cut is a Pauli string used to probe an
error's parity; its support may contain disconnected qubits. In
Figure~\ref{fig:family-geometries}(d), write $(b,i,j)$ for a qubit's block,
row and column. At $t=1$, the first seed $s^0=1001011$ selects rows
$0,3,5,6$. In column two of block zero this gives
$\kappa_2=Z_{0,0,2}Z_{0,3,2}Z_{0,5,2}Z_{0,6,2}$.
An X error at $(0,0,2)$ crosses this cut; adding X at $(0,3,2)$ cancels
its parity. Varying the column gives seven cuts in one collection.
The highlighted X check meets $\kappa_2$ twice. Its six slots contribute
$(e_2,0,e_2,0,0,0)$, where $e_2$ marks only cut $\kappa_2$;
every suffix therefore crosses at most one cut in this collection.
Three seeds (reversed for the X probes), two blocks and two probe Paulis give twelve collections.
At general $t$, each collection has $7t$ cuts. A separate code-algebra
proof shows that every logical residual crosses at least $4t$ cuts in
some collection; this is the threshold used with the one-error bound.

\paragraph{How does reduced weight establish the color-code bound?}
Panel (e) uses reduced weight in place of a cut collection. For this self-dual, one-logical-qubit
family, any two nontrivial pure-X logical supports have odd overlap;
the same holds for pure-Z supports. A disjoint collection of centralizing
probes therefore contains at most one nontrivial logical probe.
To obtain a bound that scales with $d$, we minimize error weight over
equivalent representatives modulo stabilizers~\cite{bombin2007optimal,gidney2024cultivation}. On the displayed
face, the five-qubit hook $H=Z_1Z_2Z_3Z_4Z_5$ satisfies
$H(Z_0Z_1Z_2Z_3Z_4Z_5)=Z_0$; its reduced weight is at most one.
More generally, a hook on $s$ qubits of a weight-$w$ face has reduced
weight at most $\min(s,w-s)\le3$, since $w\le6$.
Thus $\ell$ injected errors have reduced weight at most $3\ell$, whereas
a logical residual has reduced weight at least $d$. This proves the
lower bound $\lceil d/3\rceil$ under the declared contract. The purple
boundary shows a logical operator's support.

\clearpage
\section{Detailed evaluation results}
\label{app:evaluation-results}
\begin{table}[!htbp]
\centering\small
\setlength{\tabcolsep}{2pt}
\renewcommand{\arraystretch}{0.90}
\setlength{\aboverulesep}{.6pt}
\setlength{\belowrulesep}{.8pt}
\captionsetup{skip=2pt}
\caption{Certified distance guarantees and Standard Scheme checking times.}
\label{tab:intro-results}
\begin{tabular}{@{}>{\raggedright\arraybackslash}p{126pt}lccccc@{}}
\toprule
Code & Parameters & \shortstack{Standard\\Scheme} & \shortstack{Flag\\Scheme} & \shortstack{Shor\\Scheme} & \shortstack{Knill\\Scheme} & \shortstack{Std. check\\(s)} \\
\midrule
\multicolumn{7}{l}{\emph{Family results}} \\
Rotated Surface code & $[[d^2,1,d]]$ & $d$ & $d$ & $d$ & $d$ & 214 \\
Unrotated Surface code (HGP) & $[[d^2+(d-1)^2,1,d]]$ & $d$ & $d$ & $d$ & $d$ & 23.9 \\
Cartesian LP code (HGP) & $[[98t^2,18,4t]]$ & $4t$ & $4t$ & $4t$ & $4t$ & $3692^\ddagger$ \\
Toric code & $[[2d^2,2,d]]$ & $d$ & $d$ & $d$ & $d$ & 9.48 \\
Triangular color code & $[[(3d^2+1)/4,1,d]]$ & $\ge\lceil d/3\rceil$ & $\ge\lceil d/3\rceil$ & $\ge\lceil d/3\rceil$ & $\ge\lceil d/3\rceil$ & $5511^\dagger$ \\
\midrule
\multicolumn{7}{l}{\emph{Translated finite source certificate}} \\
Shor code, interleaved & $[[9,1,3]]$ & $3$ & $3$ & $3$ & $3$ & 105 \\
\midrule
\multicolumn{7}{l}{\emph{Finite target certificates}} \\
Five-qubit code & $[[5,1,3]]$ & $2$ & $3$ & $3$ & $2$ & 0.304 \\
Steane code (finite order) & $[[7,1,3]]$ & $2$ & $3$ & $3$ & $2$ & 0.382 \\
Cube code & $[[8,3,2]]$ & $1$ & $2$ & $2$ & $1$ & 0.285 \\
Shor code, block order & $[[9,1,3]]$ & $1$ & $2$ & $2$ & $1$ & 0.209 \\
Toric code (fixed order) & $[[18,2,3]]$ & $3$ & $3$ & $3$ & $3$ & 0.424 \\
Bivariate bicycle code & $[[18,4,4]]$ & $2$ & $4$ & $4$ & $2$ & 0.344 \\
Cyclic-product HGP code & $[[72,8,4]]$ & $4$ & $4$ & $4$ & $4$ & 0.536 \\
Cyclic-simplex HGP code & $[[98,18,4]]$ & $4$ & $4$ & $4$ & $4$ & 0.506 \\
\bottomrule
\end{tabular}
\par\smallskip{\footnotesize Timings average three completed measurements;
$^\ddagger$ averages two, and $^\dagger$ reports one.\par}
\end{table}

Each single-wire Pauli injection costs one error (\S~\ref{sec:qec-primer}).
Serial stabilizer measurements use fresh ancilla qubits and $d$ rounds for
families ($t$ for LP codes), or one pass with zero readout flips for finite
cases. Standard Scheme and Flag Scheme families admit all records.
For all native family certificates, the Shor Scheme requires zero edge flips
and zero XOR in each cat-readout group, and the Knill Scheme requires zero
resource and Bell flips. These are the complete benchmark predicates,
stronger than the Surface code correction-erasing conditions in
\S~\ref{sec:trust}. The Toric code also has a separate formal exactness
theorem admitting all records in every scheme. These conditions constrain
the record of relative readout flips.

\paragraph{Reading the table.}
We use the standard notation
$[[n,k,d]]$ for a code with $n$ data qubits,
$k$ logical qubits and distance
$d$~\cite{nielsen_chuang_2010,gottesman1997stabilizer,huang2025efficient}.
HGP denotes a hypergraph-product construction.
Rotated Surface and triangular color code families use odd $d\ge3$,
unrotated Surface code uses $d\ge2$, Toric code uses $d\ge3$, and
Cartesian LP code uses $t\ge1$.
Table~\ref{tab:intro-results} reports $d^{\rm circ}_K$ from
Definition~\ref{def:dcirc}. Equality in the
detector-silent setting additionally requires a matching silent witness.
The two Surface code rows use the canonical formal programs for their
four-scheme equalities. The printed rotated Surface code order additionally
has the Standard Scheme equality in \S~\ref{sec:standard-exactness}.
Their timings check native lower certificates.
The family benchmarks check symbolic certificates. The interleaved Shor
experiment translates a supplied source certificate using the supported
structural rules, while the finite target benchmarks check
supplied target certificates directly.
The native Surface package writes the same ordered schedule with the
recursive family body of the supplement and fuel $s+1$.
The timed stages
are lower-proof checking, supplied upper-witness checking for finite cases,
and circuit binding for finite targets. Circuit/proof emission,
verification-condition generation, witness search and Lean builds are
performed separately.

\paragraph{Combining the images.}
After the checker of Section~\ref{sec:verifier} recomputes the
single-injection images, backward-flow exactness
and linearity combine the images by XOR, including readout cancellations,
to establish the assertion for every execution within the declared budget.
The intermediate Hoare assertions in Figure~\ref{fig:compiled-hook-proof}
explain this reasoning. The emitted finite certificate stores image evidence
and structural proof nodes. Its expanded circuit and image table grow with $d$.

\paragraph{Comparison protocol.}
Each search instance is built in five steps. First, our tool instantiates
the family's symbolic source program at distance $d$, giving the ordered
QStab program; this is the same source file that the timed family
certificate is checked against. For the triangular color code, the ordered
stabilizers come from a generator that agrees with the CodeML definition at
every tested size. Second, our Standard Scheme compiler emits the QClifford
circuit. Third, we translate the QClifford gate list gate by gate into
Stim~\cite{gidney2021stim} operations and let Stim propagate every
pre-operation X, Y and Z injection on every operand to the end of the
circuit, which gives the final data error and the readout flips of each
single injection. Fourth, each injection is summarized by the syndrome of
its final data error against the code's stabilizers and by its parities
against a complete basis of logical operators; readout flips impose no
constraint, matching the unrestricted readouts of the Standard Scheme
contract. Identical summaries are merged, and the summaries form a check
matrix and a logical matrix. Fifth, the published encoder of
codeDistance~\cite{webster2026distancefinding} turns the question of the
fewest injections with zero syndrome and nonzero logical parity into a MaxSAT
problem, which two PySAT algorithms solve on Glucose 4: linear search (LSU),
which is the package default, and core-guided RC2. The optimum is mapped
back to concrete injections and replayed in Stim as one combined execution,
which must have exactly that many injections, zero syndrome and a
nontrivial logical residual. At the smallest instance of each family, every
Stim single-injection image also equals the image computed by our checker.
The Standard circuits use $d$
rounds for unrotated and rotated Surface codes and for the triangular color
code, and $t=d/4$ rounds for Cartesian
LP codes. Total search time includes process startup, program instantiation,
compilation, propagation, logical-basis construction, encoding, search and
witness replay. Each completed optimum is replayed as a combined execution.
Every instance runs once, in a fresh single-threaded process with a
7,200\,s limit, and each series ends at its first unsolved instance. The limit
exceeds every family-proof total. The searches ran as high-priority
processes on the evaluation host, with up to eight in parallel.
The first three family-proof totals in Figure~\ref{fig:distance-scaling} sum certificate
compilation and verification: 27.3\,s, 219.5\,s and 3715.8\,s, respectively.
The color-code panel shows the table's verification-only time of 5511\,s.
Table~\ref{tab:intro-results} reports verification alone. The reported means use three completed
runs for each family except Cartesian LP, which uses two completed runs;
its third attempt exhausted memory. The triangular color-code timing is one run.

LSU completes unrotated Surface up to $d=11$ and times out at $d=13$; rotated Surface up to $d=15$ and times out at $d=17$; Cartesian LP up to $d=4$ and times out at $d=8$; triangular color up to $d=17$ and times out at $d=19$.
RC2 completes unrotated Surface up to $d=11$ and times out at $d=13$; rotated Surface up to $d=13$; Cartesian LP up to $d=4$ and times out at $d=8$; triangular color up to $d=15$ and times out at $d=17$.
Every completed search for the Surface and LP families returns distance $d$. For the triangular color code the searches return 2, 3, 4, 5, 6, 7, 8, 9 at $d=3,5,7,9,11,13,15,17$, that is $(d+1)/2$, the exact circuit-level distance of this Standard schedule; it lies above the certified family lower bound $\lceil d/3\rceil$. Both configurations agree on every instance that both complete. The RC2 search for rotated Surface $d=15$ was stopped unsolved after 4518\,s, before the limit. LSU is the faster configuration on the larger instances, and Table~\ref{tab:baseline-measurements} lists every measurement.

\begin{table}[!htbp]
\centering\small
\setlength{\tabcolsep}{3.2pt}
\captionsetup{skip=2pt}
\caption{Wall time in seconds of every exact-distance search in
Figure~\ref{fig:distance-scaling}. A dash marks an instance that was not run
because its series had ended; $^\ast$stopped unsolved before the 7,200\,s limit.}
\label{tab:baseline-measurements}
\begin{tabular}{@{}llrrrrrrrrr@{}}
\toprule
Family & Search & $d=3$ & $d=5$ & $d=7$ & $d=9$ & $d=11$ & $d=13$ & $d=15$ & $d=17$ & $d=19$ \\
\midrule
Unrotated Surface & LSU & 1.21 & 1.63 & 6.18 & 31.3 & 493 & $>$7200 & -- & -- & -- \\
 & RC2 & 0.82 & 1.78 & 6.86 & 81.0 & 3970 & $>$7200 & -- & -- & -- \\
Rotated Surface & LSU & 0.89 & 1.81 & 2.93 & 8.00 & 30.3 & 174 & 1989 & $>$7200 & -- \\
 & RC2 & 0.88 & 1.39 & 3.10 & 22.7 & 364 & 6759 & $>$4518$^\ast$ & -- & -- \\
Triangular color & LSU & 0.80 & 1.04 & 2.16 & 4.93 & 13.1 & 51.3 & 162 & 1051 & $>$7200 \\
 & RC2 & 0.87 & 1.00 & 2.20 & 5.35 & 15.7 & 83.6 & 2036 & $>$7200 & -- \\
\quad distance found & & 2 & 3 & 4 & 5 & 6 & 7 & 8 & 9 & -- \\
\midrule
 &  & $d=4$ & $d=8$ &  &  &  &  &  &  &  \\
Cartesian LP & LSU & 5.74 & $>$7200 &  &  &  &  &  &  &  \\
 & RC2 & 7.69 & $>$7200 &  &  &  &  &  &  &  \\
\bottomrule
\end{tabular}
\end{table}

\clearpage
\label{page:appendix-end}

\fi

\ifdefined\PCQECSupplement
\input{appendices.tex}
\fi

\end{document}